\documentclass[acmsmall]{acmart}

\newcommand{\asplossubmissionnumber}{245}

\usepackage[normalem]{ulem}

\graphicspath{{figs/}}
\usepackage{xspace}
\usepackage{amsmath}
\usepackage{verbatimbox}
\usepackage{tikz}
\usepackage{cprotect}
\usepackage{bussproofs}
\usepackage{pdfpages}
\usepackage{svg}
\usepackage{newfloat}
\usepackage{siunitx}
\usepackage{listings}
\usepackage{makecell}
\usepackage{courier}
\usepackage{stmaryrd}
\usepackage[caption=false]{subfig}
\usepackage{xcolor}
\usepackage{soul}
\usepackage[skip=5pt]{caption}

\DeclareFloatingEnvironment[
    fileext=lox,
    listname={List of Rules},
    name=Rule,
    placement=tbhp,
    within=section,
]{rulefigure}
\newcommand{\refrule}[1]{Rule \ref{#1}}

\newcommand{\colorlang}{\textsc{CoolerSpace}\xspace}

\newcommand{\onnx}{\textsc{ONNX}\xspace}

\newif\ifshowptitles
\showptitlestrue

\newcommand{\revnew}[1]{#1}

\ifx\figurename\undefined \def\figurename{Figure}\fi
\renewcommand{\figurename}{Fig.}
\renewcommand{\paragraph}[1]{\textbf{#1} }

\newcommand{\Sect}[1]{Sec.~\ref{#1}}

\newcommand{\Fig}[1]{Fig.~\ref{#1}}
\newcommand{\Tbl}[1]{Tbl.~\ref{#1}}
\newcommand{\Equ}[1]{Equ.~\ref{#1}}

\newcommand{\Prog}[1]{Prog.~\ref{#1}}

\newcommand{\specialcell}[2][c]{\begin{tabular}[#1]{@{}l@{}}#2\end{tabular}} % change l to c for centering

\newcommand{\prog}[1]{\textsc{#1}\xspace}

\newcommand{\no}[1]{#1}
\renewcommand{\no}[1]{}
\newcommand{\RNum}[1]{\uppercase\expandafter{\romannumeral #1\relax}}

\newcommand{\Gammahas}{\Gamma \vdash}

\newcommand{\matmultext}{\textsf{matmul}}
\newcommand{\channelcounttext}{\textsf{channel\_count}}
\newcommand{\pathexiststext}{\textsf{path\_exists}}
\newcommand{\broadcastabletext}{\textsf{broadcastable}}

\newcommand{\mixtext}{\textsf{mix}}

\newcommand{\spectrumtext}{\text{spectrum}}
\newcommand{\lighttext}{\text{Light}}
\newcommand{\reflectancetext}{\text{Reflectance}}
\newcommand{\absorptiontext}{\text{Absorption}}
\newcommand{\scatteringtext}{\text{Scattering}}

\newcommand{\colortext}{\text{color}}
\newcommand{\tristimulustext}{\text{tristimulus}}
\newcommand{\srgbtext}{\text{sRGB}}
\newcommand{\oprgbtext}{\text{opRGB}}
\newcommand{\xyztext}{\text{XYZ}}
\newcommand{\lmstext}{\text{LMS}}

\newcommand{\perceptualtext}{\text{perceptual}}
\newcommand{\labtext}{\text{LAB}}
\newcommand{\hsvtext}{\text{HSL}}

\newcommand{\matrixtext}{\text{Matrix}}
\newcommand{\pigmenttext}{\text{Pigment}}
\newcommand{\chromaticitytext}{\text{Chromaticity}}

\newcommand{\addtext}{\textsf{add}}
\newcommand{\subtext}{\textsf{sub}}
\newcommand{\multext}{\textsf{mul}}
\newcommand{\divtext}{\textsf{div}}

\newcommand{\powtext}{\textsf{pow}}

\newcommand{\lighttype}{\tau_\lighttext}
\newcommand{\reflectancetype}{\tau_\reflectancetext}
\newcommand{\absorptiontype}{\tau_\absorptiontext}
\newcommand{\scatteringtype}{\tau_\scatteringtext}

\newcommand{\colortype}{\tau_\colortext}
\newcommand{\tristimulustype}{\tau_\tristimulustext}
\newcommand{\srgbtype}{\tau_\srgbtext}
\newcommand{\oprgbtype}{\tau_\oprgbtext}
\newcommand{\xyztype}{\tau_\xyztext}
\newcommand{\lmstype}{\tau_\lmstext}

\newcommand{\perceptualtype}{\tau_\perceptualtext}
\newcommand{\labtype}{\tau_\labtext}
\newcommand{\hsvtype}{\tau_\hsvtext}

\newcommand{\matrixtype}{\tau_\matrixtext}
\newcommand{\pigmenttype}{\tau_\pigmenttext}
\newcommand{\chromaticitytype}{\tau_\chromaticitytext}

\setcopyright{rightsretained}
\acmDOI{10.1145/3689741}
\acmYear{2024}
\acmJournal{PACMPL}
\acmVolume{8}
\acmNumber{OOPSLA2}
\acmArticle{301}
\acmMonth{10}
\acmSubmissionID{oopslab24main-p354-p}
\received{2024-04-05}
\received[accepted]{2024-08-18}

\begin{document}

% enable page numbers
\settopmatter{printfolios=true}

\title{\colorlang: A Language for Physically Correct and Computationally Efficient Color Programming}

\author{Ethan Chen}
\orcid{0009-0008-1250-769X}
\affiliation{%
  \institution{University of Rochester}
  \city{Rochester}
  \country{USA}
}
\email{echen48@ur.rochester.edu}

\author{Jiwon Chang}
\orcid{0000-0003-3945-925X}
\affiliation{%
  \institution{University of Rochester}
  \city{Rochester}
  \country{USA}
}
\email{jchang38@ur.rochester.edu}

\author{Yuhao Zhu}
\orcid{0000-0002-2802-0578}
\affiliation{%
  \institution{University of Rochester}
  \city{Rochester}
  \country{USA}
}
\email{yzhu@rochester.edu}

\date{}

% Abstract must be placed before \maketitle under this template for some reason 
\begin{abstract}
    Color programmers manipulate lights, materials, and the resulting colors from light-material interactions.
Existing libraries for color programming provide only a thin layer of abstraction around matrix operations.
Color programs are, thus, vulnerable to bugs arising from mathematically permissible but physically meaningless matrix computations.
Correct implementations are difficult to write and optimize.
 % TODO
We introduce \colorlang to facilitate physically correct and computationally efficient color programming.
\colorlang raises the level of abstraction of color programming by allowing programmers to focus on describing the logic of color physics. 
Correctness and efficiency are handled by \colorlang.
The type system in \colorlang assigns physical meaning and dimensions to user-defined objects.
The typing rules permit only legal computations informed by color physics and perception.
Along with type checking, \colorlang also generates performance-optimized programs using equality saturation.
\colorlang is implemented as a Python library and compiles to \onnx, a common intermediate representation for tensor computations.
\colorlang not only prevents common errors in color programming, but also does so without run-time overhead:
even unoptimized \colorlang programs out-perform existing Python-based color programming systems by up to 5.7 times;
our optimizations provide up to an additional 1.4 times speed-up.

\end{abstract}

\begin{CCSXML}
<ccs2012>
<concept>
<concept_id>10011007.10011006.10011050.10011017</concept_id>
<concept_desc>Software and its engineering~Domain specific languages</concept_desc>
<concept_significance>500</concept_significance>
</concept>
<concept>
<concept_id>10010147.10010371</concept_id>
<concept_desc>Computing methodologies~Computer graphics</concept_desc>
<concept_significance>500</concept_significance>
</concept>
</ccs2012>
\end{CCSXML}

\ccsdesc[500]{Software and its engineering~Domain specific languages}
\ccsdesc[500]{Computing methodologies~Computer graphics}

\keywords{language design, color science, type systems}

\maketitle % should come after the abstract
\pagestyle{plain} % should come right after \maketitle

\section{Introduction}
\label{sec:intro}

Color programming broadly refers to the programmatic manipulation of lights, materials (e.g. pigments), and the resulting color of light-material interactions.
Color programming is fundamental to almost every domain of art, science, and engineering.
Imaging and display technologies are, in essence, about capturing and reproducing colors~\cite{sharma2017color, miller2019color, rowlands2017physics};
computer graphics simulate light-material interaction and color capturing in cameras~\cite{pharr2023physically}; 
%color is central in effective visualization and communication of complex information~\cite{ware2019information};
artists use vibrant palettes of colors, both real and digital, to create their works~\cite{sochorova2021practical}, while art conservators analyze and preserve the original pigments in historical pieces~\cite{berns2016color, johnston2001color}.
%neuroscientists study how light interaction with photopigments on retinal gives rise to our color perception~\cite{wandell2022visual, isetbio};
%in geology the diverse colors of minerals provide vital clues about the Earth's history~\cite{tilley2020colour}, whereas visual ecologists study how animals use colors and lights for natural tasks and behaviors~\cite{cronin2014visual, johnsen2012optics};
%the interplay of light and material is critical to the textile/dye/printing industry in faithfully producing desired colors~\cite{broadbent2001basic, christie2014colour, billmeyer1984textbook}.

Color programmers must follow the rules of physics governing light-material interaction and the standards of different color encodings (\Sect{sec:background}).
The languages (e.g., Python) and libraries (NumPy~\cite{harris2020array} and OpenCV~\cite{opencv_library}) they use, however, are physics-agnostic: they, in large part, provide only a thin wrapper around raw tensor operations.
The physical meanings of objects (e.g., color, light power spectrum, material scattering spectrum) are not tracked.
Thus, programmers are prone to accidentally writing mathematically permissible but physically meaningless or incorrect code.
Physically correct code can be time consuming to implement and are not always computationally efficient (\Sect{sec:motivation}).

%To accelerate this difficult task, 
%Programmers often use color programming libraries to abstract the complexities of color \cite{colorjs,pycolor}. However, existing libraries typically provide only a thin wrapper around underlying vectors and matrices. This means matrices with the same dimension that represent different color encodings or entirely different physical concepts are indistinguishable. Programmers are responsible for ensuring that any specified operations in their code are physically meaningful. 

%They also need to write performant code, since imaging and graphics applications deal with large datasets, sometimes in real-time. 

%In short, programmers must juggle the complexities of light, material, color, and pigments to write bug-free and performant code. To accelerate this difficult task, programmers often utilize color programming libraries to abstract the complexities of color \cite{colorjs,pycolor}. However, existing libraries typically provide only a thin wrapper around underlying vectors and matrices. This means matrices with the same dimension that represent different color encodings or entirely different physical concepts are indistinguishable. Programmers are responsible for ensuring that any specified operations in their code are physically meaningful. 

We propose \colorlang to facilitate physically correct and computationally efficient color programming (\Sect{sec:ov}).
The core of \colorlang is a type system (\Sect{sec:lang}), which raises the level of abstraction of color programming from tensors to physical objects, such as lights, materials, and colors.
The domain-specific typing rules, which are statically checked, permit only physically-based or perceptually accurate computations.

% \colorlang is not the first type system to codify physical units as types.
% Previous research has applied type theory to other physical units of measurement ~\cite{allen_object-oriented_2004, karr_incorporation_1978, dreiheller_programming_1986}.
% \colorlang differs from these prior works, as \fixme{intro is a really bad place to put this, the differences are nuanced and can't be easily explained here. I propose next paragraph instead.}

% There are interesting parallels and differences between \colorlang's type system and those of prior works.
% We discuss these differences in detail in \Sect{sec:type_system_design_decisions} and \Sect{sec:related}.

%To address these limitations, we propose that color, light and related objects should be typed. This allows us to employ applied type theory conveniences. The additional layer of abstraction helps programmers write clear and correct code. Types also allows for more thorough and domain-specific type-checking and optimization. We choose to extend Python due to its ubiquity in color science. 

In addition to avoiding common errors, the higher level of programming abstraction also frees programmers from the burden of efficiently implementing color science algorithms.
Instead, a \colorlang program is translated to a semantically-equivalent set of tensor algebra operations (\Sect{sec:translation}), which are then optimized using equality saturation~\cite{tensat,tate2009equality} (\Sect{sec:optimization}). Translation is guided by formal translational semantics that are provably type sound.

%We also implement a framework for high-level optimization of color programs by reducing color \& light operations to tensor algebra. We then utilize a modified version of the SOTA tensor algebra optimizer to optimize the resulting program \cite{tensat,tate2009equality}.

\colorlang assists any programmers working with color and light, and can be particularly useful for a significant scientific community that might not intersect with the conventional CS community: color scientists and vision scientists.
These are domain experts who write programs to directly manipulate light and color data.
They must do so in a physically accurate way.
However, researchers in these fields are less familiar with modern programming techniques (e.g., type checking, performance optimizations). They are exactly the population \colorlang can help.

We implement \colorlang as a Python library, as color and vision scientists predominantly use high-level languages like Python and MATLAB. 
The Python program is compiled to \onnx~\cite{onnx}, an intermediate representation for tensor algebra.
The \onnx program is then executed using \onnx Runtime~\cite{onnxruntime}.
We show that \colorlang can express common algorithms in color programming and prevent common errors without runtime overhead.
Unoptimized \colorlang programs out-perform existing Python color programming systems by up to 5.6 times. \colorlang is capable of further optimizing its programs by up to 1.4 times.
% and improve performance by up to 1.4 times (up to 5.6 times compared to existing Python libraries). \fixme{Rephrase later to match abstract}

The entire \colorlang system, along with programs developed using \colorlang, will be made open-source.
Our specific contributions are as follows.
\begin{enumerate}
    \item We demonstrate a class of bugs in color programs where programmers write mathematically permissible but physically incorrect computations. 
    \item We design a type system specific to color programming. The type system codifies and enforces the fundamental principles of color physics. 
    \item We introduce \colorlang, which implements the type system and automatically generates efficient color programs through tensor algebra optimizations. 
    \item We experimentally show that \colorlang prevents common bugs in color programming while providing up to 5.6 $\times$ speed-ups against existing Python color programming libraries.
    \item We also experimentally show that \colorlang's optimizer provides an additional 1.4 $\times$ speed-up to our compiled programs.
\end{enumerate}

\section{Background: Lights, Colors, and Materials}
\label{sec:background}

Color programming involves manipulating lights, materials, and different representations of color.
%Color programming is prevalent in a number of important application domains such as image processing/computational photography, computer graphics, chemistry, and geology.
This section provides the necessary scientific background.

\begin{figure*}[t]
  \centering
  \subfloat[\small{Color perception.}]
  {
  \includegraphics[height=1in]{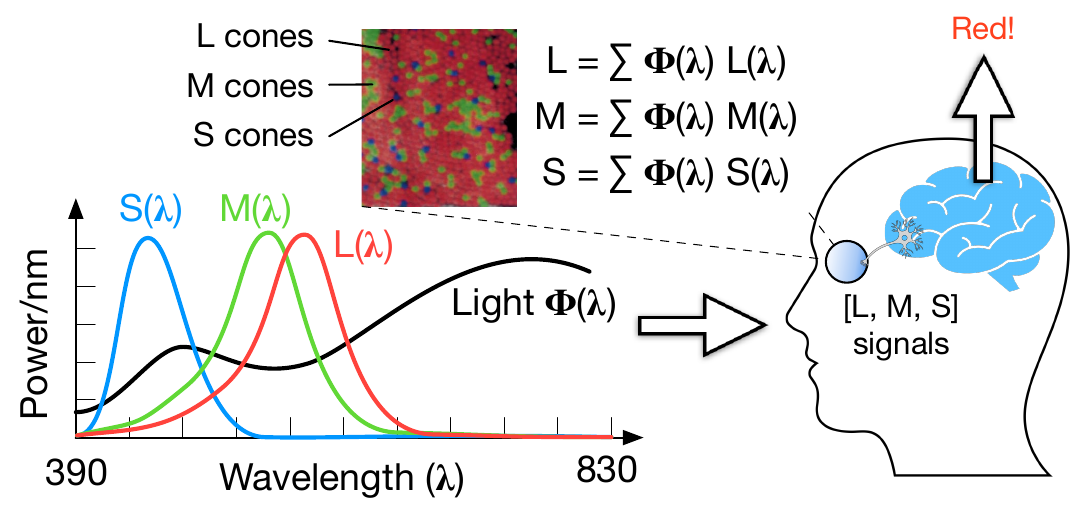}
  \label{fig:color}
  }
  \hspace{7pt}
  \subfloat[\small{Color encodings are different in different color spaces.}]
  {
  \includegraphics[height=1in]{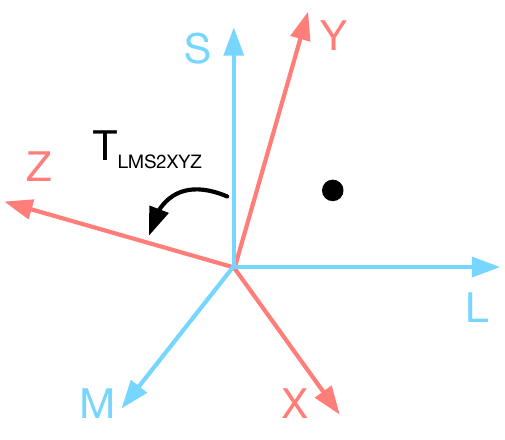}
  \label{fig:colorspace}
  }
  \hspace{7pt}
  \subfloat[\small{Material color depends on light spectrum and material reflectance.}]
  {
  \includegraphics[height=1in]{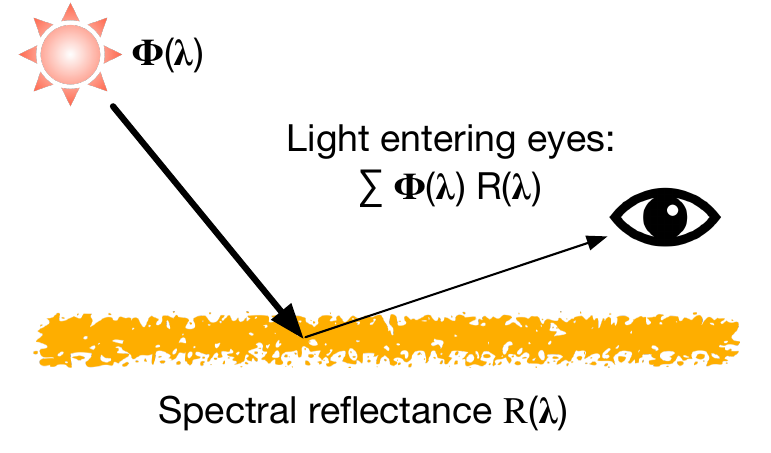}
  \label{fig:reflection}
  }
  \caption{(a): A light with a Spectral Power Distribution (SPD) $\Phi(\lambda)$ gets transformed into a triplet [L, M, S] on the retina, which represents the total responses of the Long, Medium, and Short cone photoreceptor cells.
  The brain interprets a combination of [L, M, S] as a color. (b): Geometrically, a color is a point in a 3D color space. We can change the basis of the coordinate system to derive a new color space. The same color is encoded differently in different color spaces. (c): Material color depends on both the spectrum of the incident light and the spectral reflectance of the material.}
  \label{fig:basics}
\end{figure*}

%\begin{figure}
%    \centering
%    \includegraphics[width=\columnwidth]{reflection.png}
%    \caption{Incident light spectrum reflecting on a surface produces reflected light spectrum via element-wise multiplication of vectors over visible wavelengths.}
%    \label{fig:reflection-diagram}
%\end{figure}

\paragraph{Lights.}
In the realm of color science, a light is physically represented by a Spectral Power Distribution (SPD) function, which describes the power distribution of the light over wavelengths.
The $\Phi(\lambda)$ function in \Fig{fig:color} illustrates one example.
The SPD is defined over the visible spectrum, usually between 390 \si{\nano\meter} and 830 \si{\nano\meter}, and quantized into discrete intervals for ease of computation.

\paragraph{Colors.}
Humans perceive colors from lights because photons are absorbed by cone cells on the retina, which in turn generate neural responses. These responses are interpreted by the brain as a particular color.
A cone's behavior is described by its Spectral Sensitivity Function (SSF)~\cite{wandell1995foundations}, which represents the neural response generated per unit power at a particular wavelength.

There are three kinds of cone cells that are responsible for color vision, each with a unique SSF that peaks at long, medium, and short wavelengths, respectively; these cone cells are thus called the L, M, and S cones.
$L(\lambda)$, $M(\lambda)$, and $S(\lambda)$ in \Fig{fig:color} show the SSFs of the three cone cells.

An incident light's power, after retinal processing, gets converted into three numbers, i.e., the total L, M, and S cone responses stimulated by the light.
Mathematically, this is:\footnote{The summation is sometimes also written as an integration~\cite{marschner2021physics}.
We choose summation to reflect the actual computation performed in programs.}
% 1) the quantum nature of lights where energy levels, hence wavelengths, are discrete~\cite{feynman2011feynman}, and 2) the actual computation done in color science programs.}
\begin{align}
  [L, M, S] = [\sum_{\lambda = 390}^{830} \Phi(\lambda) L(\lambda), \sum_{\lambda = 390}^{830} \Phi(\lambda) M(\lambda), \sum_{\lambda = 390}^{830} \Phi(\lambda) S(\lambda)]
  \label{eq:lms}
\end{align}
where $\Phi(\lambda)$ is the SPD of the incident light. 
This can be understood as weighting the light SPD by the cone sensitivity per wavelength and summing the weighted responses over the visible spectrum.

%\newcommand*{\horzbar}{\rule[.5ex]{5.5ex}{0.5pt}}
%\begin{align}
%\begin{bmatrix}
%L\\
%M\\
%S
%\end{bmatrix}
%=
%\begin{bmatrix}
%\horzbar & L(\lambda) & \horzbar \\
%\horzbar & M(\lambda) & \horzbar \\
%\horzbar & S(\lambda) & \horzbar \\
%\end{bmatrix}
%\times
%\begin{bmatrix}
%\bigg| \\
%\Phi(\lambda) \\
%\bigg|
%\end{bmatrix}
%\label{eq:cones}
%\end{align}

%\noindent where $L(\lambda)$, $M(\lambda)$, and $S(\lambda)$ are the SSFs of the three cone cells, respectively.

%\Fig{fig:color} (top) gives an equivalent integration form of this matrix multiplication.

%That is, a light is converted to just three numbers, the total L, M, S cone responses.
Our brain, over time, learns to associate an [L, M, S] triplet with a color.
Thus, a color can be represented as a point in a 3D (LMS) space.
This is the fundamental reason why human color perception is trichromatic.

%Importantly, color perception involves a huge dimensionality reduction: a light is a high-dimensional space vector (SPD of the light) and is reduced to a color, which is a three-dimensional space vector (cone responses).
%Therefore, there are infinitely many unique lights (i.e., unique SPDs) that can result in the same color perception, a phenomenon called metamerism~\cite{??}.
%This can be understood when we re-write \Equ{eq:lms} in the matrix form:
%
%This is an under-determined system of equations.
%It means that there are infinitely many lights $\Phi(\lambda)$ that satisfies the linear system, i.e., resulting in the color [L, M, S].

In addition to the LMS space, there are many other color spaces in which a color can be represented.
Geometrically, this amounts to changing the basis of a coordinate system and re-expressing the same color in the new space.
For instance, the most commonly used color space in color science is the CIE 1931 XYZ color space~\cite{brainard2010colorimetry}, which is a new coordinate system that is a linear transformation ($T_{LMS2XYZ}$) away from the LMS space.
This is illustrated in \Fig{fig:colorspace}.
A color in the LMS space [$L_c$, $M_c$, $S_c$] can be re-expressed as [$X_c$, $Y_c$, $Z_c$] in the XYZ space by:
\begin{align}
    [X_c, Y_c, Z_c]^T = T_{LMS2XYZ} \times [L_c, M_c, S_c]^T
\end{align}

\paragraph{Gamma.}
The LMS space is a \textit{linear} color space,
in that the channel values are proportional to light power.
For instance, if the light power is doubled across the spectrum, the resulting channel values will simply double accordingly. However, color spaces used to encode digital images are usually \textit{non-linear}.
Common examples include the popular sRGB and opRGB color spaces.
In these color spaces, channel values are proportional to \textit{perceived brightness}, which is non-linearly related to light power.
The non-linear transformation between light power and brightness is called Gamma correction/encoding~\cite{poynton2012digital}. Gamma correction is governed by a single value $\gamma$\footnote{There also exist piece-wise Gamma functions.}:
%The relationship between a gamma-applied color space and its linear counterpart is given in \Equ{eq:gamma_equation}.
\begin{equation}
    \alpha = \beta^{\frac 1 \gamma}
    \label{eq:gamma_equation}
\end{equation}

\noindent $\alpha$ represents the gamma-encoded color space, and $\beta$ represents its linear counterpart.

\paragraph{Materials.}
Much of the light entering our eyes is reflected off materials.
The apparent color of a material depends on the light striking the material and the material's physical properties. 
The simplest phenomenological model of a material is the spectral reflection function $R(\lambda)$, which describes how much light is reflected back at a given wavelength $\lambda$\footnote{For simplicity we assume the surface is diffuse, where reflection is angle insensitive; phenomenological models such as BRDF and BSSDF~\cite{pharr2023physically} used for modeling non-diffuse surfaces can be similarly supported.}.
Given an incident light with an SPD $\Phi(\lambda)$, the SPD of the reflected light is $\sum_{\lambda = 390}^{830} \Phi(\lambda) R(\lambda)$, as illustrated in \Fig{fig:reflection}.

Physically, the reason a light gets reflected back is the complicated interaction of photons being absorbed and/or scattered by particles inside the material.
The absorption and scattering behavior of a material is modeled by the spectral absorption function $K(\lambda)$ and spectral scattering function $S(\lambda)$.
%, which describe, respectively, how a material absorbs and, for the unabsorbed portion, scatters photons.
The reflectance spectrum $R(\lambda)$ is related to the absorption and scattering spectra through the Kubelka-Munk model~\cite{kubelka1931article, kubelka1948new}. The model is expressed in \Equ{eq:km_reflectance}.
%a simplification of the general radiative transfer equation~\cite{chandrasekhar2013radiative}:
\begin{equation}
   R(\lambda) = 1 + \frac{K(\lambda)}{S(\lambda)} - \sqrt{\frac{K(\lambda)^2}{S(\lambda)^2} + 2 \frac{K(\lambda)}{S(\lambda)}}
   \label{eq:km_reflectance}
\end{equation}

All functions are spectra, indicating that the scattering and absorption capability of a material (and thus the reflection) depend on the wavelength of the incident light.
The advantage of modeling materials using scattering and absorption spectra is that it allows us to easily simulate the color of pigment mixing.
The absorption and scattering coefficients of homogeneous mixtures are modeled as a weighted average of that of the constituent materials~\cite{duncan1940colour}.
Specifically, when mixing N materials, each with a spectral absorption and scattering function of $K_i(\lambda)$ and $S_i(\lambda)$, respectively, the resulting mixture has a spectral absorption and scattering function of:
\begin{align}
   K_{mix}(\lambda) = \frac 1 C \sum_i^N{K_i(\lambda) \times C_i}, ~~S_{mix}(\lambda) = \frac 1 C \sum_i^N{S_i(\lambda) \times C_i}
   \label{eq:get-scattering-absorption}
\end{align}

\noindent where $C_i$ denotes the weight concentration of the $i^{th}$ material. $C$ is the sum of all individual concentrations.
The models are the scientific bases of simulating material mixing in \colorlang.

\section{Motivations}
\label{sec:motivation}

Color programs are often written using libraries like Colour-Science~\cite{colour_developers_colour_2015} and NumPy~\cite{harris2020array}.
These libraries provide only a thin wrapper around raw matrix operations.
The physical meanings of objects (e.g., color, light power spectrum, material scattering spectrum) are not tracked by these tools ~\cite{scikit_no_types}.
Programmers are, thus, responsible for keeping track of the physical meanings manually.
This burden can lead to a variety of subtle bugs.

The consequence of such bugs is often ``silent data corruption'', as physically incorrect operations do not usually lead to program crashes.
These bugs are tricky to catch, motivating the need to type-check color programs.
Users complain when they observe an undesirable output, as discussed in a Google IO talk~\cite{googleio_understanding_color}.

\paragraph{Physically Meaningless Operations.}
Without meaningful type information, programmers are prone to defining arithmetic operations that are physically meaningless but mathematically permissible.
For instance, common libraries provide code for color space conversion, but do not check whether such conversion is performed on the correct color space.
The code snippet in \Prog{prog:incorrect_numpy_trans}, written using the popular Colour-Science Python library~\cite{colour_developers_colour_2015}, executes without complaint but is incorrect.

\begin{lstlisting}[language=Python, caption={Incorrect Translation from \texttt{sRGB} to \texttt{LAB} space.}, label={prog:incorrect_numpy_trans}]
image = open_image('srgb_image.png')  # sRGB image
colour.XYZ_to_Lab(image) # should use sRGB_to_XYZ prior to XYZ_to_Lab
\end{lstlisting}

In this example, a programmer intends to convert an image encoded in the \texttt{sRGB} color space to the CIELAB color space, but the \texttt{sRGB} color data is inadvertently, and incorrectly, treated as data encoded in the CIEXYZ color space on line 2.
Nothing prevents programmers from making this mistake.
Needless to say, translating an \texttt{sRGB} image to a \texttt{LAB} image as if the former were encoded in the \texttt{XYZ} space is not physically meaningful.
The program still executes without complaint, as the operation is mathematically permissible. 

Incorrect color space handling is a common issue reported by programmers~\cite{threejs_color_management}.
For instance, a PyTorch programmer says, ``\textit{there is no way for the library to know if the input tensor that you are passing is indeed in rgb colorspace. So you can silently get wrong results if you are not careful}''~\cite{pytorch_no_type}.
Similarly, a scikit-image user says, ``\textit{scikit-image operates on numpy arrays exclusively, so we have no way of knowing metadata about the image}''~\cite{scikit_no_types}.

% Such color interpolation is commonly used in image anti-aliasing~\cite{shirley2009fundamentals}.
% Nothing prevents programmers from directly averaging an \texttt{opRGB} color with an \texttt{sRGB} color, as shown in \Fig{fig:op+srgb}.
% Nothing prevents programmers from applying 

One might also want to simulate light reflection off a surface using the light spectrum and the material reflectance spectrum as shown in \Fig{fig:reflection}, but may accidentally multiply two light spectra, as they have the same data dimension. 
Such a multiplication makes no physical sense.

% \fixme{move references to the quotes above? perceptual uniformity ref belongs to below.}
% Searching major code repositories yields hundreds of issues related to the incorrect application of color science. Programmers often handle color spaces incorrectly \citep{noauthor_feature_2021, threejs_color_nodate} and misunderstand perceptual uniformity \citep{zuniga_how_2021}.

%\paragraph{Encoding Confusion.}
%Confusing color encodings can lead to uncaught errors. For example, OpenCV provides programmers with a function to convert the colorspace of an image. Since OpenCV does not attach encoding information to images, no type checking is done to ensure that user-specified colorspace conversions are coherent. It is possible to perform an \texttt{sRGB} to \texttt{HSV} conversion on an image stored in \texttt{LAB} space.

% Crucially, \colorlang also targets another significant scientific community that might not intersect with the conventional CS community: color scientists and vision scientists. These are domain experts who write programs to directly manipulate light and color data. They must do so in a physically accurate way.
% Color and vision scientists are less familiar with modern programming techniques (e.g., type checking, performance optimizations), and are exactly the population CoolerSpace can help.

\paragraph{Incorrect Understanding of Color Science.}
Another class of bugs arises from incorrect understanding of color science, ranging from a lack of understanding of radiometry (physics) vs. photometry (human perception)~\cite{pharr2023physically}, to mistakenly equating mixing lights with mixing materials~\cite{sochorova2021practical}.

Consider a scenario where one wants to estimate the color of  mixing two lights represented in the \texttt{sRGB} color space.
A naive programmer may simply add the \texttt{sRGB} values, as seen in \Prog{prog:incorrect_inter}.
% Even if one were to convert \texttt{opRGB} into \texttt{sRGB} before performing linear interpolation, the result would still be incorrect.

% \begin{figure}[h]
%     \begin{lstlisting}[language=Python]
% image1 = open_image('image1.png')
% image2 = open_image('image2.png')

% # Physically meaningless but permitted operation
% interpolated_img = (image1 + image2) / 2
% \end{lstlisting}
% \caption{Incorrect \texttt{sRGB} interpolation.}
% \label{fig:op+srgb}
% \end{figure}

\begin{lstlisting}[language=Python, caption={Incorrect \texttt{sRGB} light addition.}, label={prog:incorrect_inter}]
# Assuming both images are encoded in sRGB
image1 = open_image('image1.png')
image2 = open_image('image2.png')

# Physically meaningless but permitted operation
mixed_img = image1 + image2
\end{lstlisting}

% \fixme{maybe easy to just say add two lights?}
This code is incorrect, because, as discussed in \Sect{sec:background}, \texttt{sRGB} is a non-linear color space: \texttt{sRGB} channel values are not proportional to light power.
To accurately simulate the mixing of two colored lights, the addition must take place in a linear color space. 
The code is shown in \Prog{prog:correct_inter}.

% \begin{figure}[t]
% \begin{lstlisting}[language=Python]

% \end{lstlisting}
% \caption{Correct \texttt{sRGB} interpolation.}
% \label{fig:correct_op+srgb}
% \end{figure}

\begin{lstlisting}[language=Python, caption={Correct \texttt{sRGB} light addition.}, label={prog:correct_inter}]
# Convert sRGB image to linear sRGB
image1_linear = (image1 / 255) ** 2.2
image2_linear = (image2 / 255) ** 2.2

# Add in linear space
mixed_linear = image1_linear + image2_linear

# Re-apply gamma
mixed = (mixed_linear ** (1 / 2.2)) * 255
\end{lstlisting}

\begin{figure}[t]
    \centering
    \includegraphics[width=0.75\linewidth]{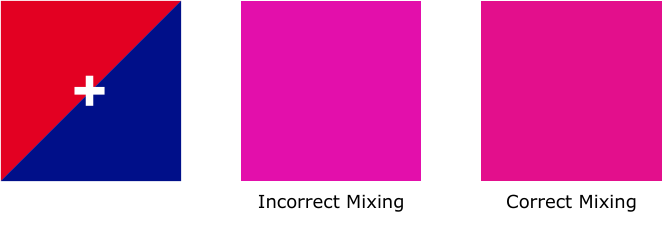}
    \caption{Mixing of red (\texttt{sRGB} [227, 0, 34]) and blue (\texttt{sRGB} [0, 15, 137]) lights in the non-linear sRGB space (incorrect) vs. in a linear color space (correct).}
    \label{fig:naive_color_interpolation}
\end{figure}

% Similar issues can occur in other color spaces like \texttt{HSV}. Unlike tristimulus color spaces, \texttt{HSV} is a polar coordinate color space. Programmers who attempt to interpolate \texttt{HSV} values ~\cite{wrong_hsv_interpolation}.

Incorrectly performing linear physics operations in a non-linear color space is a common issue of color programming in the wild~\cite{wrong_srgb_interpolation, yet_another_wrong_rgb_mixing, third_srgb_error, wrong_hsv_interpolation}.
Programmers usually have to manually track whether data are encoded in a physically linear space.
For instance, three.js warns programmers that ``\textit{it's important that the working color space be linear and the output color space be nonlinear}''~\cite{threejs_color_management}, without providing any checks.

``Silent data corruptions'' from incorrect color manipulations are often too subtle to catch.
\Fig{fig:naive_color_interpolation} compares the outputs of the naive interpolation and the correct interpolation:
the results are visually similar.
It can be hard to distinguish the output of unprincipled color programs from properly written color programs.
Bugs can easily pass human scrutiny.

\paragraph{Implementation Details and Speed Concerns.}
The correct program to add two lights represented in the \texttt{sRGB} color space, shown in \Prog{prog:correct_inter}, requires a non-trivial amount of complexity.
However, the program is semantically simple --- the program mixes two colored lights. 
The complexity of physically accurate color manipulation can and should be abstracted from programmers.

Isolating the logic of color physics from its implementation has another key advantage:
it frees programmers from optimizing for performance.
%A single full HD (1920 $\times$ 1080 pixels) image represented in \texttt{sRGB} space consumes 6 megabytes of memory. When dealing with videos or hyperspectral image data, the amount of data consumed is orders of magnitude higher.
Color programs operate on large datasets. 
For reference, an uncompressed one minute full HD (1920 $\times$ 1080 pixels) video filmed at 60 frames per second constitutes over 22 GB of data.
Due to the immense dataset sizes, color programmers are sensitive to slow run-times.
The absence of optimization can be a deal breaker.
One user of the Colour-Science python library writes, "\textit{this library is incredible but a lot of functions seem to be really slow}" ~\cite{color_science_runtime_complaint_1}.
Similar sentiments have been expressed by other developers~\cite{color_science_runtime_complaint_2, color_science_runtime_complaint_3}. 
Operating on such a large dataset means that even small optimizations can yield sizable execution time reductions.
Manually writing optimal code is not always obvious, and should be left to a compiler.

% \fixme{Position this better}

% Precise/approx types
% Precise and approximate types

% % Measurement types
% Previous work has explored the application of type theory to physical units and dimensions.
% \colorlang's type system cannot be easily mapped to 

%Color programs can be expressed in the language of tensor algebra.
%Therefore we can apply existing tensor algebra optimization techniques to the color science domain.
%Spectral and image data are efficiently represented with tensors. Since color programs can be expressed through tensor algebra, we can utilize existing tensor algebra optimization techniques to optimize color programs. Given the massive size of input data, small optimizations in color space programs can yield significant reductions in run-time. 

%\subsection{Color Reproduction Accuracy}
%Some color space transformations are lossy. For example, the transformation between reflectance spectrum and pigment types results in the loss of information. By de-prioritising lossy transformations during parse tree evaluation time, we can minimize the number of lossy transformations.

%\begin{verbatim}
%rs1 = cs.ReflectanceSpectrum([...])
%rs2 = cs.ReflectanceSpectrum([...])
%p = P(rs1) + P(rs2)
%eval(p)
%\end{verbatim}

%The above would be simplified to the following set of operations:

%\begin{verbatim}
%rs1 = cs.ReflectanceSpectrum([...])
%rs2 = cs.ReflectanceSpectrum([...])
%p = P(rs1 + rs2)
%eval(p)
%\end{verbatim}

\section{\colorlang System Overview}
\label{sec:ov}

%\begin{figure*}[t]
%	\centering
%	\begin{minipage}[t]{0.48\textwidth}
%		\includegraphics[width=0.8\textwidth]{egraph-1.png}
%		\caption{e-graph Representation of Toy Program}
%		\label{fig:egraph-1}
%	\end{minipage}
%	\begin{minipage}[t]{0.48\textwidth}
%		\includegraphics[width=0.8\textwidth]{egraph-2.png}
%		\caption{Saturated e-graph for Toy Program}
%		\label{fig:egraph-2}
%	\end{minipage}
%\end{figure*}

\begin{figure*}[t]
    \centering
    \includegraphics[width=\columnwidth]{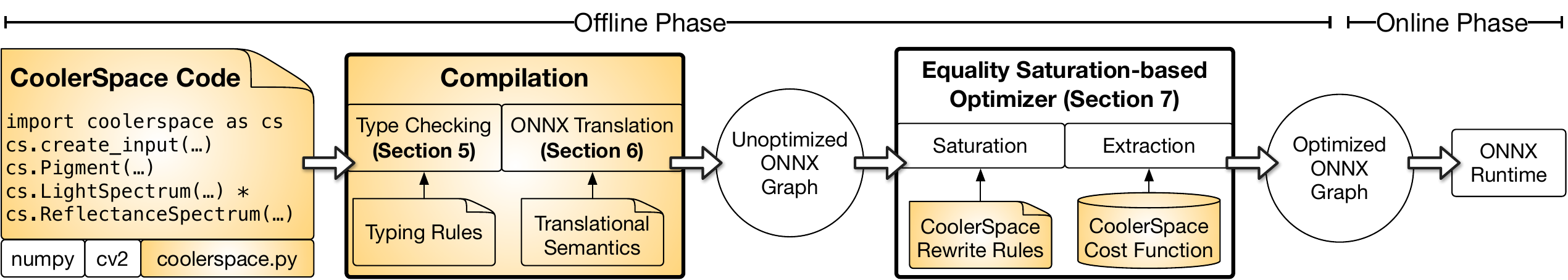}
    \caption{\colorlang overview. The colored components are introduced by \colorlang whereas the rest are existing tools.
    \colorlang is a Python-based meta-programming system: the Python program gets compiled and optimized into another program, an \onnx graph, which executes on the \onnx Runtime~\cite{onnxruntime}.
    Compilation and optimization are done once (offline phase), so they introduce only a one-time cost.}
    \label{fig:overview}
\end{figure*}

\colorlang is a Python-based meta-programming system.
%library used to generate color programs.
The key goal of \colorlang is to allow programmers to focus on the logic of color physics while relying on the programming system to guarantee correctness and to optimize for efficiency.
%where physical meanings are explicitly represented in the program and static type checking guarantees that only physically correct computations are permitted.
\Fig{fig:overview} provides an overview of the system.
%, including an offline phase and an online phase.
This section walks through the pipeline at a high level and highlights our major design decisions.

\paragraph{Language.}
From a programmer's perspective, \colorlang is a Python library.
\Prog{code:toy-program} shows a simple program written in \colorlang.
We choose Python as our host language as it is the lingua franca of color programmers, who make use of libraries such as OpenCV~\cite{opencv_library}, Colour-Science~\cite{colour_developers_colour_2015}, and NumPy~\cite{harris2020array}.

% Crucially, \colorlang also targets another significant scientific community that might not intersect with the conventional CS community: color scientists and vision scientists. These are domain experts who write programs to directly manipulate light and color data. They must do so in a physically accurate way.
% Color and vision scientists are less familiar with modern programming techniques (e.g., type checking, performance optimizations), and are exactly the population CoolerSpace can help.

% \begin{minipage}{\linewidth}
\begin{lstlisting}[language=Python, caption={A \colorlang program, where we create two sets of light spectra, which are cast to the \texttt{sRGB} type. The two sets of colored lights are then mixed. Notice how the programmer has to consciously unify the color space but is free from writing and optimizing the actual color mixing code.}, label={code:toy-program}, float, floatplacement=H]
import coolerspace as cs
light1 = cs.create_input(
    shape=[1920,1080],
    colorspace=cs.LightSpectrum
)
light2 = cs.create_input(
    shape=[1920,1080],
    colorspace=cs.LightSpectrum
)
color1 = cs.sRGB(light1)
color2 = cs.sRGB(light2)
color3 = color1 + color2
cs.create_output(color3)
\end{lstlisting}
% \end{minipage}

%\colorlang is a meta-programming system in that the Python program, when being executed, does not generate the actual results.
%Instead, it will be compiled into another program, an \onnx graph, which will be optimized to yield a performance-optimized \onnx graph, which will eventually be executed by \onnx Runtime~\cite{onnxruntime}.

%The \colorlang program goes through an offline phase and an online phase.
%The offline phase has two stage: compilation and optimization.

\paragraph{Compilation.}
The crux of \colorlang is a type system, which assigns types to user-defined objects.
The types store both the physical meaning and dimension of the data.
For example, the type of a full HD \texttt{sRGB} image would store the dimensions of the image (1920 $\times$ 1080) and the color encoding (\texttt{sRGB}).
Types are used to enforce physical and dimensional correctness through static type checking.
%Arithmetic operations between two objects of different physical representations are forbidden.
%Likewise, operations between objects with mismatching dimensions is also prohibited.
The type system is formalized in  \Sect{sec:lang}.

During compilation, the Python program is ``executed'', and any operations between \colorlang objects are intercepted, type checked, but \textit{not} evaluated.
If a program type checks, \colorlang creates a corresponding \onnx program that is equivalent to the original \colorlang program.
For instance, the second last line in \Fig{code:toy-program}, \texttt{color1+color2}, does not actually mix \texttt{sRGB} colors.
Instead, the `\texttt{+}' operator is overloaded to type check that both colors are in the same color space and of the same dimension.
Afterwards, \onnx code is generated with the arithmetic for \texttt{sRGB} color mixing.
The exact translation strategy is defined in \Sect{sec:translation}.

%A single operation in \colorlang is translated to one or more \onnx operations.
We have chosen \onnx as a compilation target, becasue \onnx is a popular format for tensor algebra.
There exists a vibrant community and ecosystem that provides cross-platform \onnx support ~\cite{onnxruntime}.
%allows us to express user programs in the language of tensor operations.
Given that color programs naturally manipulate tensors,
mapping \colorlang programs to \onnx allows us to benefit from \onnx's existing ecosystem. 

\paragraph{Optimization.}
The \onnx graph produced by the compiler faithfully reproduces the semantics of the original program but might not be optimal.
Our optimizer then converts the unoptimized \onnx graph into an optimized one.

The particular optimization strategy we use is based on equality saturation~\cite{tate2009equality, willsey2021egg}, which has been shown to be effective for optimizing tensor computations~\cite{jia2019taso, tensat}.
The technique is broadly split into two phases: saturation and extraction.
During the saturation phase, rewrite rules are used to generate a set of equivalent programs, each with a different cost governed by a cost function.
Then, the extraction phase extracts the cheapest program.
While using equality saturation for tensor optimizations is well established, our contribution lies in specifying rewrite rules and a cost function tailored to color programming. These are described in \Sect{sec:optimization}.

\paragraph{Execution.}
The optimizer outputs an optimized \onnx program, which is executed using \onnx Runtime~\cite{onnxruntime}.
Since \colorlang generates a reusable \onnx file, the compilation and optimization costs only have to be paid once per program.

\begin{table*}[t]
    \centering

    \caption{Simplified \colorlang abstract syntax. We omit trivial operations such as indexing color channels. A complete syntax is defined in the supplemental material.}
    
    \begin{tabular}{ll}
        \textsc{Arrays} & $a \in \text{floating point arrays}$ \\
    
        \textsc{Variable Names} & $x \in \text{variable names}$ \\
    
        \textsc{Tristimulus Color Types} & $\tristimulustype ::= \xyztype | \lmstype | \srgbtype | \oprgbtype $ \\
    
        \textsc{Perceptual Color Types} & $\perceptualtype ::= \hsvtype | \labtype$ \\
    
        \textsc{Color Types} & $\colortype ::= \tristimulustype | \perceptualtype$ \\
    
        \textsc{Spectral Types} & 
        $\tau_\spectrumtext ::= \lighttype | \reflectancetype | \scatteringtype | \absorptiontype | \pigmenttype$ \\
    
        \textsc{Physical Types} & 
        $\tau ::= \tau_\colortext | \tau_\spectrumtext | \chromaticitytype | \matrixtype$ \\
    
        \textsc{Dimension Types} & $d ::= \mathbb{N} | d \times d$  \\  
    
        \textsc{Shaped Types} & $s ::= (\tau, d)$ \\
    
        \textsc{Values} & 
        \makecell[l]{
            $v ::= x | \tau(a) | \tau(v) | \tau(v, v) | \mixtext(v, v, v, v) | v + v | v - v | v / v |$ \\ \hspace{0.65cm} $ v \times v | \matmultext(v, v)$ 
        }
        \\
    
        \textsc{Expressions} & $e ::= x = v$ \\
    
        \textsc{Programs} & $P ::= e; P | e$
    \end{tabular}

    \label{tab:colorlang_syntax}
\end{table*}

\section{\colorlang Type System}
\label{sec:lang}

After discussing the general principles behind our type system (\Sect{sec:lang:ov}), we will walk through the syntax by first describing the supported types (\Sect{sec:lang:types}), followed by describing the permissible operations and the typing rules that govern these operations (\Sect{sec:lang:rules}).

\subsection{Overview}
\label{sec:lang:ov}

We show the abstract syntax of \colorlang in \Tbl{tab:colorlang_syntax}.
A program in \colorlang consists of a set of expressions, each of which represents a physical operation that manipulates colors and/or spectra. 
Critically, each value in an expression is typed, permitting static type checking and promoting physics-aware color programming.

\colorlang's type system is designed to capture common computations used in color programming. \colorlang's types capture the underlying physical qualities of lights and materials, digital encodings of color, and models of human color perception. \colorlang allows users to operate on lights, materials, and the colors that result from light-material interactions.

We acknowledge that our type system, and arguably any type system, is inherently opinionated, as it stipulates as set of concrete rules.
\colorlang's type system is designed to follow the rules of fundamental physics, or wherever applicable, of standards defined by bodies such as the International Commission on Illumination (CIE).
There may exist scenarios where programmers would prefer to break our rules in favor of other considerations like speed.
In these scenarios, we provide users with an escape from our type system (see the usage of the \texttt{Matrix} type in \Sect{sec:lang:rules}).

% \fixme{Second, the coding idioms in these algorithms exercise the entirety of CoolerSpace’s type system, which we argue is representative of the space of color programming, because CoolerSpace allows specifying the operations on three fundamental aspects of color science: light spectra, material properties, and the colors resulting from light-material interactions.}

%\subsection{Abstract Syntax}
%\label{sec:lang:syntax}
%A complete version of the abstract syntax tree and typing rules of \colorlang can be found in our supplemental materials.  

\subsection{Types in \colorlang}
\label{sec:lang:types}

The type system is the most important component of \colorlang.
All type-checked values in \colorlang are of a Shaped Type $(\tau, d)$, which is a product type of a Physical Type $\tau$ and a Dimension Type $d$.
Physical Types represent physical properties, e.g., colors, light, and material; we will describe them in detail later.
Dimension Types represent the tensor dimension of a value/object.
The Dimension Types are expressed as a product of natural numbers, representing the shape of a matrix.
For example, a full HD image ($1920 \times 1080$ in resolution) encoded in the \texttt{sRGB} color space would have the Shaped Type $(\srgbtype, 1920 \times 1080)$.

Both Physical Types and Dimension Types are important because permissible computations in \colorlang should be performed on inputs that are physically meaningful and of correct dimension.
For instance, adding lights with colors is physically meaningless, and adding two color objects with mismatching dimensions is mathematically meaningless.

%All type checked values have a shaped type. We need to specify both type and dimension type for all values, as type checking operations relies on both type and dimension information. 

The Physical Types supported by \colorlang can be largely split into three broad categories: Tristimulus Color Types, Perceptual Color Types, and Spectral Types.
%There are three miscellanious types that don't fit cleanly into the three categories.

\paragraph{Tristimulus Color Types.}
Tristimulus Color Types, $\tristimulustype$, represent color spaces in which any color is represented by three channels, i.e., the tristimulus values.
These color spaces are defined by the choice of three primary colors and a white color.
A color of these types is internally encoded as a linear combination of the three primaries.
Thus, a specific color can be encoded differently across different tristimulus color spaces. 

\colorlang currently supports four Tristimulus Color Types: \texttt{opRGB}, \texttt{sRGB}, \texttt{LMS}, and \texttt{XYZ}.
Extending to other color spaces is straightforward.
The \texttt{LMS} and \texttt{XYZ} spaces are linear color spaces, in the sense that a color encoded in these spaces has channel values proportional to the \textit{power} of the corresponding light.
In contrast, \texttt{sRGB} and \texttt{opRGB} color spaces are non-linear. Channel values are proportional to \textit{perceived brightness}, as discussed in \Sect{sec:background}.

Linear and non-linear color spaces have different uses in color programming and are both important to support in \colorlang.
Linear color spaces are usually used when colors are initially captured or produced because of its direct relationship to light power.
%For instance, a color in the \texttt{LMS} color space represents the retinal responses of the three photoreceptor cells stimulated by a given light.
By contrast, non-linear color spaces are usually used when encoding and storing colors; in fact, most image file formats encode colors in the \texttt{sRGB} color space by default.
A classic workflow in the graphics pipeline is to render pixel colors in a linear space and store the image in a non-linear space.

\paragraph{Perceptual Color Types.}
\colorlang also supports a set of Perceptual Color Types, $\perceptualtype$, each corresponding to a perceptual color space.
Unlike tristimulus color spaces, perceptual color spaces do not represent colors as a mixture of primary colors;
instead, they represent colors by modeling how humans subjectively perceive colors.

For example, the CIELAB color space (abbreviated as \texttt{LAB}) models the opponent process of the human visual system~\cite{stockman2010color}. \texttt{LAB} represents a color by lightness (perceived brightness), red-green opponency, and yellow-blue opponency.
The \texttt{HSV} (also called HSL or HSB) color space represents a color as its lightness, hue, and saturation.

\colorlang provides the perceptual color spaces to support use-cases where subjective assessments of colors are involved.
For instance, both the \texttt{HSV} and \texttt{LAB} color spaces are commonly used to compare colors; in fact, \texttt{LAB} is the most common color space to quantify color differences (known as the CIE Delta E metric~\cite{sharma2017color}), a key task in color programming.
The \texttt{HSV} color space is commonly used to design color pickers in digital applications.

\paragraph{Spectral Types.}
A Spectral Type, $\tau_\spectrumtext$, represents a physical property that is dependent on wavelength.
For instance, the \texttt{Reflectance} type represents the reflectance of a surface/material over wavelength;
similarly, the \texttt{Light} type represents the spectral power distribution of lights.

Two other important Spectral Types are the \texttt{Scattering} and \texttt{Absorption} types, which represent the spectral scattering and spectral absorption functions of a surface/material, respectively.
The \texttt{Pigment} type, which represents materials (e.g., pigments like phthalo blue), is a product type between \texttt{Scattering} and \texttt{Absorption}.
This is because both scattering and absorption spectra are required to accurately model the mixture of two materials~\cite{kubelka1931article}.

Internally, spectral types are represented as histograms across the visible spectrum, defined between 390 \si{\nano\meter} and 830 \si{\nano\meter} in \colorlang.
We uniformly quantize this visible spectrum into 89 unique bands (i.e., the Spectral Type has a channel count of 89) at a 5 \si{\nano\meter} interval, but more fine-grained quantization schemes can be trivially implemented.

% \begin{enumerate}
%     \item \textbf{Light}: Represents the spectral power distribution of a light across the visible spectrum.
%     \item \textbf{Reflectance}: Represents the spectral reflectance of a surface across the visible spectrum.
%     \item \textbf{Scattering}: Represents the scattering spectrum of a material across the visible spectrum.
%     \item \textbf{Absorption}: Represents the absorption spectrum of a material across the visible spectrum.
% \end{enumerate}

\paragraph{Matrix Type.}
The \texttt{Matrix} type allows programmers to specify a numerical tensor in \colorlang.
These tensors are usually used for geometrically or arithmetically manipulating colors and spectra.
For instance, a tristimulus color can be seen as a point in a 3D Euclidean space, and a color science programmer might want to project the color to a plane, e.g., when simulating color vision deficiency~\cite{brettel1997computerized}.
%Similarly, a programmer may wish to scale the color channels of an image to, e.g., implement a tone mapping algorithm~\cite{??}. 
Projection (and in fact any linear transformation) is mathematically a matrix multiplication, hence the need for a \texttt{Matrix} type.

% \begin{enumerate}
%     \item \textbf{Chromaticity}: Represents the chromaticity of a color independent of the color's brightness.
%     \item \textbf{Matrix}: Represents a tensor of arbitrary shape. The Matrix type does not hold any semantic meaning. Used to perform unsupported operations.
%     \item \textbf{Pigment}: Represents a pigment. Product type of Scattering and Absorption. Can be used to simulate the mixture of pigments.
% \end{enumerate}

\subsection{Typing Rules}
\label{sec:lang:rules}

The typing rules allows only physically meaningful arithmetic operations.
Each class of the typing rules is, thus, designed to allow expressing a particular set of physical operations.
Our description below focuses on how \colorlang helps implement physically correct color programs.

\paragraph{Mixing Lights.}
The first class of typing rules expresses mixing two lights, which is perhaps the single most widely used operation in color science where one, for instance, mixes multiple lights in order to produce a target color.

Light mixing can be done in either the spectral or tristimulus domains.
Both are expressed by the `$+$' operator in the syntax.
When mixing two lights in the spectral space, the intention is to calculate the spectrum of the resulting light.
This is expressed by the \refrule{rule:light_add}, 
which stipulates that the result of mixing two set of light spectra is another set of light spectra.

%Mixing light spectra is permissible in \colorlang because mixing two light sources is essentially physically accumulating photons from two lights and thus the power spectra can be added together.

\begin{rulefigure}[h]
    \begin{prooftree}
        \AxiomC{$\Gammahas v_1: (\lighttype, d)$}
        \AxiomC{$\Gammahas v_2: (\lighttype, d)$}
        \RightLabel{\textsc{LightAdd}}
        \BinaryInfC{$\Gammahas v_1 + v_2 : (\lighttype, d)$}
    \end{prooftree}

    \caption{Light addition rule.}
    \label{rule:light_add}
\end{rulefigure}

Mixing lights using their colors is also permitted.
Programmers can mix tristimulus colors additively with the intention to calculate the color of the mixture of the original lights. 
This is represented in \refrule{rule:rgb_add}, which stipulates that both input colors must belong to the same tristimulus color space.
Then, the resulting color of the light will be presented in the same color space.
Note both rules also enforce that the dimensions of the two operands must match.

\begin{rulefigure}[h]
    \begin{prooftree}
        \AxiomC{$\Gammahas v_1: (\tristimulustype, d)$}
        \AxiomC{$\Gammahas v_2: (\tristimulustype, d)$}
        \RightLabel{\textsc{TristimulusAdd}}
        \BinaryInfC{$\Gammahas v_1 + v_2 : (\tristimulustype, d)$}
    \end{prooftree}
    \caption{Tristimulus addition rule.}
    \label{rule:rgb_add}
\end{rulefigure}

\paragraph{Mixing Perceptual Colors.} Color mixing can be done in either a physically uniform or perceptually uniform manner.
The rules above are intended for the physically uniform mixture of colors -- the linearity of addition in the spectral power domain is preserved. 
However, the human visual system perceives color non-uniformly.
For example, a linear increase in LMS cone responses does not correspond to a linear increase in perceived brightness.
Perceptually uniform color spaces, represented by $\perceptualtype$, are color spaces designed to represent the range of human-perceivable colors uniformly.
The distance between any two colors in a perceptually uniform color spaces like \texttt{LAB} are representative of the perceived difference between the two colors~\citep{sharma2017digital}.

% \fixme{}

Programmers may want to mix colors in a perceptually uniform color space when creating gradients or when conducting psychophysical experiments ~\citep{fairchild1995time}.
\colorlang allows programmers to add colors in a perceptually uniform manner, provided that the inputs to the addition operation are of a perceptual color space.
This is represented in \refrule{rule:perceptual_add}. Like in \refrule{rule:rgb_add}, the addition is only permitted if the operands are of the same specific perceptual type.

\begin{rulefigure}[h]
\begin{prooftree}
    \AxiomC{$\Gammahas v_1: (\perceptualtype, d)$}
    \AxiomC{$\Gammahas v_2: (\perceptualtype, d)$}
    \RightLabel{\textsc{PerceptualAdd}}
    \BinaryInfC{$\Gammahas v_1 + v_2 : (\perceptualtype, d)$}
\end{prooftree}
\caption{Perceptual addition rule.}
\label{rule:perceptual_add}
\end{rulefigure}

\paragraph{Light Reflection.} 
\colorlang also allows expressing reflecting light off of a surface.
This operation allows programmers to calculate the color of an object under a particular illuminant.
Physically, such calculation must be done in the spectral space, where light SPD and material reflectance are defined (\Fig{fig:reflection}).
%is expressed as a power distribution over the spectrum and the surface is described by the reflectance over the spectrum.
Syntactically, reflection is expressed with `$\times$'.
\refrule{rule:reflect} describes the corresponding typing rule.

\begin{rulefigure}[h]
    \begin{prooftree}
        \AxiomC{$\Gammahas v_1: (\lighttype, d)$}
        \AxiomC{$\Gammahas v_2: (\reflectancetype, d)$}
        \RightLabel{\textsc{Reflect}}
        \BinaryInfC{$\Gammahas v_1 \times v_2 : (\lighttype, d)$}
    \end{prooftree}

    \caption{Type rule for reflection.}
    \label{rule:reflect}
\end{rulefigure}

\paragraph{Mixing Materials.}
%In addition to mixing lights, 
Color programming also involves mixing materials, e.g., pigments.
For instance, painters routinely mix their primary paints to produce new colors. This process must be faithfully implemented in any digital painting software~\cite{sochorova2021practical}.
Print and dye industries also investigate how to properly mix inks and dyes to produce the target material quality~\cite{broadbent2001basic}.
Syntactically, mixing material is expressed by $\mixtext(\cdot)$, which takes four parameters: two pigment objects and their corresponding concentrations.

\begin{rulefigure}[h]
    \begin{prooftree}
        \def\defaultHypSeparation{\hskip.1in}
        \AxiomC{$\Gammahas v_1: (\absorptiontype, d)$}
        \AxiomC{$\Gammahas v_2: (\scatteringtype, d)$}
        \RightLabel{\textsc{PgmtInit}}
        \BinaryInfC{$\Gammahas \pigmenttype(v_1, v_2): (\pigmenttype, d)$}
    \end{prooftree}

    \begin{prooftree}
        \def\defaultHypSeparation{\hskip.1in}
        \AxiomC{$\Gammahas v_1, v_2 : (\pigmenttype, d)$}
        \AxiomC{$\Gammahas v_3, v_4 : (\matrixtype, d)$}
        \RightLabel{\textsc{PgmtMix}}
        \BinaryInfC{$\Gammahas \mixtext(v_3, v_1, v_4, v_2) : (\pigmenttype, d)$}
    \end{prooftree}
    
    \caption{\texttt{Pigment} rules.}
    \label{rule:pigment}
\end{rulefigure}

Mixing pigments is both physically and mathematically different from mixing lights and warrants its own typing rules.
Two rules in \colorlang govern pigment-related operations.
First, we allow initializing a pigment type from an absorption and a scattering type, as enshrined by the \textsc{PgmtInit} rule in \refrule{rule:pigment}.
This reflects the fact that a \texttt{Pigment} is internally described by the absorption and scattering spectra of the material.
Second, the \textsc{PgmtMix} rule in \refrule{rule:pigment} expresses the mixing the of two pigment objects.
The rule states that the output of mixing two \texttt{Pigment} objects of matching dimensions is another \texttt{Pigment} object of the same dimension.
% The rule states that when mixing two pigment objects, the result is another pigment object so long as the dimensions between both pigments and their concentrations all match. 

\paragraph{Transforming Colors.}
Programmers can scale each channel of a tristimulus color through element-wise multiplication.
Syntactically, these operations are expressed as $v_{rgb} \times v_{matrix}$.
Rule \ref{rule:color_matrix_mul} governs this operation.

\begin{rulefigure}[h]
    \begin{prooftree}
        \AxiomC{$\Gammahas v_1 : (\tristimulustype, d)$}
        \AxiomC{$\Gammahas v_2 : (\matrixtype, 3)$}
        \RightLabel{\textsc{TriScale}}
        \BinaryInfC{$\Gammahas v_1 \times v_2 : (\tristimulustype, d)$}
    \end{prooftree}
    \caption{Elementwise scaling of a tristimulus color.}
    \label{rule:color_matrix_mul}
\end{rulefigure}

%We also allow applying a transformation matrix to a tristimulus color using $\matmultext(v_{rgb}, v_{matrix})$,
%essentially manipulating colors as geometric objects.
%This is useful in applications such as simulating color blindness~\cite{brettel1997computerized, vienot1999digital}.

%\ref{rule:color_matrix_matmul}
%in which $\channelcounttext(\tau)$ returns the channel count of a specific color type.
%For instance, the \texttt{sRGB} type would have a channel count of 3, as the sRGB type has 3 channels (red, green, and blue).

%\begin{rulefigure*}[t]
%    \begin{prooftree}
%        \AxiomC{$\Gammahas v_1 : (\tristimulustype, d_1 \times \channelcounttext(\tristimulustype))$}
%        \AxiomC{$\Gammahas v_2 : (\matrixtype, \channelcounttext(\tristimulustype), d_2)$}
%        \RightLabel{\textsc{ColorMatrixMatMul}}
%        \BinaryInfC{$\Gammahas \matmultext(v_1, v_2) : (\tristimulustype, d_1 \times d_2)$}
%    \end{prooftree}
%    \caption{Matrix multiplication between a tristimulus color and a matrix.}
%    \label{rule:color_matrix_matmul}
%\end{rulefigure*}

\paragraph{Type Casting.}
If one wants to mix, for instance, an \texttt{sRGB} color with a \texttt{XYZ} color, one must cast the \texttt{sRGB} type to the \texttt{XYZ} type (or vice versa).
Syntactically, casting is expressed by $\tau(v)$, where $\tau$ is the target type and $v$ is the object to cast.

\begin{figure}[t]
    \centering
    \includegraphics[width=0.75\columnwidth]{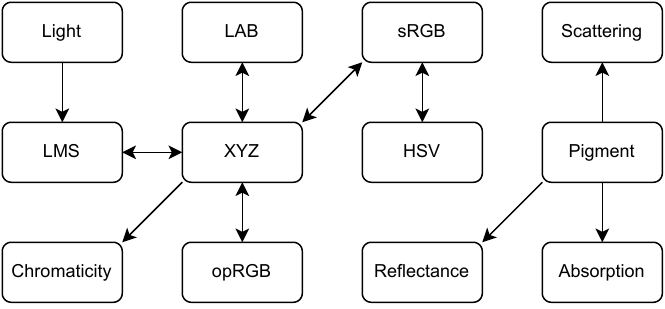}
    \caption{The graph of all permissible castings. \colorlang currently only supports a small number of commonly used color spaces. This is not an inherent limitation of \colorlang. With more engineering effort, \colorlang's interface can be extended to support other color spaces.}
    \label{fig:casting_graph}
\end{figure}

Principles in color science dictate the set of legal castings, which is illustrated in \Fig{fig:casting_graph}.
A casting between any origin and destination type is allowed if there exists a path from the former to the latter in \Fig{fig:casting_graph}.
The legal castings are expressed in \refrule{fig:casting_rule}. The $\pathexiststext(\tau_1, \tau_2)$ function type checks if there is a path from type $\tau_1$ to type $\tau_2$ in \Fig{fig:casting_graph}.

\begin{rulefigure}[h]
    \begin{prooftree}
        \AxiomC{$\Gammahas v : (\tau_1, d)$}
        \AxiomC{$\pathexiststext(\tau_1, \tau_2)$}
        \RightLabel{\textsc{Cast}}
        \BinaryInfC{$\Gammahas \tau_2(v) : (\tau_2, d)$}
    \end{prooftree}
    \caption{Casting rule}
    \label{fig:casting_rule}
\end{rulefigure}

Our casting rules prevent mathematically ill-posed castings.
For instance, casting a \texttt{Light} type to an \texttt{LMS} type is permitted, but casting an \texttt{LMS} type to a \texttt{Light} type is not.
This is because converting a light spectrum to an LMS color is a dimensionality reduction (\Equ{eq:lms}), so the inversion is mathematically ill-posed. There are many light spectra that correspond to the same LMS color.
%This is evident by examining \Equ{eq:cones}:
%solving for $\Phi(\lambda)$ given the [L, M, S] responses is an under-determined system and, thus, has infinitely many solutions.

%\footnote{The only exception is that we do not allow casting the \texttt{Pigment} type to the \texttt{Matrix} type even though path exists. This is because a pigment carries both spectral absorption and spectral scattering information; mixing the two in one resulting matrix is confusing.}.

%\fixme{this feels like a subtle detail. maybe we can still draw matrix in the figure and explain the exception in the footnote. also i wonder if we should draw scattering/absorption spectra and say pigments are allowed to be cast to them (lossy) but the other way is not allowed.}
%A separate rule is defined for casting to and from the \texttt{Matrix} type. We separate the \texttt{Matrix} casting rules from the general casting rule as the transitive property of casting does not apply when casting to and from the \texttt{Matrix} type.
%For example, the \texttt{Pigment} type is not castable to and from the \texttt{Matrix} type.
%However, it is possible to cast the \texttt{Pigment} type to the \texttt{Reflectance} type. If we were to apply the casting rule in figure \ref{fig:casting_rule}, casting \texttt{Pigment} to \texttt{Matrix} would be permissible, as a path would exist from \texttt{Pigment} to \texttt{Matrix} by way of \texttt{Reflectance}. The matrix casting rules are in figure \ref{fig:matrix_casting_rules}.

The \texttt{Pigment} type casts to the \texttt{Scattering} type or the \texttt{Absorption} type, because the former is defined as a product type of the latter two; these two castings are lossy and cannot be reversed.
Importantly, we allow the \texttt{Pigment} type to cast to the \texttt{Reflectance} type.
This represents the physical reality that the reflectance spectrum of a material can be derived from the material's scattering and absorption spectra, which are carried in the \texttt{Pigment} type.

% Finally, we do not allow casting to the \texttt{Matrix} type from any other type.
% This is to prevent a programmer from escaping the typing rules to perform physically meaningless computations.
% If we allowed such castings, one might, for instance, cast an \texttt{sRGB} type to a \texttt{Matrix} type and then multiply the matrix with a \texttt{XYZ} color --- multiplying two colors is meaningless.

%Finally, we allow casting to and from the \texttt{Matrix} type from most other types.
%This is to allow programmers to define operations that may not be permitted under our strict typing rules.
%Casting to and from the \texttt{Matrix} should be done with caution. 

% When casting other types to and from the \texttt{Matrix} type, the latter will carry the correct dimension information.
% This is enshrined by the \textsc{MatrixCastTo} and \textsc{MatrixCastFrom} rules in \Rule{fig:matrix_casting_rules}, in which $\channelcounttext(\tau)$ returns the channel count of a specific color type.
% For instance, the \texttt{sRGB} type would have a channel count of 3, as the sRGB type has 3 channels (red, green, and blue);
% %When a type is cast to the \texttt{Matrix} type, the dimension of the original tensor is used as the matrix object's dimension.
% As a specific instance of the \textsc{MatrixCastTo} rule, a full HD \texttt{sRGB} image has a shaped type of $(\srgbtype, 1920 \times 1080)$.
% If cast to the \texttt{Matrix} type, the resulting shaped type would be $(\matrixtype, 1920 \times 1080 \times 3)$.

\paragraph{Using the Matrix Type.}
%\modified{New section to show off the Matrix type, and also to demonstrate that our type system is opinionated.}
We allow casting to and from the \texttt{Matrix} type from most other types.
By casting objects to the \texttt{Matrix} type, programmers can escape our strict type system and define arithmetic operations that would not normally be permissible.
For example, programmers can define arbitrary addition operations using \refrule{rule:matrix_subtraction}. If one wishes to arithmetically add two \texttt{sRGB} colors, as seen in \Prog{prog:incorrect_inter} of \Sect{sec:motivation}, they may do so by first casting the objects to the \texttt{Matrix} type before the addition operation.
% \fixme{For instance, if one wishes to arithmetically interpolate between two sRGB colors, they could cast them first into the \texttt{Matrix} type and average two matrices. specifically reference that early example prog.}

\begin{rulefigure}[h]
    \begin{prooftree}
        \AxiomC{$\Gammahas v_1: (\matrixtype, d)$}
        \AxiomC{$\Gammahas v_2: (\matrixtype, d)$}
        \RightLabel{\textsc{MatrixAdd}}
        \BinaryInfC{$\Gammahas v_1 + v_2 : (\matrixtype, d)$}
    \end{prooftree}
    \caption{Simplified matrix type addition.}
    \label{rule:matrix_subtraction}
\end{rulefigure}

\paragraph{Other Rules.}
\colorlang also defines other operations and their associated typing rules.
For instance, programmers can retrieve individual channels of an object through the syntax $v.c$, where $c$ represents the channel to be accessed.
We also allow the application of transformation matrices to tristimulus colors using $\matmultext(v_{rgb}, v_{matrix})$.
%This operation type checks only if the channel $c$ matches the physical type of the object.
The supplemental material contains a comprehensive list of \colorlang's typing rules.

\section{\colorlang to ONNX Translation}
\label{sec:translation}

% Brief introduction to ONX
% INTERESTING examples 

Programs written in \colorlang are translated into \onnx.
%An operation in \colorlang may map to one or more \onnx operations.
The decision to use ONNX as a compilation target is justified in \Sect{sec:ov}.
We introduce our translation strategy with formal translational semantics in \Sect{sec:translation:semantics}.
We then prove that our translation is sound in \Sect{sec:translation:proof}:
any type checked value in \colorlang translates to a type checked value in \onnx.
\revnew{Finally, in \Sect{sec:type-soundness-proof}, we show that \colorlang is type sound in addition to being translationally sound.}
%if our translational strategy is followed. 
%Given the assumption that \onnx is type sound, \colorlang must also be type sound.

\subsection{Translational Semantics}
\label{sec:translation:semantics}

\begin{table}[t]
    \centering

    \caption{Subset of \onnx syntax.}
    
    \begin{tabular}{ll}
        \textsc{Natural Numbers} & $\mathbb{N} \in \text{natural numbers}$ \\
        \textsc{Arrays} & $a \in \text{floating point arrays}$ \\
        \textsc{Dimension Types} & $d ::= \mathbb{N} | d \times d$\\
        \textsc{Values} & \makecell{$u::= a | \addtext(u, u) | \divtext(u, u) | \multext(u, u)$ \\$ | \subtext(u, u) | \matmultext(u, u)|\powtext(u, u)$}
    \end{tabular}

    \label{tab:onnx_syntax}
\end{table}

\begin{table*}[t]
    \centering

    \caption{Subset of \colorlang to \onnx translational semantics. M$_1$, M$_2$, and M$_3$ represent the LMS Cone Fundamentals, XYZ to LMS transformation matrix, and XYZ to RGB transformation matrix, respectively. They are constant matrices that can be found in standard color science texts~\cite{wyszecki2000color}, and are omitted here.}
    
    %\resizebox{1\columnwidth}{!}{
    \renewcommand*{\arraystretch}{1}
    \renewcommand*{\tabcolsep}{3pt}
    \begin{tabular}{ll}
        $\llbracket \mathbb{N}\rrbracket \triangleq \mathbb{N}$ &  \textbf{T-Nat} \\
        $\llbracket v_1 + v_2\rrbracket \triangleq \Phi(+, \tau_1, \tau_2)(v_1, v_2)$, \hspace{.1cm} $v_1 : (\tau_1, d_1)$, $v_2 : (\tau_2, d_2)$ & \textbf{T-Add} \\
        
        $\llbracket a\rrbracket \triangleq a$ & \textbf{T-Array} \\
        $\llbracket v_1 \times v_2\rrbracket \triangleq \Phi(\times, \tau_1, \tau_2)(v_1, v_2)$, \hspace{.1cm} $v_1 : (\tau_1, d_1)$, $v_2 : (\tau_2, d_2)$ & \textbf{T-Mul} \\
        
        $\llbracket \tau(a)\rrbracket \triangleq a$ & \textbf{T-Init} \\
        $\llbracket \tau_d(v_o)\rrbracket \triangleq \Psi(\tau_o, \tau_d)(v_o)$, \hspace{.1cm} $v_o : (\tau_o, d_o)$ & \textbf{T-Cast} \\
        
        $\llbracket d\rrbracket \triangleq d$ & \textbf{T-Dim} \\
        $\llbracket (\tau, d)\rrbracket \triangleq d \times \channelcounttext(\tau)$ & \textbf{T-Type} \\

        $\Phi(+, \xyztype, \xyztype)(v_1, v_2) \triangleq \addtext \left( \llbracket v_1 \rrbracket, \llbracket v_2 \rrbracket \right)$ \\
        
        $\Phi(+, \lmstype, \lmstype)(v_1, v_2) \triangleq \addtext \left( \llbracket v_1 \rrbracket, \llbracket v_2 \rrbracket \right)$ \\
        
        \multicolumn{2}{l}{
        \specialcell{$\Phi(+, \srgbtype, \srgbtype) ( v_1, v_2 )  \triangleq \multext (\powtext(\addtext(\powtext(\divtext(\llbracket v_1 \rrbracket, 
            \begin{bmatrix}
              255
            \end{bmatrix} 
        ), 
            \begin{bmatrix}
              2.2
            \end{bmatrix}
        ),$ \\ \hspace{3.9cm} $\divtext(\powtext(\llbracket v_2 \rrbracket, 
            \begin{bmatrix}
                255
            \end{bmatrix}
        ) 
            \begin{bmatrix}
                2.2
            \end{bmatrix}
        ) ), 
            \begin{bmatrix}
                0.455
            \end{bmatrix}
        ), 
            \begin{bmatrix}
            255
            \end{bmatrix}
        )$}} \\

        $\Phi(\times, \lighttype, \reflectancetype)(v_1, v_2) \triangleq \multext(\llbracket v_1 \rrbracket, \llbracket v_2 \rrbracket)$ \\
        
        $\Psi(\lighttype, \lmstype)(v) \triangleq \matmultext \left (
                \llbracket v \rrbracket,
                \text{M}_1
            \right )$ & \\
 
        $\Psi(\lmstype, \xyztype)(v) \triangleq \matmultext 
            \left ( 
            \llbracket v \rrbracket,
            \text{M}_2
            \right )
        $ & \\
        
        \multicolumn{2}{l}{$\Psi(\xyztype, \srgbtype)(v) \triangleq
        \powtext \left (
            \multext \left (
                \matmultext \left ( 
                    \llbracket v \rrbracket,
                    \text{M}_3
                \right ),
                \begin{bmatrix}
                    255
                \end{bmatrix}
            \right ),
            \begin{bmatrix}
                2.2
            \end{bmatrix}
        \right )
        $} \\
    \end{tabular}
    %}

    \label{tab:color2onnx}
\end{table*}

The translation process is guided by our translational semantics.
The translational semantics must be understood in conjunction with the abstract syntax of \onnx, our target language.
To our best knowledge, there is no formal syntax of \onnx. Completely formalizing \onnx is out of our scope.
We do, however, formalize a subset of \onnx pertaining to \colorlang, which is shown in \Tbl{tab:onnx_syntax}.

Unlike \colorlang, \onnx does not assign Physical Types to values, because ONNX is designed to express tensor algebra.
Therefore, values in \onnx are assigned only Dimension Types.
%\footnote{\onnx does support multiple different element types including \texttt{UINT64} and \texttt{BOOL}. We do not encode this type information in our version of \onnx's abstract syntax, as we only translate \colorlang arrays into \texttt{DOUBLE} element types.}
\addtext, \divtext, \multext, \subtext{}, and \powtext{} perform element-wise operations between two tensors;
\matmultext{} is matrix multiplication.
All expressions support dimension broadcasting, which is discussed in \Sect{sec:translation:proof}

Given the abstract syntax of ONNX, \Tbl{tab:color2onnx} shows the formal translational semantics for converting a subset of the \colorlang abstract syntax to \onnx's abstract syntax.
Due to space constraints, we show only a subset of the translational semantics, focusing on those that highlight important properties of this translation.

Immediately clear is that the Physical Type information in \colorlang is lost during translation.
As shown by the \textbf{T-Type} rule, the Physical Type $\tau$ in \colorlang gets reduced to only its dimension information in \onnx.
For instance, after translating to \onnx, objects of the same dimension in \texttt{LMS} and \texttt{sRGB} space are no longer differentiated, as both \texttt{LMS} and \texttt{sRGB} have three channels.
% For instance, the \texttt{LMS} and \texttt{sRGB} types are not differentiated after translating to \onnx, as both have three channels.

The translations of the actual expressions are encoded in the $\Phi$ and $\Psi$ lookup tables, which represent, respectively, the \onnx implementation of each operation and casting in \colorlang.
For instance, $\Phi(+, \srgbtype, \srgbtype)(v_1, v_2)$ encodes the \onnx implementation of \texttt{sRGB}+\texttt{sRGB}, and $\Psi(\lighttype, \lmstype)(v)$ encodes how we cast a \texttt{Light} spectrum to a color in the \texttt{LMS} color space.

Two interesting properties of $\Phi$ and $\Psi$ are worth discussing.
First, arithmetic operations type checked by the same type rule do not necessarily translate to the same set of ONNX operations in $\Phi$.
%\Fig{fig:xyz_add} shows an example of adding two \texttt{XYZ} objects and \Fig{fig:srgb_add} shows an example of adding two \texttt{sRGB} objects.
For instance, adding two \texttt{XYZ} objects and adding two \texttt{sRGB} objects are both expressed with the `$+$' syntax and are type-checked by \refrule{rule:rgb_add}.
However, \texttt{sRGB} addition is translated to a much more complicated sequence of ONNX operations, as seen in the corresponding entry in $\Phi$ (\Tbl{tab:color2onnx}).
The reason is that the \texttt{sRGB} color space has a defined gamma curve, which must be removed before addition and reapplied after addition.

%\begin{figure}
%	\begin{lstlisting}[language=Python]
%import colorspace as cs
%srgb1 = cs.create_input(name="sRGB1",
%    shape=[1920,1080],
%    colorspace=cs.sRGB
%)
%srgb2 = cs.create_input(name="sRGB2",
%    shape=[1920,1080],
%    colorspace=cs.sRGB
%)
%srgb3 = srgb1 + srgb2
%cs.create_output(srgb3)
%\end{lstlisting}
%	\caption{\texttt{sRGB} addition in \colorlang.}
%	\label{code:srgb_addition}
%\end{figure}
%
%\begin{figure}
%    \includegraphics[width=\columnwidth]{pictures/srgb_addition_onnx.pdf}
%    \caption{\texttt{sRGB} addition in ONNX.}
%    \label{onnx:srgb_addition}
%\end{figure}

% TODO: Modify to join together and minimize page space

Second, casting operations between types that are not adjacent in \Fig{fig:casting_graph} are intentionally \textit{not} defined in $\Psi$.
Casting operations between non-adjacent types are expanded into a series of individual casting operations representing the shortest path between the origin type and the destination type.
For example, \texttt{LMS} and \texttt{sRGB} are not connected by an edge.
The casting between the two, $\Psi(\lmstype, \srgbtype)(v)$, is converted to $\Psi(\xyztype, \srgbtype)(\Psi(\lmstype, \xyztype)(v))$.
This cascaded translation is sub-optimal: it involves multiplying two constant matrices.
We rely upon equality saturation to optimize these inefficient translations (\Sect{sec:optimization}) while maintaining a small $\Phi$. Note that there is no ambiguity in non-adjacent casting: there exists at most one path between any pair of types, because the casting graph in \Fig{fig:casting_graph} is a directed forest.

\subsection{Translational Soundness}
\label{sec:translation:proof}

We now prove translational soundness: any well-typed value in \colorlang remains well-typed after being translated to \onnx.
%A proof of translational soundness is equivalent to a proof of type safety for \colorlang \textit{assuming} \onnx is type safe.
Translational soundness indicates that our translation preserves typeability: \colorlang's type safety is as strong as that of \onnx. 
This strategy is inspired by Gator~\cite{gator}. 
However, proving \onnx's type safety is beyond the scope of this paper.
%We will present the overall proof strategy and show a representative subcase while referring interested readers to the Supplemental Material for the complete proof.

\begin{figure}[t]
\begin{prooftree}
    \AxiomC{}
    \RightLabel{\textsc{TrivialBroadcast}}
    \UnaryInfC{$\broadcastabletext(d, d)$}
\end{prooftree}

\begin{prooftree}
    \AxiomC{}
    \RightLabel{\textsc{ScalarBroadcast}}
    \UnaryInfC{$\broadcastabletext(1, d)$}
\end{prooftree}

\begin{prooftree}
    \AxiomC{$d_1 = d_2 \times d_3$}
    \RightLabel{\textsc{SubsetBroadcast}}
    \UnaryInfC{$\broadcastabletext(d_3, d_1)$}
\end{prooftree}

\begin{prooftree}
    \AxiomC{$u_1 : d_1$}
    \AxiomC{$u_2 : d_2$}
    \AxiomC{$\broadcastabletext(d_1, d_2)$}
    \RightLabel{\textsc{OnnxAddR}}
    \TrinaryInfC{$\addtext(u_1, u_2) : d_2$}
\end{prooftree}

%\begin{prooftree}
%    \AxiomC{$u_1 : d_1$}
%    \AxiomC{$u_2 : d_2$}
%    \AxiomC{$\broadcastabletext(d_1, d_2)$}
%    \RightLabel{\textsc{OnnxSubR}}
%    \TrinaryInfC{$\subtext(u_1, u_2) : d_2$}
%\end{prooftree}
%
%\begin{prooftree}
%    \AxiomC{$u_1 : d_1$}
%    \AxiomC{$u_2 : d_2$}
%    \AxiomC{$\broadcastabletext(d_1, d_2)$}
%    \RightLabel{\textsc{OnnxMulR}}
%    \TrinaryInfC{$\multext(u_1, u_2) : d_2$}
%\end{prooftree}
%
%\begin{prooftree}
%    \AxiomC{$u_1 : d_1$}
%    \AxiomC{$u_2 : d_2$}
%    \AxiomC{$\broadcastabletext(d_1, d_2)$}
%    \RightLabel{\textsc{OnnxDivR}}
%    \TrinaryInfC{$\divtext(u_1, u_2) : d_2$}
%\end{prooftree}
%
%\begin{prooftree}
%    \AxiomC{$u_1 : d_1$}
%    \AxiomC{$u_2 : d_2$}
%    \AxiomC{$\broadcastabletext(d_1, d_2)$}
%    \RightLabel{\textsc{OnnxPowR}}
%    \TrinaryInfC{$\powtext(u_1, u_2) : d_2$}
%\end{prooftree}
%
%\begin{prooftree}
%    \AxiomC{$u_1 : d_l \times n$}
%    \AxiomC{$u_2 : n \times d_r$}
%    \RightLabel{\textsc{OnnxMatMul}}
%    \BinaryInfC{$\matmultext(u_1, u_2) : d_l \times d_r$}
%\end{prooftree}

\caption{Subset of \onnx typing rules.}
\label{fig:onnx_rules}
\end{figure}

\paragraph{ONNX Typing Rules.}
The translational soundness proof depends on the formal typing rules of \onnx.
\Fig{fig:onnx_rules} shows a set of key typing rules for ONNX.
% The rules ensure that tensor dimensions check in an operation, which is complicated because \onnx, like many tensor libraries (e.g., numpy), permits flexible dimension \textit{broadcasting}.
The rules ensure that the tensor dimensions of operands are valid.
This is complicated because \onnx, like many tensor libraries (e.g., NumPy), permits flexible dimension \textit{broadcasting}.
Arithmetic operations are valid even if the Dimension Types of two input tensors do not match, so long as the dimension of the smaller tensor can be ``broadcast'', or expanded, to be compatible with that of a larger tensor.

Many \colorlang expressions rely on broadcasting in \onnx to implement.
For instance, removing and applying gamma from a tensor of \texttt{sRGB} objects requires broadcasting a scalar (gamma) to an entire tensor of \texttt{sRGB} colors.
Therefore, the \powtext{} function in \onnx, which is used to implement gamma, must broadcast a scalar (2.2) to every element in a tensor in an element-wise manner.
See the $\Phi(+, \srgbtype, \srgbtype)(v_1, v_2)$ entry in the translational semantics (\Tbl{tab:color2onnx}) for an example.
%\Fig{fig:srgb_add} shows a concrete example when adding two \texttt{sRGB} tensors.
% The translated \onnx program
% %The $\Phi(+, \srgbtype, \srgbtype)(v_1, v_2)$ entry in the translational semantics (\Tbl{tab:color2onnx})
% uses, among others, the \powtext{} function in \onnx, which can broadcast a scalar (255) to every tensor element in an element-wise manner.
% \divtext{} and \multext{} in 
%\Fig{fig:srgb_add}
% \Tbl{tab:color2onnx} broadcast, too.
%Broadcasting is also used to support scaling channels of Tristimulus Color Types, e.g., increasing the saturation channel of an \texttt{HSV}-typed color.

To our best knowledge, broadcasting has not been formally specified in \onnx.
We formalize a subset of broadcasting rules that are relevant to \colorlang.
These rules are defined in the rules \textsc{TrivialBroadcast}, \textsc{ScalarBroadcast}, and \textsc{SubsetBroadcast}.
Here, \broadcastabletext($d_1$, $d_2$) indicates that a dimension $d_1$ can be broadcast to a dimension $d_2$. 
\textsc{TrivialBroadcast} is the usual case where both inputs have the same dimensions.
\textsc{ScalarBroadcast} allows a scalar input to be broadcast to a tensor in an element-wise manner.
\textsc{SubsetBroadcast} allows a smaller tensor dimension $d_1$ to be broadcast to $d_2$ if $d_1$ is a \textit{right-aligned} subset of $d_2$ (i.e. $1080 \times 3$ is broadcastable to $1920 \times 1080 \times 3$, but $1920 \times 1080$ is not).

The rest of the rules codify legal tensor operations using the broadcast rules.
For instance, \textsc{OnnxAddR} specifies that adding two tensors is allowed so long as the first tensor dimension can be broadcast to that of the second.
Other rules governing \subtext{}, \multext{}, \divtext{}, \powtext{}, and \matmultext{} are omitted here, but can be found in Section 2.2 of the Supplementary Material.

\paragraph{Theorem.}
Formally, translational soundness states:
%If the translation from \colorlang to \onnx is sound, every well-typed value in \colorlang must translate to a well-typed value in \onnx. This property is formalized in \refrule{rule:translational_soundness}.

\begin{rulefigure}[h]
    \begin{prooftree}
        \AxiomC{$\llbracket v : (\tau, d) \rrbracket$}
        \UnaryInfC{$\llbracket v\rrbracket : \llbracket (\tau, d)\rrbracket$}
    \end{prooftree}
    %\caption{Translational soundness theorem.}
\end{rulefigure}

\paragraph{The Inductive Hypothesis.}
We use structural induction to prove translational soundness. 
The inductive hypothesis is given below.
%in \refrule{rule:inductive_hypothesis}.
The inductive hypothesis states that all sub-values of a well-typed value in \colorlang translate to well-typed values in \onnx. We use $v_i \subset v$ to mean that $v_i$ is an immediate sub-value of $v$.

\begin{rulefigure}[h]
    \begin{prooftree}
        \AxiomC{$\llbracket v_i \subset v\rrbracket$}
        \AxiomC{$\llbracket v_i: (\tau_i, d_i)\rrbracket$}
        \AxiomC{$\llbracket v : (\tau, d)\rrbracket$}
        \RightLabel{\textsc{IndHyp}}
        \TrinaryInfC{$\llbracket v_i\rrbracket : \llbracket (\tau_i, d_i)\rrbracket$}
    \end{prooftree}
    %\caption{The inductive hypothesis.}
    \label{rule:inductive_hypothesis}
\end{rulefigure}

\paragraph{Proof.}
We prove translational soundness by covering all cases of well-typed values.
We show a representative case, where the value $v$ is of the form $v_1 + v_2$ and belongs to the \texttt{XYZ} type. Other cases are similar in form.
See Section 2 of the Supplementary Material for the comprehensive proof.

%Our miniaturized proof will be show that \colorlang is translationally sound for the \texttt{Light}, \texttt{LMS}, \texttt{XYZ}, and \texttt{sRGB} types. We will only be evaluating the translational soundness of \colorlang values.

%We can prove translational soundness by showing that, for every type rule in \colorlang, any value that can be type checked under said rule is translated to a well-typed \onnx value.
%\Fig{fig:proof_coloradd_xyz} shows a proof for the \textsc{ColorAdd} rule for the \texttt{XYZ} + \texttt{XYZ} subcase.

%Our proof begins with the application of the rule \textsc{Conv-ColorAdd}. This is a converse of the \textsc{ColorAdd} rule. In order for \textsc{Conv-ColorAdd} to be a valid rule, the set of expressions that are type checked by \textsc{ColorAdd} must only be type check-able by \textsc{ColorAdd}. In other words, $v_1 + v_2 : (\colortype, d)$ must type check if and only if the premises for \textsc{ColorAdd} are satisfied. We know this to be true, as there is no other rule that type checks addition between color types.

\begin{figure*}[t]
{\scriptsize
    \begin{prooftree}
        \def\defaultHypSeparation{\hskip.05in}
        \AxiomC{$\llbracket v_1 + v_2 : (\xyztype, d)\rrbracket$}
        \RightLabel{\textsc{Conv-TriAdd}}
        \UnaryInfC{$\llbracket v_1, v_2: (\xyztype, d)\rrbracket$}
    
        \AxiomC{$\llbracket v_1, v_2 \subset v_1 + v_2\rrbracket$}
        \RightLabel{\textsc{IndHyp}}
        \BinaryInfC{$\llbracket v_1, v_2\rrbracket: \llbracket (\xyztype, d)\rrbracket$}

        \AxiomC{}
        \RightLabel{\textsc{TrivBroadcast}}
        \UnaryInfC{$\broadcastabletext(\llbracket (\xyztype, d)\rrbracket, \llbracket (\xyztype, d)\rrbracket)$}        
    
        \RightLabel{\textsc{OnnxAddR}}
        \BinaryInfC{$\addtext(\llbracket v_1\rrbracket, \llbracket v_2\rrbracket): \llbracket (\xyztype, d)\rrbracket$}
    \end{prooftree}
    }

{\scriptsize
\begin{prooftree}
    \AxiomC{$\addtext(\llbracket v_1\rrbracket, \llbracket v_2\rrbracket): \llbracket (\xyztype, d)\rrbracket$}

    \AxiomC{$\llbracket v_1, v_2: (\xyztype, d)\rrbracket$}
    \RightLabel{\textsc{T-Add}}
    \UnaryInfC{$\llbracket v_1 + v_2\rrbracket \triangleq \addtext(\llbracket v_1\rrbracket, \llbracket v_2\rrbracket)$}

    \RightLabel{\textsc{Subst}}
    \BinaryInfC{$\llbracket v_1 + v_2\rrbracket: \llbracket (\xyztype, d)\rrbracket$}
    
\end{prooftree}
}

    \caption{Proof of translational soundness for \texttt{XYZ} $+$ \texttt{XYZ} under \textsc{ColorAdd}}
    \label{fig:proof_coloradd_xyz}
\end{figure*}

\Fig{fig:proof_coloradd_xyz} shows the proof tree.
If the value $v_1 + v_2$ has the type \texttt{XYZ}, from the typing rules (\refrule{rule:rgb_add} \textsc{TristimulusAdd}) we know that $v_1$ and $v_2$ are both of type \texttt{XYZ}; this is represented by \textsc{Conv-TriAdd} in the proof tree. 
%From the application of the \textsc{Conv-ColorAdd} rule, we can conclude that $v_1$ and $v_2$ must be of physical type \texttt{XYZ}.
%Since $v_1$ and $v_2$ are typed in \colorlang and are subsets of $v_1 + v_2$, we can assume that $v_1$ and $v_2$ are typed as $\llbracket (\xyztype, d) \rrbracket$ in \onnx after translation due to the inductive hypothesis. 
By the inductive hypothesis,  $v_1$ and $v_2$ are typed as $\llbracket (\xyztype, d) \rrbracket$ in \onnx after translation. 
%We now know that $\llbracket v_1 \rrbracket$ and $\llbracket v_2 \rrbracket$ share the same Dimension Type in \onnx.
From \textsc{OnnxAddR}, we know that $\addtext(\llbracket v_1 \rrbracket, \llbracket v_2 \rrbracket)$ must also be typed as $\llbracket (\xyztype, d) \rrbracket$.

%$\llbracket v_1 + v_2 \rrbracket$ is translated to $\addtext(\llbracket v_1 \rrbracket, \llbracket v_2 \rrbracket)$ when $v_1$ and $v_2$ are of type \texttt{XYZ} according to our translational semantics. This translation is represented as \textsc{T-Add} in the proof tree.

From the translational semantics \textsc{T-Add}, we know that $\llbracket v_1 + v_2 \rrbracket$ translates to $\addtext(\llbracket v_1 \rrbracket, \llbracket v_2 \rrbracket)$.
Given that $\addtext(\llbracket v_1 \rrbracket, \llbracket v_2 \rrbracket)$ is typed as $\llbracket (\xyztype, d) \rrbracket$ in \onnx,
using substitution we can conclude that $\llbracket v_1 + v_2 \rrbracket$ is typed as $\llbracket (\xyztype, d) \rrbracket$ in \onnx.
Therefore, the translational soundness theorem is satisfied for \textsc{ColorAdd} when $v_1 + v_2$ is of type \texttt{XYZ}.

% Relegate type soundess proof to appendix, explain reasoning instead
\subsection{Type Soundness}
\label{sec:type-soundness-proof}

\revnew{
Our proof of translational soundness demonstrates that any well-typed value in \colorlang is guaranteed to translate to a well-typed \onnx value.
However, translational soundness does not imply type safety.
For example, translational soundness cannot ensure that \colorlang types are preserved after evaluation.
Consider the following hypothetical and faulty reflection type rule for reflection operations:
}

\begin{rulefigure}
    \begin{prooftree}
        \AxiomC{$\Gammahas v_1 : (\lighttype, d)$}
        \AxiomC{$\Gammahas v_2 : (\reflectancetype, d)$}
        \RightLabel{\textsc{WrongReflect}}
        \BinaryInfC{$\Gammahas v_1 \times v_2 : (\reflectancetype, d)$}
    \end{prooftree}
    \caption{\revnew{A faulty type rule for reflection operations. The correct type rule is \refrule{rule:reflect} (\textsc{Reflect}).}}
    \label{rule:faulty-reflect}
\end{rulefigure}

% \begin{rulefigure}
%     \begin{prooftree}
%         \AxiomC{$\lighttype(a_1) : (\lighttype, d)$}
%         \AxiomC{$\reflectancetype(a_2) : (\reflectancetype, d)$}
%         \RightLabel{\textsc{E-Reflect}}
%         \BinaryInfC{$\lighttype(a_1) \times \reflectancetype(a_2) \to \reflectancetype((a_1) \times (a_2))$}
%     \end{prooftree}
%     \caption{A faulty evaluation rule for reflection operations. The corresponding type rule is \refrule{rule:reflect} (\textsc{Reflect}).}
%     \label{rule:faulty-reflect-eval}
% \end{rulefigure}

\begin{rulefigure}
    \centering
    \begin{tabular}{c c}
        $\lighttype(a_1) \times \reflectancetype(a_2) \to \lighttype((a_1) \times (a_2))$ & \textsc{E-Reflect} \\
    \end{tabular}
    \caption{\revnew{The evaluation rule for reflection operations. The corresponding type rule is \refrule{rule:reflect} (\textsc{Reflect}).}}
    \label{rule:e-reflect}
\end{rulefigure}

\revnew{
In \refrule{rule:faulty-reflect}, reflection operations are mistakenly given the type \texttt{Reflectance}. 
The evaluation rule \textsc{E-Reflect} (\refrule{rule:e-reflect}) evaluates the physical type of $\lighttype(a_1) \times \reflectancetype(a_2)$ to \texttt{Light}.
Thus, the type of the expression $\lighttype(a_1) \times \reflectancetype(a_2)$ changes after evaluation. Preservation is violated.
Importantly, the translational soundness proof would not be able to catch this faulty type rule.
$v_1 \times v_2$ is translated to $\multext(\llbracket v_1 \rrbracket, \llbracket v_2 \rrbracket)$, according to the translational semantics in \Tbl{tab:color2onnx}.
$\multext(\llbracket v_1 \rrbracket, \llbracket v_2 \rrbracket)$ would still type check in \onnx, as the dimension types of $v_1$ and $v_2$ match.
}

\revnew{
We prove type soundness of \colorlang in addition to translational soundness.
Our approach is a straightforward proof of progress and preservation for each rule.
We have defined a set of evaluation rules to aid the proof.
The entire type soundness proof can be found in Sections 3 and 4 of the Supplementary Material.
}

\section{Type System Design Decisions}
\label{sec:type_system_design_decisions}
There were several considerations that informed our design decisions for \colorlang.
We detail them in this section.

\paragraph{Handling of Non-linear Tristimulus Color Spaces.} 
The \texttt{sRGB} and \texttt{opRGB} color spaces are non-linear tristimulus color spaces (\textbf{Gamma} paragraph of \Sect{sec:background}).
In \colorlang, interpolations between two \texttt{sRGB} or two \texttt{opRGB} objects are done in linear space (see the $\Phi(+, \srgbtype, \srgbtype) ( v_1, v_2 )$ entry in \Tbl{tab:color2onnx}).
We convert \texttt{sRGB} and \texttt{opRGB} values to linear space prior to interpolation as interpolation in linear spaces is uniform with respect to physical luminance.

An alternative is to perform non-linear tristimulus interpolation in a perceptual space instead.
This is because non-linear tristimulus color spaces are roughly uniform in \textit{perceived brightness}; programmers may expect interpolation between non-linear tristimulus colors to be perceptually uniform.
However, non-linear tristimulus color spaces are not uniform in chromaticity.
Principled perceptual interpolation must be done in a perceptually uniform space like \texttt{LAB}.

\colorlang could convert \texttt{sRGB} values to \texttt{LAB} values prior to interpolation.
However, there are many competing models of uniform color perception like CIELUV \cite{sharma2017digital}, CIELAB \cite{sharma2017digital}, and CAM16~\cite{cam16}.
Interpolation operations done in different perceptually uniform color spaces will yield different results.
We do not want to make any assumptions on what model of perceptual uniformity the programmer prefers. We therefore reject this design.

\revnew{
\paragraph{Tristimulus Addition Restrictions.} 
\refrule{rule:rgb_add} (\textsc{TristimulusAdd}) stipulates that the operands of a tristimulus addition must be of the same tristimulus type.
For example, addition between two \texttt{LMS} objects is valid, but addition between an \texttt{LMS} and an \texttt{XYZ} object is invalid.
However, there is nothing wrong, in principle, with interpolating two tristimulus values of different spaces.
% As discussed in \Prog{prog:phys-interp-space-agnostic} and \Fig{fig:translated-interpolation-xyz-lms}, physical interpolation is tristimulus color space agnostic.
One can imagine a set of translational semantics that, given an addition between an \texttt{LMS} or an \texttt{XYZ} object, first converts the \texttt{LMS} value to \texttt{XYZ} space or converts  the \texttt{XYZ} value to \texttt{LMS} space.
However, the output type of an operation between operands of differing spaces would be unspecified.
}

% Having an unspecified tristimulus color space as an output would not be problematic if all operations on tristimulus color spaces were space agnostic. 
% However, certain operations, like \refrule{rule:color_matrix_mul} (\textsc{TriScale}) are color space dependent.

\revnew{
To address that issue we could ideally use bidirectional type checking ~\cite{strict_bidirectional} to derive the expected color space of the tristimulus addition operation. 
For example, in \Prog{prog:bidirectional-example}, the output of the operation between the \texttt{XYZ} and \texttt{LMS} is assigned to the \texttt{mixed} variable. 
The \texttt{mixed} variable is annotated with the Python type hint \texttt{: cs.XYZ}, indicating that the programmer expects the \texttt{mixed} variable to be of the \texttt{XYZ} type.
Using the \texttt{mixed} variable's type hint, \colorlang could infer, through bidirectional type checking, that the programmer expects \texttt{cs.XYZ(...) + cs.LMS(...)} to output an object of type \texttt{XYZ}.
}
% This is because the Python type hint for the \texttt{mixed} variable indicates that the expected value of the addition operation is an \texttt{XYZ} object.

\begin{lstlisting}[language=Python, caption={\revnew{Bidirectional typing example. The cs.LMS(...) + cs.XYZ(...) operation is inferred to have the output type of cs.XYZ, as the \texttt{mixed} object that the operation output is assigned to has the \texttt{XYZ} type.}}, label={prog:bidirectional-example}]
mixed: cs.XYZ = cs.LMS(...) + cs.XYZ(...)
\end{lstlisting}

\revnew{
Unfortunately, implementing bidirectional type checking in a Python library is not feasible.
Python is a dynamically typed interpreter language.
It is impossible for \colorlang to obtain the type hint of the \texttt{mixed} variable prior to variable assignment.
Therefore, the output type of \texttt{cs.LMS(...) + cs.XYZ(...)} cannot be known when evaluating the addition operation.
}

\revnew{
It's important to note that while bidirectional typing would alleviate the stringent type restrictions for tristimulus space addition operations, perceptual addition (\refrule{rule:perceptual_add}) would remain the same.
This is because the outputs of arithmetic operations performed in different perceptual color spaces are not equivalent.
The user must still specify the perceptual color space used for interpolation.
}

\revnew{
\paragraph{Casting.}
Readers familiar with physical unit types (measurement types) literature~\cite{allen_object-oriented_2004, karr_incorporation_1978, dreiheller_programming_1986} will notice some parallels between our type system and those of previous works in the field of physical unit types.
However, not all assumptions of physical unit type systems are applicable to \colorlang.
This difference informs the design of our casting rule (\refrule{fig:casting_rule}). 
% However, this is one key difference, which informs an important design decision.
% Reference measurement type theory here
% Check if "measurement type theory" is good verbiage
% Explicitly give an example in measurement types that illustrates this point
}

\revnew{
In physical unit types literature, units can be easily converted into other units of measurement representing the same dimension\footnote{The dimension terminology here is not to be confused with dimension types in \colorlang. Dimensions in \colorlang refer to matrix dimensions.}, but not to units of different dimensions. 
Dimensions indicate the physical quantity being measured, and units indicate the standard of measurement~\cite{varkor_types_2018, allen_object-oriented_2004}.
Examples of dimensions include time, length, and mass.
Corresponding examples of units include seconds, meters, and grams.
A programmer can convert a Fahrenheit measurement to a Celsius measurement using the syntax of ~\citep{allen_object-oriented_2004} in \Prog{prog:ext-convert}. The conversion from Fahrenheit to Celsius is possible as both units share the same dimension type --- temperature.
However, conversion from Fahrenheit to, say, meters is not permitted:
the dimension types are mismatched.
}

\begin{lstlisting}[caption={\revnew{Conversion from Fahrenheit to Celsius in ~\citep{allen_object-oriented_2004}.}}, label={prog:ext-convert}]
fahrenheitValue.inUnit<CelsiusDegrees>()
\end{lstlisting}

\revnew{
There are seven dimensions represented in \colorlang: reflectance spectra, absorption spectra, scattering spectra, pigments, light spectra, color, and chromaticity.
Unlike measurement types, units in \colorlang can be converted to units of other dimensions.
For example, \texttt{Light} is a unit of the light spectra dimension, and \texttt{XYZ} is a unit of the color dimension.
\texttt{Light} values can be coerced into \texttt{XYZ} values.
However, the reverse is not true.
\texttt{XYZ} values cannot be coerced into \texttt{Light} values.
This is because the operation is under-determined; there exist multiple light spectra that correspond to a single color.
A similar relationship exists between units of the color dimension and the \texttt{Chromaticity} unit of the chromaticity dimension.
}

\revnew{
Since casting is not always bidirectional in \colorlang, we designed the \texttt{path\_exists} function in the \textsc{Cast} rule (\refrule{fig:casting_rule}) to enable the type checker to automatically determine if a casting is possible from one physical type to another.
Additionally, the edges of the graph in \Fig{fig:casting_graph} represent implemented casting algorithms.
Therefore, if the type checker confirms that an object is cast-able into another type, there must exist a corresponding series of castings that convert the object into the desired type.
The \texttt{path\_exists} rule and the casting graph are also designed to facilitate the addition of new types into \colorlang.
New types can be inserted as nodes into \Fig{fig:casting_graph}, and edges can be defined that correspond to implemented casting algorithms.
No modifications to the casting type rule needs to be made.
}

\paragraph{Tristimulus Pigment Mixing.}
% Refer back to occurrence of rules
% Specify that mix operator is unique to pigment mixing!
% Real intention: allow mixing pigments represented as tristimulus colors
The \texttt{mix} function is utilized to simulate the mixture of pigments, as specified in \refrule{rule:pigment} (\textsc{PgmtMix}).
\texttt{mix} is only applicable to \texttt{Pigment} objects.
We had originally planned for the \texttt{mix} operator to simulate pigment mixing for colors encoded in tristimulus color spaces as well.
However, this problem is ill-posed --- there exists an infinite number of pigment mixtures and ambient lighting conditions that are able to generate any given color.
There are algorithms of pigment mixing that operate on tristimulus colors like MixBox~\cite{sochorova2021practical}.
These algorithms make several assumptions about the constraints of pigment mixture and the ambient lighting.
The outputs are not principled, even though they approximate artists' expectations.
We cannot use these algorithms in a physically rigorous programming language like \colorlang.

\section{Optimizing \colorlang Programs}
\label{sec:optimization}

%\begin{figure}[t]
%	\begin{verbatim}
%(Add
%  (MatMul 
%    (input Spectrum1@F@1920_1080_65)
%    (input spectrum_to_xyz@T@65_3))
%  (MatMul
%    (input Spectrum2@F@1920_1080_65)
%    (input spectrum_to_xyz@T@65_3)))
%	\end{verbatim}
%	\caption{Toy Program in \tnsrlang}
%	\label{code:toy-tnsrlang}
%\end{figure}
%
%\begin{figure}[t]
%	\begin{verbatim}
%(MatMul
%  (Add
%    (input Spectrum1@F@1920_1080_65)
%    (input Spectrum2@F@1920_1080_65))
%  (input spectrum_to_xyz@T@65_3))
%	\end{verbatim}
%	\caption{Optimized Toy Program}
%	\label{code:toy-tnsrlang-optimized}
%\end{figure}

The translated \onnx program is optimized using equality saturation~\cite{tate2009equality, willsey2021egg}. 
The technique consists of a saturation phase, where a set of equivalent programs are enumerated using \textit{rewrite rules},
each of which replaces an expression with a semantically equivalent one;
the two expressions might have different run-time costs.
Then, in the extraction phase, the program with the cheapest cost is chosen. 

The rationale of using such an optimization strategy is discussed in \Sect{sec:ov}.
While equality saturation as an optimization technique is established, this section focuses on describing the specific design decisions we made in applying the technique to optimize \colorlang programs.
%The optimizer then generates an optimized string, which will be converted back to an \onnx program that will be executed using the standard \onnx Runtime.
%We utilize the equality saturation library \verb|egg| ~\cite{willsey2021egg} to optimize the \onnx binary.
%\verb|egg| takes as input a string representation of a program.
%In order to convert the \onnx file into a format acceptable by \verb|egg|, the \onnx file is serialized into our own lisp-like string syntax called \tnsrlang.
%Our implementation is heavily inspired by TENSAT \cite{tensat}.
%\Fig{code:toy-tnsrlang} shows the \tnsrlang representation of the toy program.

%\paragraph{Input Encoding.}
%We serialize the \onnx program as a LISP-like string as the input to the optimizer, as with common equality saturation tools~\cite{willsey2021egg}.
%The input encodes the dimension information, which allows us to estimate the cost of an operation.
%%recall equality saturation relies on user-specified cost models to extract optimized programs.
%For instance, if the element-wise addition operation is performed on two tensors with sizes $1920 \times 1080 \times 65$,
%we can deduce that the element-wise addition operation will require 134,784,000 individual additions and can weigh the cost accordingly.
%The dimension information only has to be annotated for each leaf tensor, from which dimensions of other tensors can be inferred.

\paragraph{Rewrite Rules.}
%As aforementioned, the optimizer explores the set of equivalent programs using rewrite rules. 
We design a small set of tensor algebra rewrite rules.
The rules are designed for the \onnx operations used in the translational semantics (\Tbl{tab:color2onnx}).
A list of rewrite rules can be found in the supplemental material.

Part of the rules are adapted from TASO~\cite{jia2019taso}, which investigates rewrite rules for tensor algebra containing up to four operators. 
Those rules are not only overkill for our purposes (because \colorlang uses a subset of tensor algebra), but also do not support operations that are unique to color programming.
Specifically, we include rewrite rules for the tensor exponentiation operator, which is important for implementing non-linear color space transformations.
%many color space transformations that involve non-linear color spaces such as \texttt{sRGB} and \texttt{LAB}.

%We find that it is sufficient to use a small subset of these rules, which yields  reduces optimization time while still obtaining good results for color programs. 
%For instance, we avoid rewrite rules that increase the number of operations.

\paragraph{Cost Function.}
In the extraction phase we need to compare the different programs yielded by the saturation phase.
We implement a cost function that estimates the run-time cost of a given program;
we do so empirically by calculating the total number of operations the program performs.
%different operations are weighted differently. 
We will show in \Sect{sec:exp} that even with these coarse-grain estimates we can still get statistically significant speedups. Better estimates will further improve performance.

Our cost function also enables constant propagation, which color programs commonly benefit from. 
Although the popular ONNX Runtime library~\cite{onnxruntime} performs constant propagation, it does so only when an \textit{entire} sub-expression consists of only computations on constants.
This is not always the form un-optimized \onnx programs are in.

We adjust our cost function to accommodate ONNX Runtime's constant propagation requirements.
During equality saturation, we flag each tensor in the input program to indicate whether it is a constant. 
For instance, M$_1$, M$_2$, and M$_3$ in \Tbl{tab:color2onnx} are constant matrices, and will be flagged as such.
%when encoded as part of the input to the optimizer. 
During extraction, our cost function assigns a cost of 0 to any sub-expression that can be pre-computed using only constant values.
The cheapest programs will be those that have sub-expressions consisting of only computations on constants.
This matches ONNX Runtime's requirements for constant propagation, and enables more aggressive constant propagation.

\paragraph{Implementation.}
We implement our optimizer using egg~\cite{willsey2021egg}, an e-graph-based equality saturation tool. 
% mewo bewo :3
We extend egg's interface to implement the cost function and rewrite rules, which are heavily modified from TENSAT's ~\citep{tensat} usage of egg. 
We also implement converters between egg's LISP-like string input and \onnx programs.

\section{Experimental Setup}
\label{sec:exp}

\paragraph{Benchmarking Programs.}
There is no standard benchmark for evaluating \colorlang, so we design six programs that are commonly seen in color programming to assess the overhead and optimization capabilities of \colorlang.
These programs are also tailored to exercise the entirety of \colorlang's type system.
We briefly describe how they exercise different syntactic features, typing rules, translational semantics, and  optimization cases in \colorlang.
The supplemental material contains the source code of all the programs.
% \colorlang's type system is representative of the color programming space as a whole. 

\textsc{\bfseries Color Space Conversion (SpaceConv)} is a program that converts an input image in \texttt{sRGB} space to an image in \texttt{opRGB} space.
\colorlang abstracts away the complexity of applying and removing gamma from \texttt{sRGB} and \texttt{opRGB}.
\texttt{sRGB} and \texttt{opRGB} are non-adjacent in \Fig{fig:casting_graph};
therefore, this casting requires an intermediate step in \onnx that can be eliminated by our optimizer.

%In \Sect{sec:translation:semantics}, we explained that casting between non-adjacent types is translated into a series of individual casting operations between intermediate types. These translations can be non-optimal, as multiple individual casting operations can often be folded into a single casting operation.

\begin{figure}[h]
    \centering
    \includegraphics[width=\linewidth]{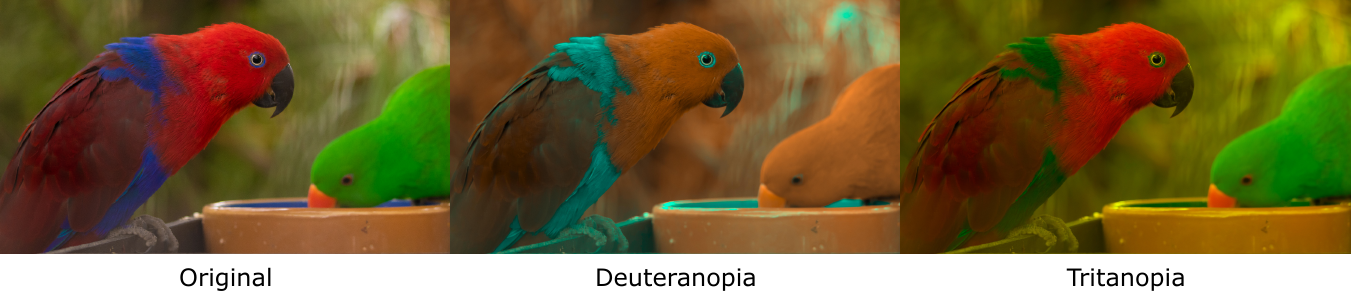}
    \caption{Color blindness simulation. Original image courtesy of Simon Amarasingham~\cite{amarasingham_red_2019}.}
    \label{fig:blind}
\end{figure}

\textsc{\bfseries Color Blindness Simulation (ColorBlindness)} simulates dichromatic color vision.
Example outputs of the program can be found in \Fig{fig:blind}.
% The program takes as input an image in \texttt{sRGB} space, which is first cast to \texttt{LMS} space.
While most images are originally encoded in the \texttt{sRGB} space, principled color blindness simulation must be done in the \texttt{LMS} space.
\colorlang automatically handles the implementation logic of casting from \texttt{sRGB} to \texttt{LMS} and back.
In the LMS space, the image is projected using a transformation matrix corresponding to a particular color blindness type~\cite{vienot1999digital}.
% The image is converted back to \texttt{sRGB} to be rendered.
%\Fig{fig:colorblind_sim} shows the output of this program for simulating Deuteranopia, a particular kind of ``red-green blindness'' due to the missing of M cones on the retina.
%As expected, red and green shades appear almost identical in the simulated image.
% This program allows us to exercise the \texttt{Matrix} type and its associated rules.
The program also demonstrates \colorlang's ability to treat colors as geometric objects and to cast them with a transformation matrix of the \texttt{Matrix} type.

% Humans with color blindness are missing one or more cone cell types (refer to \Sect{sec:background} for more information on cone cells). 
% Deuteranopes are missing M cones and tritanopes are missing S cones.
% As a result, deuteranopes are unable to distinguish between reds and greens, and tritanopes cannot discriminate blues from greens.

% People with only two cone cell types are called dichromats.
% Trichromats have a three dimensional sense of color -- one dimension for each cone cell type.
% Dichromats' sense of color is only two dimensional.
% In order to simulate color blindness, we must project three dimensional colors onto a two dimensional plane in LMS space ~\cite{burrus_understanding_2021, vienot1999digital}.

%\begin{figure}[t]
%    \centering
%    \includegraphics[width=.9\columnwidth]{colorblindness}
%    \caption{\textsc{\bfseries Color Blindness Simulation} converts an image seen by normal trichromatic vision (left) to what a deuteranopia would see (right). Original image courtesy of Simon Amarasingham ~\cite{amarasingham_red_2019}.}
%    \label{fig:colorblind_sim}
%\end{figure}

\begin{figure}[h]
    \centering
    \includegraphics[width=.8\linewidth]{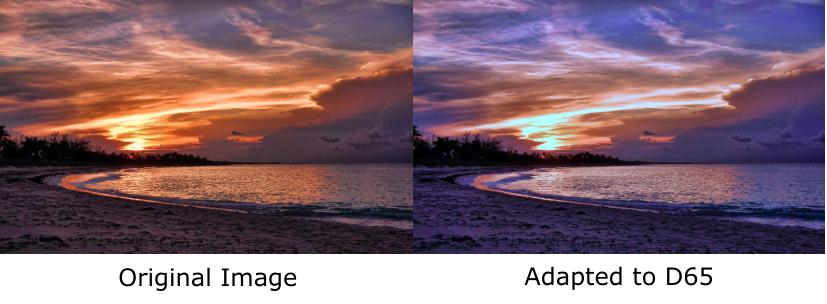}
    \caption{Chromatic adaptation simulation. Original image courtesy of Trish Hartman~\cite{hartmann_eleuthera_2012}.}
    \label{fig:adaptation}
\end{figure}

\textsc{\bfseries Chromatic Adaptation (Adaptation)}
is a program that simulates how the visual system adapts to the illuminant of a scene and preserves constant color perception across different illuminants~\cite{stockman2010color}.
Chromatic adaptation is the basis of white balancing in the camera image processing pipeline~\cite{rowlands2020color}.

Our implementation takes as input two \texttt{Light} spectra representing the original and target illuminants, as well as an input image in \texttt{sRGB} space.
It then applies the classic von Kries transformation matrix~\cite{rowlands2020color} in \texttt{LMS} space to calculate the adapted image.
%\Fig{fig:adaptation_sim} shows an example, where the left image is chromatically adapted/white balanced to the right image.
This program exercises the casting between the \texttt{Light} type and the Tristimulus Color Types.
The output of the program is shown in \Fig{fig:adaptation}, where the original image, captured under the CIE Standard Illuminant D35 (estimated), is adapted to the CIE Standard Illuminant D65 (typical daylight).
%Humans are exposed to a wide variety of different lighting conditions in daily life.
%Sunrises and sunsets produce reddish, high color temperature illuminants.
%The mid-day sun creates blueish, low color temperature illuminants.
%
%Despite the wide swathe of different illuminants we are exposed to, our color perception remains static. A white object will appear white regardless of the illuminant the object is placed under. Our ability to adjust to different illuminants is known as \textit{chromatic adaptation}. 

%\begin{figure}[t]
%    \centering
%    \includegraphics[width=.9\columnwidth]{adaptation}
%    \caption{\textsc{\bfseries Chromatic Adaptation} program essentially perform white balancing as commonly seen in camera raw processing pipeline. Original image courtesy of Trish Hartmann ~\cite{hartmann_eleuthera_2012}.}
%    \label{fig:adaptation_sim}
%\end{figure}

\textsc{\bfseries Color Interpolation (Interpolation)}
is a program that linearly interpolates the colors of two different \texttt{sRGB} images.
As mentioned in \Sect{sec:motivation}, programmers often attempt to interpolate colors in the \texttt{sRGB} color space, which is neither perceptually nor physically linear. 
This program demonstrates that \colorlang performs arithmetic operations on non-linear and non-perceptual color spaces in linear space for physical accuracy.
% It is the proper implementation of \Fig{fig:naive_color_interpolation}.

\textsc{\bfseries Pigment Mixing (Mixing)}
uses the \texttt{Pigment} type to simulate mixing two sets of pigments under typical daylight~\cite{henderson1963spectral}.
The mixing algorithm follows the Kubelka-Munk model~\citep{kubelka1931article, kubelka1948new} introduced in \Sect{sec:background}.
Pigment mixing is commonly simulated in digital painting apps~\cite{sochorova2021practical}.
Pigment mixing is a complex phenomenon to accurately model.
\colorlang abstracts away the complexity of the K-M model and allows the programmer to simulate pigment mixing through the \texttt{mix}($\cdot$) function and the \texttt{Pigment} type.
The \texttt{mix}($\cdot$) function and the \texttt{Pigment} type ensure that the pigment mixing simulation is done in the spectral space, with the correct spectral types as inputs.
%After mixing, the program calculates the resulting colors under the D65 illuminant. These colors are converted into an sRGB image. This program exercises our language's pigment mixture functionality.

\paragraph{LAB to HSV Conversion (LAB2HSV)}
%In order to demonstrate \colorlang's support for perceptual color spaces, our final program
converts an image in \texttt{LAB} space to an image in \texttt{HSV} space. \texttt{LAB} and \texttt{HSV} are complex, perceptual color spaces.
Translating images to and from these spaces involve multiple expensive and non-linear operations~\cite{lindbloom_xyz_2017}.
This program demonstrates \colorlang's ability to handle complicated programs.

\paragraph{Experimental Environment.}
The programs are compiled and run on two machines.  
Machine 1 has two Nvidia GeForce RTX 2080 GPU (\SI{8}{GB} VRAM each), an Intel Xeon Silver 4114 CPU, and \SI{64}{GB} DRAM. 
Machine 2 is equipped with two Nvidia GeForce RTX 4090 GPUs (\SI{24}{GB} VRAM each), an AMD Ryzen 9 7900X3D 12-Core Processor CPU, and \SI{128}{GB} of DRAM.
While both machines have two GPUs, only one is utilized during testing.

Python 3.11, ONNX Runtime 1.16, CUDA 11.8, and egg 0.9.4 are used during execution.
The compiler and optimizer are run on a single core of the CPU.
The optimized ONNX programs are run on either the CPU or the GPU, depending on the exact comparison being made.

%For each program, we recorded the average compilation time, optimization time, unoptimized time, and optimized time. 
To get statistically meaningful results, we compile, optimize, and run each \colorlang program 100 times. Both the unoptimized and optimized ONNX files are executed.
The one-tailed t-test~\cite{lakens2017equivalence} is used to test the statistical significance of our speed-ups.

\paragraph{Comparison Against Existing Solutions.}
We also benchmark \colorlang against existing Python solutions.
\colorlang's performance across the six benchmark programs is compared to equivalent programs written with the Colour-Science library ~\cite{colour_developers_colour_2015} and Numba~\cite{numba}. Colour-Science 0.4.4, Numba 0.59, and NumPy 1.25.2 are used.

Colour-Science is chosen as a benchmarking target as it is a commonly used Python library in the color science community.
However, it naturally comes with the Python interpreter overhead.
To construct a stronger baseline, we also choose to benchmark \colorlang against Numba, a compiler capable of translating Python and Numpy code into machine code ~\cite{numba}. Numba lacks the overhead of the Python interpreter.

% The purpose of benchmarking \colorlang against existing Python solutions is not to demonstrate \colorlang's optimization capabilities. Any performance gain is more a product of the differences between the programs' software stacks than the optimization that \colorlang provides. The purpose of this benchmark is to demonstrate that \colorlang can provide type safety guarantees without additional computation overhead. Therefore, we have elected to compare our \textit{unoptimized} ONNX executable runtimes against Numba and Colour-Science.

\section{Results}
\label{sec:res}

We perform an empirical analysis on two \colorlang programs to demonstrate how \colorlang's type system enforces correctness (\Sect{sec:res:example}).
\colorlang is capable of type checking, compiling, and optimizing a program with a minimal amount of overhead (\Sect{sec:res:oh}).
Even unoptimized \colorlang programs are faster than existing python solutions (\Sect{sec:res:existing}), and the optimizations bring up to a further 1.4 $\times$ speed-up (\Sect{sec:res:perf}).

\subsection{Case Study on the Type System}
\label{sec:res:example}

We provide an empirical study of two \colorlang programs.
We show \colorlang statically prevents programmers from specifying physically or perceptually unprincipled operations; \colorlang is also able to use type information to accurately translate user code.

\paragraph{\prog{ColorBlindness}.}
\Prog{prog:cooler_blind} shows the source code for the \prog{ColorBlindness} program. \Prog{prog:colour_blind} is the source code for the corresponding program written in Colour-Science and NumPy.
In the Colour-Science program, the \texttt{colorblind} function takes as input an image and a colorblind matrix of the NumPy array type. 
The user makes the assumption that the input image is of the \texttt{sRGB} type, and manually invokes specific operations to translate the image to the XYZ space (line 3), then to the LMS space (line 5). 
The operations on lines 3 and 5 take raw NumPy arrays as input. 
No information on the color space of the input image is recorded.
As a result, Colour-Science and NumPy are unable to validate the encoding of their input.

\begin{lstlisting}[language=Python, caption={Colour-Science \textsc{ColorBlindness} code.}, label={prog:colour_blind}]
def colorblind(image: np.ndarray, colorblind_matrix: np.ndarray):
    # Convert image from sRGB to XYZ
    xyz_image = colour.sRGB_to_XYZ(image)
    # Convert image from XYZ to LMS
    lms_image = xyz_image @ xyz_to_lms
    # Apply single-plane color blindness transformation
    lms_image_modulated = lms_image @ colorblind_matrix
    # Convert image back to sRGB and return
    xyz_image_modulated = lms_image_modulated @ lms_to_xyz
    return colour.XYZ_to_sRGB(xyz_image_modulated)
\end{lstlisting}

On lines 2 and 3 of the \colorlang program (\Prog{prog:cooler_blind}), the programmer specifies the color space and dimensions of the input data (\texttt{sRGB});
on line 5, the programmer also specifies an \texttt{sRGB} to \texttt{LMS} casting, which \colorlang types check to confirm that the \texttt{image} is indeed in the \texttt{sRGB} space.
%Using this information, \colorlang knows that the \texttt{image} object is a \texttt{sRGB} image.
%\colorlang can infer that the user intends to cast an \texttt{sRGB} object to a \texttt{LMS} object on line 5.
This transformation is then implemented correctly behind the scenes.

\begin{lstlisting}[language=Python, caption={\colorlang  \textsc{ColorBlindness} code.},
label={prog:cooler_blind}]
# Inputs
image = cs.create_input("image", [1080, 1920], cs.sRGB)
colorblind_matrix = cs.create_input("colorblind_matrix", [3, 3], cs.Matrix)
# Convert image to LMS
image_lms = cs.LMS(image)
# Apply single-plane color blindness transformation
colorblind_image_lms = cs.matmul(image_lms, colorblind_matrix)
# Convert back
colorblind_image = cs.sRGB(colorblind_image_lms)
\end{lstlisting}
%# Compilation
%cs.create_output(colorblind_image)
%cs.compile(path)

In line 7 of \Prog{prog:colour_blind} and \Prog{prog:cooler_blind}, the programmers apply the single-plane color blindness transformation matrix to the \texttt{LMS} image.
In the Colour-Science implementation, the matrix multiplication has no guarantee that the user-input \texttt{colorblind\_matrix} is $3 \times 3$. 
The NumPy program may crash during execution if an incorrect array is provided.
Since the dimension of \texttt{colorblind\_matrix} is specified in line 3 of \Prog{prog:cooler_blind}, the \colorlang program is guaranteed to run successfully.

\paragraph{\prog{Interpolation}.}
\Prog{prog:cooler_inter} is the \colorlang implementation of the \texttt{sRGB} interpolation. It is similar in function to the \texttt{sRGB} light addition program (\Prog{prog:correct_inter}) from \Sect{sec:motivation}.
%A corresponding NumPy implementation is shown in \Prog{prog:numpy_inter}.
The operations in line 5 of \Prog{prog:cooler_inter} are interpreted as physical operations, as \texttt{sRGB} is not a perceptual color space. 
% \fixme{do you have sRGB by scalar rule? should it be srgb add srgb?}
% NOTE: I do have an sRGB by scalar rule.
As a result, \colorlang performs the addition and multiplication operations in a linear, gamma-removed space behind the scenes.

\begin{lstlisting}[language=Python, caption={\colorlang interpolation code.}, label={prog:cooler_inter}]
# Inputs
image1 = cs.create_input("image1", [1080, 1920], cs.sRGB)
image2 = cs.create_input("image2", [1080, 1920], cs.sRGB)
# Interpolate between the two images 50/50
mixed = image1 * 0.5 + image2 * 0.5
\end{lstlisting}
%# Compilation
%cs.create_output(mixed)
%cs.compile(path)

To achieve an equivalent program in NumPy, as seen in \Prog{prog:numpy_inter}, considerably more code is required.
The additional code is error-prone.
The programmer needs to manually specify the gamma removal and application procedures in lines 3, 4, and 9.
Gamma values vary by color space.
In NumPy, there is no guarantee that the input images are represented in \texttt{sRGB} space. There is not even a guarantee that the two input images are of the same encoding.
%Therefore, it is possible for this interpolation function to be misused. 
The end result would be an interpolated image that is physically and perceptually inaccurate.

\begin{lstlisting}[language=Python, caption={NumPy interpolation code.}, label={prog:numpy_inter}]
def interpolate(image1: np.ndarray, image2: np.ndarray):
    # Remove gamma of sRGB color space
    image1_linear = (image1 / 255) ** 2.2
    image2_linear = (image2 / 255) ** 2.2
    # Interpolate 50/50
    image_avg = image1_linear * 0.5 + image2_linear * 0.5
    # Re-apply gamma
    return image_avg ** (1 / 2.2) * 255
\end{lstlisting}

\subsection{Compilation and Optimization Time}
\label{sec:res:oh}

%In the offline phase, we compile an input \colorlang program into an unoptimized \onnx program (and perform type checking along the way), and then execute an optimizer to produce an optimized \onnx program.

\begin{figure}[t]
    \centering
    \subfloat[\small{Machine 1}] {
        \includegraphics[width=0.48\linewidth]{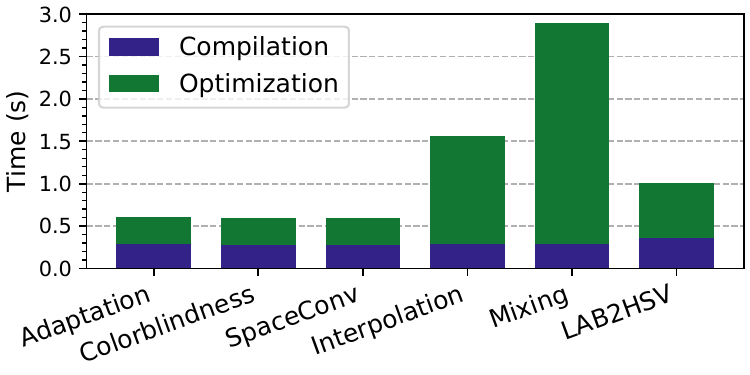}
    }
    \subfloat[\revnew{\small{Machine 2}}] {
        \includegraphics[width=0.48\linewidth]{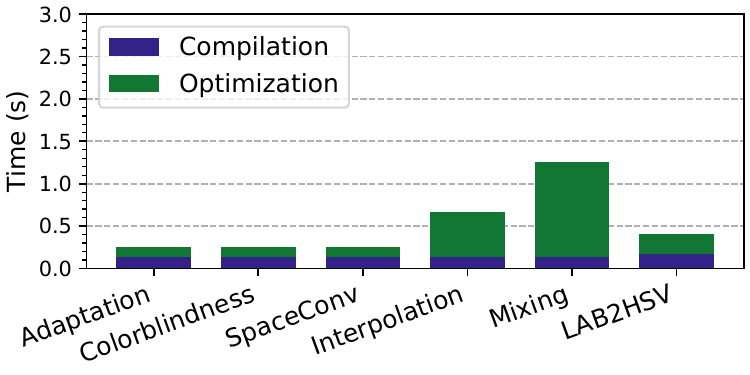}
    }
    
    \caption{Compilation and optimization time by program.}
    
    \label{fig:compilation_optimization_time}
\end{figure}

%We compile and optimize each program 100 times, and 
Compilation and optimization are both one-time costs.
Still, we show that these one-times costs are minimal.
\Fig{fig:compilation_optimization_time} compares the average compilation and optimization times for each program, which are all less than 3 seconds with low variance one machine 1.
The standard deviations of the two processes are below \SI{0.014}{\milli\second} and \SI{0.077}{\milli\second}, respectively.
\revnew{
On machine 2, compilation and optimization are faster. 
The average compilation time is less than 1.5 seconds with low variance.
The standard deviations of compilation and optimization are \SI{0.001}{\milli\second} and \SI{0.23}{\milli\second} respectively. 
}

The \prog{LAB2HSV} program has a notably higher compilation time in comparison to the other five programs on both machines. This is because the compiled \onnx program has a significantly higher \onnx operation count: 67, as opposed to about 15 in others. 
%there are 67 \onnx operations in the LAB to HSV transformation, whereas there are only about 15 operations in other programs.

%Interestingly, \colorlang operations are not equally expensive to compile.
%For instance, while \prog{LAB2HSV} requires a 57 \onnx operations, it only uses 1 \colorlang operation (casting);
%by contrast, \prog{Adaptation} uses 5 \colorlang operations, but its corresponding \onnx graph uses 14 \onnx operations.
%The fact that \onnx programs have far more operations than the \colorlang programs shows that \colorlang is capable of abstracting complex operations from programmers. \colorlang hides implementation details and allows programmers to focus on expressing color physics.

The optimization time is generally longer than compilation time, but still below 3 seconds even for the worst test case scenario.
\prog{Mixing} and \prog{Interpolation} are more expensive to optimize as they have a higher number of e-nodes and e-classes in their saturated e-graphs. 
%Therefore, saturation time and extraction time are considerably higher for these two programs. 
Note that equality saturation is a worst-case exponential time algorithm, and a timeout is usually used to limit the optimization time.
In our experiments, we place no such limits.
%In practice, very large \colorlang programs may take a long time to fully saturate and extract.
%This is an inherent limitation of the method. For practical usage, we would define a timeout or iteration limit for equality saturation. This means that equality saturation may not be guaranteed to saturate for sufficiently large programs. 

\subsection{Comparison with Existing Libraries}
\label{sec:res:existing}

\begin{figure}[t]
    \centering
    \subfloat[\small{Machine 1}] {
        \includegraphics[width=0.48\linewidth]{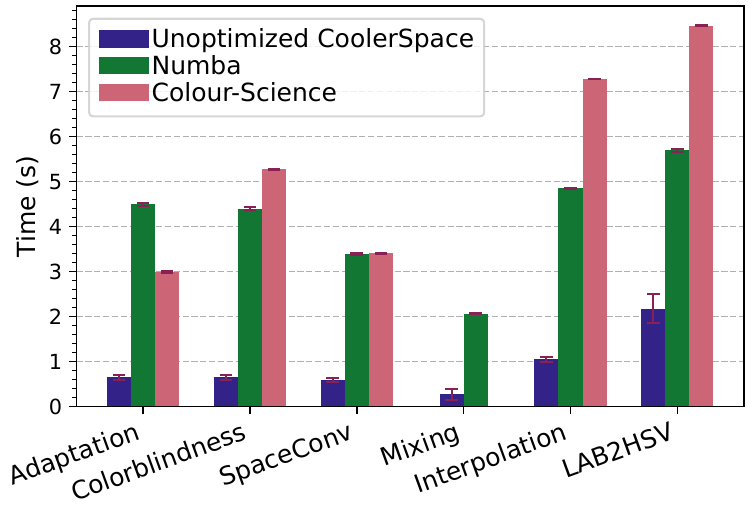}
    }
    \subfloat[\revnew{\small{Machine 2}}] {
        \includegraphics[width=0.48\linewidth]{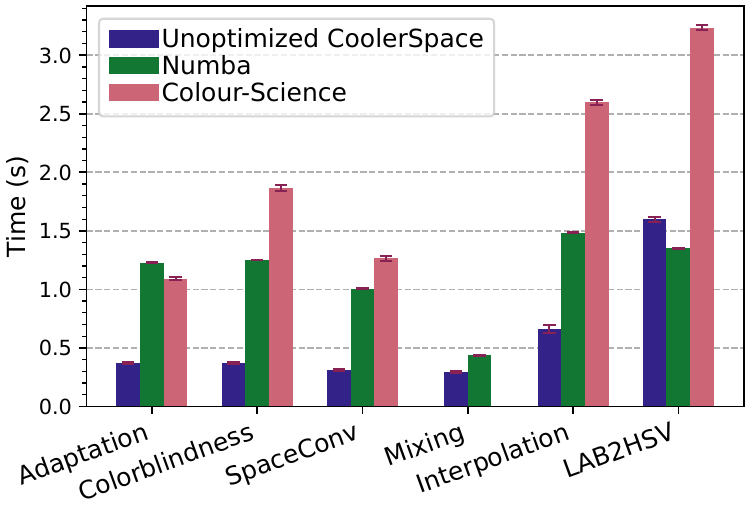}
    }
    
    \caption{\colorlang average program execution time on CPU benchmarked against other Python solutions. \prog{Adaptation}, \prog{Colorblindness}, \prog{SpaceConv}, \prog{Interpolation}, and \prog{LAB2HSV} were run on 5K images. \prog{Mixing} was run on input sizes of $600\times 600\times 89$. The error bars are 3 times the standard deviation of each set of tests. Colour-Science does not have a mixing program benchmark, as the library does not implement the Kubelka-Munk model. All Numba programs were compiled following the best practices recommended by Numba (e.g., with the \texttt{nopython} flag) for best performance. The JIT compile time costs of Numba are excluded from the time measurements.}
    \label{fig:existing_comparison}
\end{figure}

Even unoptimized \colorlang programs are faster than programs written using existing libraries by several times.
\Fig{fig:existing_comparison} compares the performance of \colorlang's unoptimized ONNX executables with Numba and Colour-Science. 
We report the CPU comparison results here since both baseline implementations are CPU-based.
\colorlang's has a $4.4 \times$ geomean speed-up over Numba, and a $5.7 \times$ geomean speed-up over Colour-Science on machine 1.
\revnew{
On machine 2, \colorlang has $3.4 \times$ and $2.1 \times$ geomean speedups over Colour-Science and Numba. All speed-ups are significant ($p < 0.01$), with the exception of \texttt{LAB2HSV} on machine 2. 
}

%All programs were run on large input sizes to reduce the relative impact of any static runtime costs on the overall measured runtime. The exact dimensions can be found in the caption of \Fig{fig:existing_comparison}. The unoptimized \colorlang executable was used as a baseline for measurement to isolate the impact of the optimization process. 

These results do not demonstrate \colorlang's optimization capabilities.
Rather, the performance gains are a product of the differences between the libraries' software stacks.
This evaluation is meant to show that, compared to existing color programming systems, \colorlang not only provides type safety guarantees but also does so without additional run-time overhead -- in fact, we provide a performance improvement.

\subsection{Optimization Effects}
\label{sec:res:perf}
\begin{figure}[t]
    \centering
    \subfloat[\small{Machine 1}]{
        \includegraphics[width=0.48\linewidth]{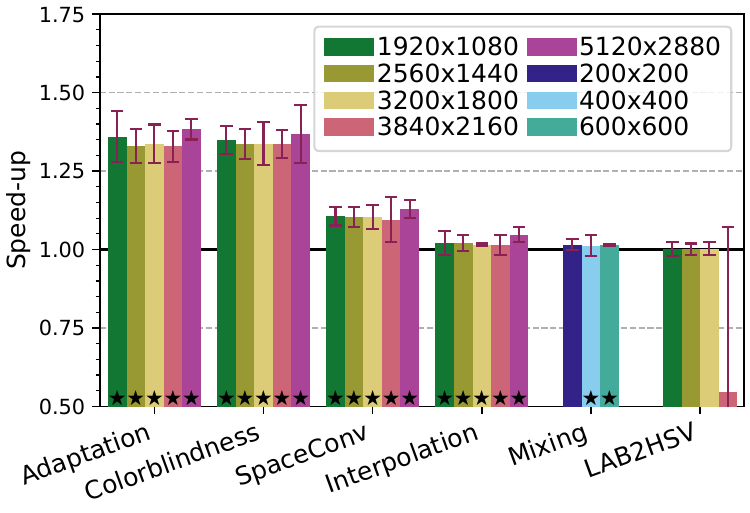}
    }
    \subfloat[\revnew{\small{Machine 2}}]{
        \includegraphics[width=0.48\linewidth]{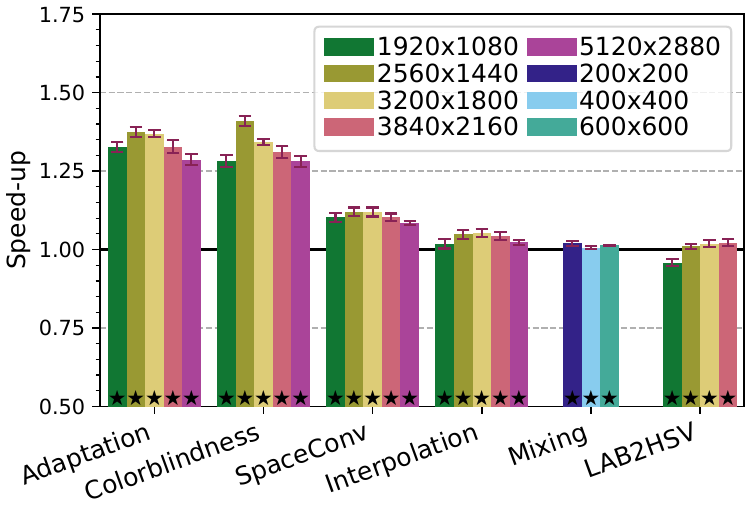}
    }
    
    \caption{Speed-ups by program under different resolutions. The error bars represent one standard deviation. A star above the bar indicates that the corresponding difference in runtime is statistically significant ($p < 0.01$). }
    \label{fig:runtime}
\end{figure}

We have also benchmarked the GPU execution times of both our optimized and unoptimized ONNX files. 
% \fixme{a nice segway into GPU results now.}
\Fig{fig:runtime} shows the speed-up of each GPU-run program under five different image resolutions.
The only exceptions are \prog{Mixing} and \prog{LAB2HSV}, which use a smaller set of resolutions to prevent out-of-memory issues during the run time.
% For instance, \prog{Mixing} deals with spectral data and is significantly more memory-consuming than other programs: each pigment is associated with multiple spectra, each of which requires 89 channels.
% Due to the high memory requirement, \prog{LAB2HSV} is unable to run for input sizes of $5120 \times 2880$ on our hardware. 
Our optimization yields a speed-up of about 12\% (geometric mean) across all programs and all resolutions on machine 1.
\revnew{
The speed-up is about 15\% on machine 2.
}
% The speed-up is insensitive to image resolution.
%This is because the programs are relatively short to run, so the kernel launch overhead is significant.
A star in \Fig{fig:runtime} 
indicates that the corresponding speedup is statistically significant.
%The input sizes of the pigment mixing and LAB to HSV programs differ from the other four programs, as these two programs are vastly more memory intensive. The lab machine is unable to run these programs at larger input sizes.
\revnew{
All programs show statistically significant speedups on some resolutions except for \prog{LAB2HSV} on machine 1.
}

%The magnitude of speedup varied by program and by input size. 

The speed-ups for \prog{ColorBlindness} and \prog{Adaptation} are attributed to the reduction in operations.
This is because the original algorithms operate in the \texttt{LMS} space whereas the input images are in the \texttt{sRGB} space, so a color space transformation is necessary;
our optimization identifies a reformulation of the arithmetics to operate directly in the \texttt{sRGB} space.
%programs, the input images were first converted to \texttt{LMS} space before the relevant transformations were applied. In the optimized ONNX executables, the number of operations done on the input images are minimized. The operations are re-arranged to operate on the smaller input.
The speed-ups for \prog{SpaceConv}, \prog{Mixing}, and \prog{Interpolation} can be mostly attributed to constant folding.
\prog{SpaceConv}'s execution time is dominated by a sequence of constant matrix multiplications, which are optimized away.

\prog{LAB2HSV} at the highest resolution on machine 1 shows a significant slowdown with high variance after optimization.
Further observation reveals a patch of 20 consecutive abnormally high run-times in the optimized \prog{LAB2HSV} results.
\revnew{
Transient memory management issues likely are the culprit, as this slowdown is not present on Machine 2.
Machine 2 has \SI{24}{GB} of VRAM, as opposed to machine 1's \SI{8}{GB} of VRAM.
}

\paragraph{Further Optimizations.}
We choose to compile to ONNX because of practical considerations:
ONNX is a convenient IR that is also cross-platform.
However, \onnx and ONNX Runtime are not necessarily the most speed-efficient options.
Our experiments show that equivalent CuPy~\cite{cupy} code is $1.2 \times$ faster than corresponding \colorlang programs.
% This is not a limitation of our type system or our optimizer. 
% In principle, we could further translate from \onnx IR into any other format.
% \colorlang does not perform any low-level optimizations. 
The difference in execution time is a product of the different implementations of CuPy and ONNX Runtime and, potentially, the GPU code generated by CuPy and ONNX Runtime.
In principle, we could directly compile to a more efficient target such as CuPy (or further translate from \onnx IR to that).
However, such backend-specific and/or device-specific optimizations are out of the scope of the current paper, which focuses on enforcing color program correctness without additional run-time overhead. 

\section{Related Works}
\label{sec:related}

\paragraph{Color Programing Libraries.}
%\colorlang is a Python library for color programming.
Commonly used Python libraries by color scientists include numpy~\cite{harris2020array}, OpenCV~\cite{opencv_library}, Pillow~\cite{clark_pillow_2023},
and Colour-Science~\cite{colour_developers_colour_2015}.
Unlike \colorlang, however, Numpy, OpenCV, and Colour-Science provide only a wrapper for tensor operations and color science algorithms without (type) checking physical correctness.
Pillow has a small set of informal ``types'' (referred to as ``modes'') to track image representation and to validate operations, but is much weaker than \colorlang.
%However, Pillow's types (referred to as "modes") are focused on digital representation.
For instance, Pillow can distinguish between \texttt{RGB} and \texttt{RGBA}, but not between actual color spaces: \texttt{opRGB} and \texttt{sRGB} images are treated identically.
Pillow also lacks types for other important physical objects such as light and material properties.
%Colour science~\cite{colour_developers_colour_2015} is another popular and useful python library for color programming. The library provides a wide range of useful color science functions like chromatic adaptation and generating illuminants from color temperature. However, these functions take as input array-like objects. No type checking is done to ensure that defined operations are physically meaningful.
%Unlike prior works, \colorlang provides programmers with physically meaningful typing information, and applies strict type rules to user programs.
Finally, unlike all existing libraries, \colorlang is a meta-programming library, which compiles a Python program into an optimized ONNX file. Other Python libraries perform no performance optimization.

\paragraph{Domain-Specific Languages.}
\colorlang is a programming system for color science.
Several domain-specific languages exist for the field of visual computing. All raise the level of programming abstraction.
Gator~\cite{gator} introduces a type system for expressing coordinate systems in rendering.
%SafeGI~\cite{ou2010safegi} also uses type checking to enforce physical correctness in rendering algorithms.
Simit~\cite{kjolstad2016simit} is a language for physics simulation.
Scenic~\cite{fremont2019scenic} is a language for probabilistically modeling a virtual scene.
None of these domain specific languages target color programming.
\colorlang uses tensor shape information to check the validity of operations.
Similar static type checking systems for tensor shape are seen in array programming languages~\cite{joisha2006algebraic, slepak2014array}.

%On the efficiency side of the coin, research has been done on compiling visual computing code written in dynamic languages \cite{yang2016vizgen} and image processing pipeline optimization \cite{ragan2013halide}. 

%In \colorlang, we prove that our lanugage is \textit{translationally sound} by showing that any type-checked expression in \colorlang translates to a type-checked expression in \onnx.
%By proving translational soundness, we can show that \colorlang is type sound given the assumption that \onnx is type sound.
%Our proof of translational soundness is inspired by GATOR~\cite{gator}. 

\paragraph{Tensor Representations and Optimizations.} 
\colorlang compiles user programs into tensor algebra represented by \onnx~\cite{onnx}.
We chose \onnx because it is cross-platform and has a vibrant user community ~\cite{jin2020compiling,danopoulos2021utilizing}.
%\onnx is a file format for machine learning models.
%\colorlang does not produce such models, but we find such formats useful as they allow us to express user programs in the language of tensor algebra.
%Inter-operable neural network standards other than \onnx have been created.

\colorlang uses equality saturation to optimize tensor algebra~\cite{tate2009equality}.
Other tensor optimization techniques are in principle applicable too~\cite{vasilache2018tensor, kjolstad2017tensor, chen2018tvm, susungi2018meta}.
Our rewrite rules borrow from previous works on tensor optimization~\cite{jia2019taso, tensat} but include color specific rules.
Our cost function is based on a first-order estimation of operation counts;
while empirically effective, future work can consider integrating hardware-aware models~\cite{ahrens2022autoscheduling, liu2022verified, anderson2021efficient}.
Our implementation is based on the egg library~\cite{willsey2021egg}
%(due to its clean programming interface)
with an extension to support constant propagation.
While it is possible to use egg to implement constant propagation, it requires serializing a large amount of constant values, which might increase memory usage and optimization time.
Our approach, by contrast, is symbolic.

\paragraph{Physical Unit Types.}
% Researchers have previously explored the application of type theory to physical units and dimensions~\cite{allen_object-oriented_2004, karr_incorporation_1978, dreiheller_programming_1986}.
% % This line of research encodes both physical units (i.e., millimeters, liters, kilograms) [\fixme{change two example lists to be parallel}] and dimensions (i.e., volume, area, mass) as types\footnote{the dimensions here are not to be confused with dimension in our system.}. 
% % A dimension can be expressed in multiple physical units, which are castable to each other, but dimensions are not castable.
% % TODO: Refer back to design decisions section
% % Path exists rule here too
% This line of research encodes both physical units (i.e., meters, liters, kilograms) and dimensions\footnote{The dimension terminology here is not to be confused with dimension types in \colorlang. Dimensions in \colorlang refer to matrix dimensions.} (i.e., length, volume, mass) as types.
% To our best knowledge, \colorlang is the first such system for color programming. 
% \colorlang encodes multiple different physical dimensions, including light spectra, reflectance spectra, and color.
% \colorlang also keeps track of  different color spaces (e.g. sRGB and XYZ), which can be seen as units of the color dimension.
Researchers have previously explored the application of type theory to physical units and dimensions~\cite{allen_object-oriented_2004, karr_incorporation_1978, dreiheller_programming_1986}.
This line of research encodes both physical units (i.e., meters, liters, kilograms) and dimensions\footnote{The dimension terminology here is not to be confused with dimension types in \colorlang. Dimensions in \colorlang refer to matrix dimensions.} (i.e., length, volume, mass) as types.
\revnew{
In physical unit types literature (also known as measurement types), a measurement can generally be converted to units of the same dimension, but not to units of other dimensions.
For example, 1 minute can be converted to 60 seconds, as minutes and seconds are both units of the time dimension.
1 minute cannot be converted to grams.
Such a conversion is nonsensical, as units of the time dimension cannot be converted to units of the mass dimension.
\colorlang also contains multiple dimensions, but unlike measurement types, \colorlang allows conversion between units of different dimensions.
This topic is discussed further in the \textbf{Casting} paragraph of \Sect{sec:type_system_design_decisions}.
}
% \paragraph{Approximate Data Types.}
% Unlike EnerJ ~\cite{enerj}, \colorlang excludes any operations that are approximate or lossy.
% Additionally, EnerJ utilizes bidirectional type checking to inform the compiler if a given operation should be approximate or precise.
% As discussed in \Sect{sec:type_system_design_decisions}, Python is unable to support bidirectional type checking.
% This limitation puts the burden on the user to specify which color space tristimulus addition operations should be performed in.
% The specification is ultimately wasted, as the output of tristimulus addition is color space agnostic.

% \paragraph{Information Flow Types.}
% Information flow types research ~\cite{jflow, info-flow-survey} utilizes type systems to statically enforce security policies on the information flow within programs.
% Non-confidentially typed data can influence confidentially typed data, but not vice versa.
% Similarly, \colorlang allows high cardinality typed values to be coerced into lower cardinality types, but not vice versa.

% Information flow types disallow programs that violate confidentiality.
% However, \colorlang's casting restrictions stem from fundamental mathematical constraints.
% It is feasible, but not advisable, to openly share password hashes.
% It is impossible to get the light spectrum of an RGB value.
% This topic is discussed in more detail in \Sect{sec:type_system_design_decisions}.

\revnew{
\paragraph{Approximate Data Types and Information Flow Types.}
Both approximate~\cite{enerj} and information flow~\cite{jflow, info-flow-survey} types enforce unidirectional information flow.
In EnerJ~\cite{enerj}, precise to approximate data flow is allowed, but the reverse is prohibited.
Similarly, information flow types prevent confidential data from affecting non-confidential outputs. Non-confidential data can affect confidential data.
\colorlang also restricts data flow in a unidirectional manner.
\texttt{Light} values can be coerced to \texttt{sRGB} values, but \texttt{sRGB} values can't be coerced to \texttt{Light} values.
}

\revnew{
In approximate data types and information flow type literature, the one directional information flow restrictions are designed to enforce best practices:
it is feasible, if inadvisable, to openly share password hashes.
By contrast, the restrictions on information flow in \colorlang are informed by mathematics and physics: it is mathematically impossible to derive light spectrum data from an \texttt{sRGB} value because there are infinitely many physical light spectra that correspond to the same \texttt{sRGB} color.
This topic is discussed further in the \textbf{Casting} paragraph of \Sect{sec:type_system_design_decisions}.
}

% \paragraph{MixBox.}
% MixBox~\cite{sochorova2021practical} 

% \fixme{to our best knowledge, \colorlang is the first such system for color programming.}
% \fixme{specifically, light spectrum vs. material property vs. color are different dimensions, and we identify physical units specific to color programming, e.g., RGB/XYZ spaces, in that they are representations of the same dimension, i.e., color.}

% \colorlang's type system encodes just the dimension information but not the physical unit information for colors.
% This is because color and vision scientists usually normalize their data. 
% For example, cone sensitivity response functions are normalized to peak at unity ~\citep{stockman2000spectral, STOCKMAN19992901}.
% Additionally, standardized color spaces like CIEXYZ 1931 do not have defined physical units.
% CIEXYZ coordinates are usually normalized such that the brightest color in the dataset has a $Y$ value of 1. \fixme{[Citation needed]}

% \colorlang is a domain-specific implementation of such a type system. Multiple dimension types are specified for different spectral distribution data (Reflectance, Scattering, etc.).
% Color is a single dimension. The different encodings of color represented in \colorlang like sRGB and CIEXYZ are akin to physical units of color.

%without having to serialize large constants.
%This is infeasible, as constant values can be large in color programming. A single full HD image in string form would be 93 million characters in size. Performing string manipulations on such a large amount of data would be prohibitively expensive. 

\section{Conclusion}
\label{sec:conclusion}

%\colorlang is a type-checked meta-programming system for color programming. 
\colorlang's type system prevents mathematically permissible but physically meaningless or incorrect computations.
\colorlang also automatically generates performance-optimized color science programs using equality saturation.
We see \colorlang as the first step, rather than the final work, in raising the level of programming abstraction for physical sciences.
Languages should empower domain experts to express the physical meaning of their programs. Correctness guarantees and performance optimizations should be left to the compiler and run-time system.

\section{Acknowledgements}
We would like to thank Professor Sreepathi Pai of the University of Rochester's Computer Science department for his feedback on our type system and soundness proofs.
We would also like to thank the anonymous reviewers of ASPLOS'24 and OOPLSA'24 for their constructive criticism and helpful insight.
The project is partially funded by NSF Award \#2225860.

\section{Data-Availability Statement}
\colorlang has four artifacts: the \colorlang library~\cite{coolerspace_github}, the \onnx optimizer~\cite{onneggs_github}, a wrapper for the equality saturation rust library egg~\cite{eggwrap_github, willsey2021egg}, and a set of benchmarking programs~\cite{benchmarker_github}. The programs are open-source and freely available on GitHub. Additionally, the all artifacts are available in Zenodo. The artifacts are also available on Zenodo~\cite{coolerspace_zenodo}.

\includepdf[pages=-]{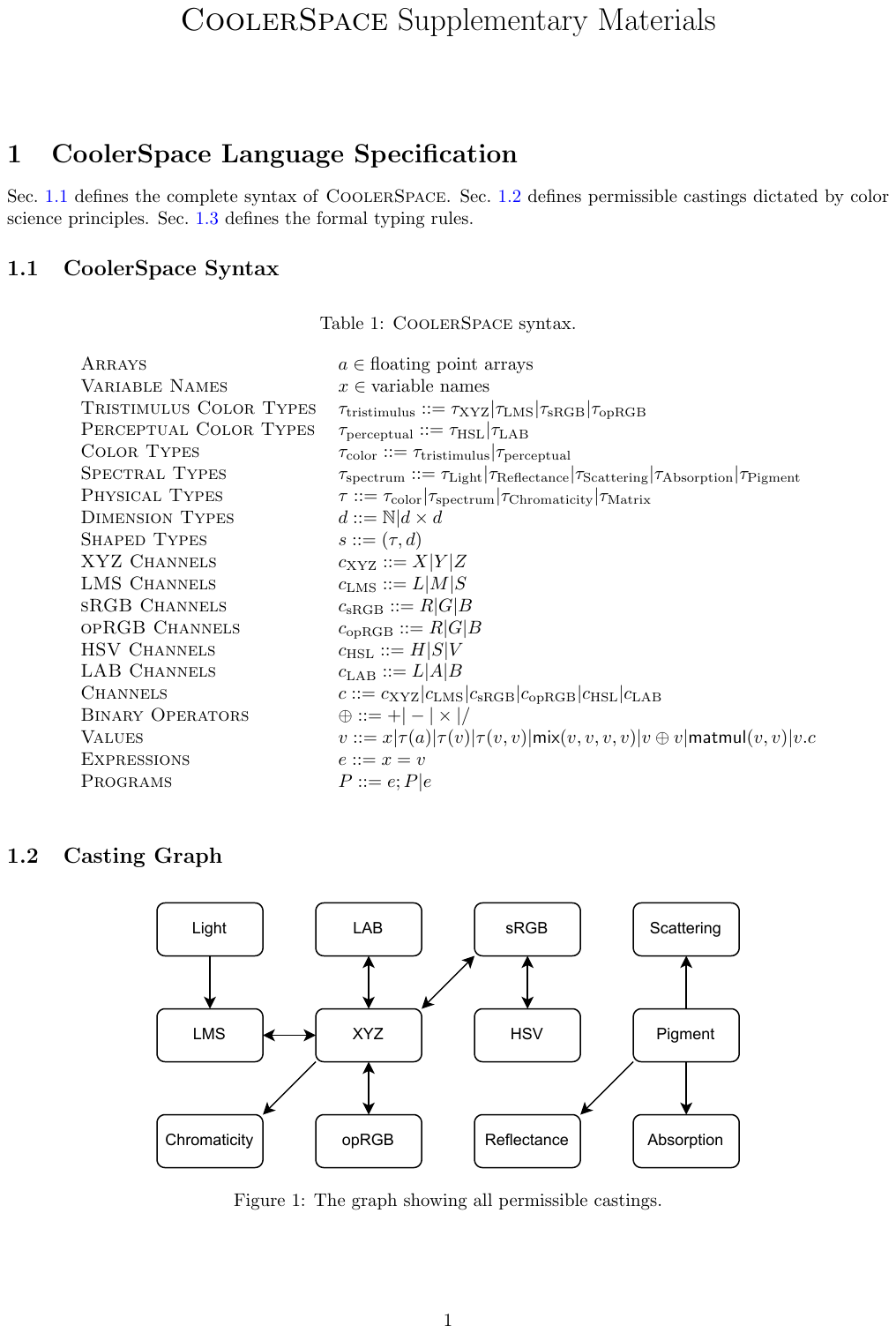}

% Split experimental setup, results, and evaluation. Results should include some form of analysis.

% All figures need to be vectorized

\bibliographystyle{ACM-Reference-Format}
\bibliography{new-tex/new-references}


\begin{thebibliography}{78}

%%% the Web address.

\ifx \showCODEN    \undefined \def \showCODEN     #1{\unskip}     \fi
\ifx \showDOI      \undefined \def \showDOI       #1{#1}\fi
\ifx \showISBNx    \undefined \def \showISBNx     #1{\unskip}     \fi
\ifx \showISBNxiii \undefined \def \showISBNxiii  #1{\unskip}     \fi
\ifx \showISSN     \undefined \def \showISSN      #1{\unskip}     \fi
\ifx \showLCCN     \undefined \def \showLCCN      #1{\unskip}     \fi
\ifx \shownote     \undefined \def \shownote      #1{#1}          \fi
\ifx \showarticletitle \undefined \def \showarticletitle #1{#1}   \fi
\ifx \showURL      \undefined \def \showURL       {\relax}        \fi
% invisible to TeX
\providecommand\bibfield[2]{#2}
\providecommand\bibinfo[2]{#2}
\providecommand\natexlab[1]{#1}
\providecommand\showeprint[2][]{arXiv:#2}

\bibitem[yet(2011)]%
        {yet_another_wrong_rgb_mixing}
 \bibinfo{year}{2011}\natexlab{}.
\newblock \bibinfo{title}{Answer to "adding/mixing colors in {HSV} {Space}"}.
\newblock \bibinfo{howpublished}{\url{https://stackoverflow.com/a/7388476}}.
\newblock


\bibitem[wro(2014)]%
        {wrong_srgb_interpolation}
 \bibinfo{year}{2014}\natexlab{}.
\newblock \bibinfo{title}{Answer to "{Interpolate} from one color to another"}.
\newblock \bibinfo{howpublished}{\url{https://stackoverflow.com/a/21010385}}.
\newblock


\bibitem[wro(2016)]%
        {wrong_hsv_interpolation}
 \bibinfo{year}{2016}\natexlab{}.
\newblock \bibinfo{title}{Weird interpolation between colors in hsv?}
\newblock \bibinfo{howpublished}{\url{https://stackoverflow.com/q/37471461}}.
\newblock


\bibitem[thi(2021)]%
        {third_srgb_error}
 \bibinfo{year}{2021}\natexlab{}.
\newblock \bibinfo{title}{How to calculate (a physical) ratio of colors to achieve a target color?}
\newblock \bibinfo{howpublished}{\url{https://math.stackexchange.com/q/4335003}}.
\newblock


\bibitem[adriahf(2016)]%
        {color_science_runtime_complaint_2}
\bibfield{author}{\bibinfo{person}{adriahf}.} \bibinfo{year}{2016}\natexlab{}.
\newblock \bibinfo{title}{Increase the velocity of the calculations. · {Issue} \#302 · colour-science/colour}.
\newblock
\newblock
\urldef\tempurl%
\url{https://github.com/colour-science/colour/issues/302}
\showURL{%
\tempurl}


\bibitem[Ahrens et~al\mbox{.}(2022)]%
        {ahrens2022autoscheduling}
\bibfield{author}{\bibinfo{person}{Willow Ahrens}, \bibinfo{person}{Fredrik Kjolstad}, {and} \bibinfo{person}{Saman Amarasinghe}.} \bibinfo{year}{2022}\natexlab{}.
\newblock \showarticletitle{Autoscheduling for sparse tensor algebra with an asymptotic cost model}. In \bibinfo{booktitle}{\emph{Proceedings of the 43rd ACM SIGPLAN International Conference on Programming Language Design and Implementation}}. \bibinfo{pages}{269--285}.
\newblock
\urldef\tempurl%
\url{https://doi.org/10.1145/3519939.3523442}
\showDOI{\tempurl}


\bibitem[Allen et~al\mbox{.}(2004)]%
        {allen_object-oriented_2004}
\bibfield{author}{\bibinfo{person}{Eric Allen}, \bibinfo{person}{David Chase}, \bibinfo{person}{Victor Luchangco}, \bibinfo{person}{Jan-Willem Maessen}, {and} \bibinfo{person}{Guy~L. Steele}.} \bibinfo{year}{2004}\natexlab{}.
\newblock \showarticletitle{Object-oriented units of measurement}. In \bibinfo{booktitle}{\emph{Proceedings of the 19th annual {ACM} {SIGPLAN} conference on Object-oriented programming, systems, languages, and applications}} (New York, {NY}, {USA}, 2004-10-01) \emph{(\bibinfo{series}{{Oopsla} '04})}. \bibinfo{publisher}{Association for Computing Machinery}, \bibinfo{pages}{384--403}.
\newblock
\showISBNx{9781581138313}
\urldef\tempurl%
\url{https://doi.org/10.1145/1028976.1029008}
\showDOI{\tempurl}


\bibitem[Amarasingham(2019)]%
        {amarasingham_red_2019}
\bibfield{author}{\bibinfo{person}{Simon Amarasingham}.} \bibinfo{year}{2019}\natexlab{}.
\newblock \bibinfo{title}{Red and green {Eclectus} {Parrots}}.
\newblock
\newblock
\urldef\tempurl%
\url{https://www.flickr.com/photos/22896868@N05/49257147352/}
\showURL{%
\tempurl}


\bibitem[Anderson et~al\mbox{.}(2021)]%
        {anderson2021efficient}
\bibfield{author}{\bibinfo{person}{Luke Anderson}, \bibinfo{person}{Andrew Adams}, \bibinfo{person}{Karima Ma}, \bibinfo{person}{Tzu-Mao Li}, \bibinfo{person}{Tian Jin}, {and} \bibinfo{person}{Jonathan Ragan-Kelley}.} \bibinfo{year}{2021}\natexlab{}.
\newblock \showarticletitle{Efficient automatic scheduling of imaging and vision pipelines for the GPU}.
\newblock \bibinfo{journal}{\emph{Proceedings of the ACM on Programming Languages}} \bibinfo{volume}{5}, \bibinfo{number}{Oopsla} (\bibinfo{year}{2021}), \bibinfo{pages}{1--28}.
\newblock
\urldef\tempurl%
\url{https://doi.org/10.1145/3485486}
\showDOI{\tempurl}


\bibitem[Berns(2016)]%
        {berns2016color}
\bibfield{author}{\bibinfo{person}{Roy~S Berns}.} \bibinfo{year}{2016}\natexlab{}.
\newblock \bibinfo{booktitle}{\emph{Color science and the visual arts: a guide for conservators, curators, and the curious}}.
\newblock \bibinfo{publisher}{Getty Publications}.
\newblock
\urldef\tempurl%
\url{https://doi.org/10.17613/d08z-ga34}
\showDOI{\tempurl}


\bibitem[Bradski(2000)]%
        {opencv_library}
\bibfield{author}{\bibinfo{person}{G. Bradski}.} \bibinfo{year}{2000}\natexlab{}.
\newblock \showarticletitle{The OpenCV Library}.
\newblock \bibinfo{journal}{\emph{Dr. Dobb's Journal of Software Tools}} (\bibinfo{year}{2000}).
\newblock


\bibitem[Brainard and Stockman(2010)]%
        {brainard2010colorimetry}
\bibfield{author}{\bibinfo{person}{David~H Brainard} {and} \bibinfo{person}{Andrew Stockman}.} \bibinfo{year}{2010}\natexlab{}.
\newblock \showarticletitle{Colorimetry}.
\newblock \bibinfo{journal}{\emph{The Optical Society of America Handbook of Optics}}  \bibinfo{volume}{3} (\bibinfo{year}{2010}), \bibinfo{pages}{10--1}.
\newblock


\bibitem[Brettel et~al\mbox{.}(1997)]%
        {brettel1997computerized}
\bibfield{author}{\bibinfo{person}{Hans Brettel}, \bibinfo{person}{Fran{\c{c}}oise Vi{\'e}not}, {and} \bibinfo{person}{John~D Mollon}.} \bibinfo{year}{1997}\natexlab{}.
\newblock \showarticletitle{Computerized simulation of color appearance for dichromats}.
\newblock \bibinfo{journal}{\emph{Josa a}} \bibinfo{volume}{14}, \bibinfo{number}{10} (\bibinfo{year}{1997}), \bibinfo{pages}{2647--2655}.
\newblock
\urldef\tempurl%
\url{https://doi.org/10.1364/josaa.14.002647}
\showDOI{\tempurl}


\bibitem[Broadbent(2001)]%
        {broadbent2001basic}
\bibfield{author}{\bibinfo{person}{Arthur~D Broadbent}.} \bibinfo{year}{2001}\natexlab{}.
\newblock \bibinfo{booktitle}{\emph{Basic principles of textile coloration}}. Vol.~\bibinfo{volume}{132}.
\newblock \bibinfo{publisher}{Society of Dyers and Colorists Bradford}.
\newblock


\bibitem[Chang and Chen(2024)]%
        {eggwrap_github}
\bibfield{author}{\bibinfo{person}{Jiwon Chang} {and} \bibinfo{person}{Ethan Chen}.} \bibinfo{year}{2024}\natexlab{}.
\newblock \bibinfo{title}{eggwrap}.
\newblock
\newblock
\urldef\tempurl%
\url{https://github.com/horizon-research/eggwrap}
\showURL{%
\tempurl}


\bibitem[Chen(2024a)]%
        {coolerspace_github}
\bibfield{author}{\bibinfo{person}{Ethan Chen}.} \bibinfo{year}{2024}\natexlab{a}.
\newblock \bibinfo{title}{{CoolerSpace}}.
\newblock
\newblock
\urldef\tempurl%
\url{https://github.com/horizon-research/CoolerSpace}
\showURL{%
\tempurl}


\bibitem[Chen(2024b)]%
        {benchmarker_github}
\bibfield{author}{\bibinfo{person}{Ethan Chen}.} \bibinfo{year}{2024}\natexlab{b}.
\newblock \bibinfo{title}{{CoolerSpace} {Benchmarker}}.
\newblock
\newblock
\urldef\tempurl%
\url{https://github.com/horizon-research/CoolerSpaceBenchmarker}
\showURL{%
\tempurl}


\bibitem[Chen and Chang(2024)]%
        {onneggs_github}
\bibfield{author}{\bibinfo{person}{Ethan Chen} {and} \bibinfo{person}{Jiwon Chang}.} \bibinfo{year}{2024}\natexlab{}.
\newblock \bibinfo{title}{onneggs}.
\newblock
\newblock
\urldef\tempurl%
\url{https://github.com/horizon-research/onneggs}
\showURL{%
\tempurl}


\bibitem[Chen et~al\mbox{.}(2024)]%
        {coolerspace_zenodo}
\bibfield{author}{\bibinfo{person}{Ethan Chen}, \bibinfo{person}{Jiwon Chang}, {and} \bibinfo{person}{Yuhao Zhu}.} \bibinfo{year}{2024}\natexlab{}.
\newblock \bibinfo{booktitle}{\emph{CoolerSpace Artifacts}}.
\newblock
\urldef\tempurl%
\url{https://doi.org/10.5281/zenodo.13621721}
\showDOI{\tempurl}


\bibitem[Chen et~al\mbox{.}(2018)]%
        {chen2018tvm}
\bibfield{author}{\bibinfo{person}{Tianqi Chen}, \bibinfo{person}{Thierry Moreau}, \bibinfo{person}{Ziheng Jiang}, \bibinfo{person}{Lianmin Zheng}, \bibinfo{person}{Eddie Yan}, \bibinfo{person}{Haichen Shen}, \bibinfo{person}{Meghan Cowan}, \bibinfo{person}{Leyuan Wang}, \bibinfo{person}{Yuwei Hu}, \bibinfo{person}{Luis Ceze}, {et~al\mbox{.}}} \bibinfo{year}{2018}\natexlab{}.
\newblock \showarticletitle{$\{$TVM$\}$: An automated $\{$End-to-End$\}$ optimizing compiler for deep learning}. In \bibinfo{booktitle}{\emph{13th USENIX Symposium on Operating Systems Design and Implementation (OSDI 18)}}. \bibinfo{pages}{578--594}.
\newblock
\urldef\tempurl%
\url{https://doi.org/10.5555/3291168.3291211}
\showDOI{\tempurl}


\bibitem[Chlipala et~al\mbox{.}(2005)]%
        {strict_bidirectional}
\bibfield{author}{\bibinfo{person}{Adam Chlipala}, \bibinfo{person}{Leaf Petersen}, {and} \bibinfo{person}{Robert Harper}.} \bibinfo{year}{2005}\natexlab{}.
\newblock \showarticletitle{Strict bidirectional type checking}. In \bibinfo{booktitle}{\emph{Proceedings of the 2005 ACM SIGPLAN International Workshop on Types in Languages Design and Implementation}} (Long Beach, California, USA) \emph{(\bibinfo{series}{Tldi '05})}. \bibinfo{publisher}{Association for Computing Machinery}, \bibinfo{address}{New York, NY, USA}, \bibinfo{pages}{71–78}.
\newblock
\showISBNx{1581139993}
\urldef\tempurl%
\url{https://doi.org/10.1145/1040294.1040301}
\showDOI{\tempurl}


\bibitem[Clark(2023)]%
        {clark_pillow_2023}
\bibfield{author}{\bibinfo{person}{Jeffrey Clark}.} \bibinfo{year}{2023}\natexlab{}.
\newblock \bibinfo{title}{Pillow}.
\newblock \bibinfo{howpublished}{\url{https://github.com/python-pillow/Pillow}}.
\newblock


\bibitem[Danopoulos et~al\mbox{.}(2021)]%
        {danopoulos2021utilizing}
\bibfield{author}{\bibinfo{person}{Dimitrios Danopoulos}, \bibinfo{person}{Christoforos Kachris}, {and} \bibinfo{person}{Dimitrios Soudris}.} \bibinfo{year}{2021}\natexlab{}.
\newblock \showarticletitle{Utilizing cloud FPGAs towards the open neural network standard}.
\newblock \bibinfo{journal}{\emph{Sustainable Computing: Informatics and Systems}}  \bibinfo{volume}{30} (\bibinfo{year}{2021}), \bibinfo{pages}{100520}.
\newblock
\urldef\tempurl%
\url{https://doi.org/10.1016/j.suscom.2021.100520}
\showDOI{\tempurl}


\bibitem[Developers(2015)]%
        {colour_developers_colour_2015}
\bibfield{author}{\bibinfo{person}{Colour Developers}.} \bibinfo{year}{2015}\natexlab{}.
\newblock \bibinfo{title}{Colour {Science} for {Python}}.
\newblock \bibinfo{howpublished}{\url{https://www.colour-science.org/}}.
\newblock


\bibitem[Developers(2021)]%
        {onnxruntime}
\bibfield{author}{\bibinfo{person}{ONNX~Runtime Developers}.} \bibinfo{year}{2021}\natexlab{}.
\newblock \bibinfo{title}{ONNX Runtime}.
\newblock \bibinfo{howpublished}{\url{https://onnxruntime.ai/}}.
\newblock
\newblock
\shownote{Version: 1.16}.


\bibitem[Dreiheller et~al\mbox{.}(1986)]%
        {dreiheller_programming_1986}
\bibfield{author}{\bibinfo{person}{A Dreiheller}, \bibinfo{person}{B Mohr}, {and} \bibinfo{person}{M Moerschbacher}.} \bibinfo{year}{1986}\natexlab{}.
\newblock \showarticletitle{Programming pascal with physical units}.
\newblock \bibinfo{journal}{\emph{ACM SIGPLAN Notices}} \bibinfo{volume}{21}, \bibinfo{number}{12} (\bibinfo{date}{Dec.} \bibinfo{year}{1986}), \bibinfo{pages}{114--123}.
\newblock
\showISSN{0362-1340}
\urldef\tempurl%
\url{https://doi.org/10.1145/15042.15048}
\showDOI{\tempurl}


\bibitem[Duncan(1940)]%
        {duncan1940colour}
\bibfield{author}{\bibinfo{person}{DR Duncan}.} \bibinfo{year}{1940}\natexlab{}.
\newblock \showarticletitle{The colour of pigment mixtures}.
\newblock \bibinfo{journal}{\emph{Proceedings of the Physical Society}} \bibinfo{volume}{52}, \bibinfo{number}{3} (\bibinfo{year}{1940}), \bibinfo{pages}{390}.
\newblock


\bibitem[Ebenezer(2021)]%
        {color_science_runtime_complaint_1}
\bibfield{author}{\bibinfo{person}{Joshua Ebenezer}.} \bibinfo{year}{2021}\natexlab{}.
\newblock \bibinfo{title}{Computational speed for conversions · {Issue} \#788 · colour-science/colour}.
\newblock
\newblock
\urldef\tempurl%
\url{https://github.com/colour-science/colour/issues/788}
\showURL{%
\tempurl}


\bibitem[Fairchild and Reniff(1995)]%
        {fairchild1995time}
\bibfield{author}{\bibinfo{person}{Mark~D Fairchild} {and} \bibinfo{person}{Lisa Reniff}.} \bibinfo{year}{1995}\natexlab{}.
\newblock \showarticletitle{Time course of chromatic adaptation for color-appearance judgments}.
\newblock \bibinfo{journal}{\emph{Josa A}} \bibinfo{volume}{12}, \bibinfo{number}{5} (\bibinfo{year}{1995}), \bibinfo{pages}{824--833}.
\newblock
\urldef\tempurl%
\url{https://doi.org/10.1016/s0042-6989(00)00050-x}
\showDOI{\tempurl}


\bibitem[Fremont et~al\mbox{.}(2019)]%
        {fremont2019scenic}
\bibfield{author}{\bibinfo{person}{Daniel~J Fremont}, \bibinfo{person}{Tommaso Dreossi}, \bibinfo{person}{Shromona Ghosh}, \bibinfo{person}{Xiangyu Yue}, \bibinfo{person}{Alberto~L Sangiovanni-Vincentelli}, {and} \bibinfo{person}{Sanjit~A Seshia}.} \bibinfo{year}{2019}\natexlab{}.
\newblock \showarticletitle{Scenic: a language for scenario specification and scene generation}. In \bibinfo{booktitle}{\emph{Proceedings of the 40th ACM SIGPLAN Conference on Programming Language Design and Implementation}}. \bibinfo{pages}{63--78}.
\newblock
\urldef\tempurl%
\url{https://doi.org/10.1145/3314221.3314633}
\showDOI{\tempurl}


\bibitem[Geisler et~al\mbox{.}(2020)]%
        {gator}
\bibfield{author}{\bibinfo{person}{Dietrich Geisler}, \bibinfo{person}{Irene Yoon}, \bibinfo{person}{Aditi Kabra}, \bibinfo{person}{Horace He}, \bibinfo{person}{Yinnon Sanders}, {and} \bibinfo{person}{Adrian Sampson}.} \bibinfo{year}{2020}\natexlab{}.
\newblock \showarticletitle{Geometry types for graphics programming}.
\newblock \bibinfo{journal}{\emph{Proceedings of the ACM on Programming Languages}} \bibinfo{volume}{4}, \bibinfo{number}{Oopsla} (\bibinfo{year}{2020}), \bibinfo{pages}{1--25}.
\newblock
\urldef\tempurl%
\url{https://doi.org/10.1145/3428241}
\showDOI{\tempurl}


\bibitem[Guy(2017)]%
        {googleio_understanding_color}
\bibfield{author}{\bibinfo{person}{Romain Guy}.} \bibinfo{year}{2017}\natexlab{}.
\newblock \bibinfo{title}{Understanding color}.
\newblock
\newblock
\urldef\tempurl%
\url{https://www.youtube.com/watch?v=r8NeG0wmFXM}
\showURL{%
\tempurl}


\bibitem[Harris et~al\mbox{.}(2020)]%
        {harris2020array}
\bibfield{author}{\bibinfo{person}{Charles~R. Harris}, \bibinfo{person}{K.~Jarrod Millman}, \bibinfo{person}{St{\'{e}}fan~J. van~der Walt}, \bibinfo{person}{Ralf Gommers}, \bibinfo{person}{Pauli Virtanen}, \bibinfo{person}{David Cournapeau}, \bibinfo{person}{Eric Wieser}, \bibinfo{person}{Julian Taylor}, \bibinfo{person}{Sebastian Berg}, \bibinfo{person}{Nathaniel~J. Smith}, \bibinfo{person}{Robert Kern}, \bibinfo{person}{Matti Picus}, \bibinfo{person}{Stephan Hoyer}, \bibinfo{person}{Marten~H. van Kerkwijk}, \bibinfo{person}{Matthew Brett}, \bibinfo{person}{Allan Haldane}, \bibinfo{person}{Jaime~Fern{\'{a}}ndez del R{\'{i}}o}, \bibinfo{person}{Mark Wiebe}, \bibinfo{person}{Pearu Peterson}, \bibinfo{person}{Pierre G{\'{e}}rard-Marchant}, \bibinfo{person}{Kevin Sheppard}, \bibinfo{person}{Tyler Reddy}, \bibinfo{person}{Warren Weckesser}, \bibinfo{person}{Hameer Abbasi}, \bibinfo{person}{Christoph Gohlke}, {and} \bibinfo{person}{Travis~E. Oliphant}.} \bibinfo{year}{2020}\natexlab{}.
\newblock \showarticletitle{Array programming with {NumPy}}.
\newblock \bibinfo{journal}{\emph{Nature}} \bibinfo{volume}{585}, \bibinfo{number}{7825} (\bibinfo{date}{Sept.} \bibinfo{year}{2020}), \bibinfo{pages}{357--362}.
\newblock
\urldef\tempurl%
\url{https://doi.org/10.1038/s41586-020-2649-2}
\showDOI{\tempurl}


\bibitem[Hartmann(2012)]%
        {hartmann_eleuthera_2012}
\bibfield{author}{\bibinfo{person}{Trish Hartmann}.} \bibinfo{year}{2012}\natexlab{}.
\newblock \bibinfo{title}{Eleuthera {Sunset}}.
\newblock
\newblock
\urldef\tempurl%
\url{https://openverse.org/image/d693e7e7-d7aa-4801-96c5-56674e5715c6}
\showURL{%
\tempurl}


\bibitem[Henderson and Hodgkiss(1963)]%
        {henderson1963spectral}
\bibfield{author}{\bibinfo{person}{ST Henderson} {and} \bibinfo{person}{D Hodgkiss}.} \bibinfo{year}{1963}\natexlab{}.
\newblock \showarticletitle{The spectral energy distribution of daylight}.
\newblock \bibinfo{journal}{\emph{British Journal of Applied Physics}} \bibinfo{volume}{14}, \bibinfo{number}{3} (\bibinfo{year}{1963}), \bibinfo{pages}{125}.
\newblock
\urldef\tempurl%
\url{https://doi.org/10.1088/0508-3443/15/8/310}
\showDOI{\tempurl}


\bibitem[Jia et~al\mbox{.}(2019)]%
        {jia2019taso}
\bibfield{author}{\bibinfo{person}{Zhihao Jia}, \bibinfo{person}{Oded Padon}, \bibinfo{person}{James Thomas}, \bibinfo{person}{Todd Warszawski}, \bibinfo{person}{Matei Zaharia}, {and} \bibinfo{person}{Alex Aiken}.} \bibinfo{year}{2019}\natexlab{}.
\newblock \showarticletitle{TASO: optimizing deep learning computation with automatic generation of graph substitutions}. In \bibinfo{booktitle}{\emph{Proceedings of the 27th ACM Symposium on Operating Systems Principles}}. \bibinfo{pages}{47--62}.
\newblock
\urldef\tempurl%
\url{https://doi.org/0.1145/3341301.3359630}
\showDOI{\tempurl}


\bibitem[Jin et~al\mbox{.}(2020)]%
        {jin2020compiling}
\bibfield{author}{\bibinfo{person}{Tian Jin}, \bibinfo{person}{Gheorghe-Teodor Bercea}, \bibinfo{person}{Tung~D Le}, \bibinfo{person}{Tong Chen}, \bibinfo{person}{Gong Su}, \bibinfo{person}{Haruki Imai}, \bibinfo{person}{Yasushi Negishi}, \bibinfo{person}{Anh Leu}, \bibinfo{person}{Kevin O'Brien}, \bibinfo{person}{Kiyokuni Kawachiya}, {et~al\mbox{.}}} \bibinfo{year}{2020}\natexlab{}.
\newblock \showarticletitle{Compiling onnx neural network models using mlir}.
\newblock \bibinfo{journal}{\emph{arXiv preprint arXiv:2008.08272}} (\bibinfo{year}{2020}).
\newblock
\urldef\tempurl%
\url{https://doi.org/10.48550/arXiv.2008.08272}
\showDOI{\tempurl}


\bibitem[Johnston-Feller(2001)]%
        {johnston2001color}
\bibfield{author}{\bibinfo{person}{Ruth Johnston-Feller}.} \bibinfo{year}{2001}\natexlab{}.
\newblock \bibinfo{booktitle}{\emph{Color science in the examination of museum objects: nondestructive procedures}}.
\newblock \bibinfo{publisher}{Getty Publications}.
\newblock
\urldef\tempurl%
\url{https://doi.org/10.1002/col.10107}
\showDOI{\tempurl}


\bibitem[Joisha and Banerjee(2006)]%
        {joisha2006algebraic}
\bibfield{author}{\bibinfo{person}{Pramod~G Joisha} {and} \bibinfo{person}{Prithviraj Banerjee}.} \bibinfo{year}{2006}\natexlab{}.
\newblock \showarticletitle{An algebraic array shape inference system for MATLAB{\textregistered}}.
\newblock \bibinfo{journal}{\emph{ACM Transactions on Programming Languages and Systems (TOPLAS)}} \bibinfo{volume}{28}, \bibinfo{number}{5} (\bibinfo{year}{2006}), \bibinfo{pages}{848--907}.
\newblock
\urldef\tempurl%
\url{https://doi.org/10.1145/1152649.1152651}
\showDOI{\tempurl}


\bibitem[Karr and Loveman(1978)]%
        {karr_incorporation_1978}
\bibfield{author}{\bibinfo{person}{Michael Karr} {and} \bibinfo{person}{David~B. Loveman}.} \bibinfo{year}{1978}\natexlab{}.
\newblock \showarticletitle{Incorporation of units into programming languages}.
\newblock  \bibinfo{volume}{21}, \bibinfo{number}{5} (\bibinfo{year}{1978}), \bibinfo{pages}{385--391}.
\newblock
\showISSN{0001-0782}
\urldef\tempurl%
\url{https://doi.org/10.1145/359488.359501}
\showDOI{\tempurl}


\bibitem[Kjolstad et~al\mbox{.}(2017)]%
        {kjolstad2017tensor}
\bibfield{author}{\bibinfo{person}{Fredrik Kjolstad}, \bibinfo{person}{Shoaib Kamil}, \bibinfo{person}{Stephen Chou}, \bibinfo{person}{David Lugato}, {and} \bibinfo{person}{Saman Amarasinghe}.} \bibinfo{year}{2017}\natexlab{}.
\newblock \showarticletitle{The tensor algebra compiler}.
\newblock \bibinfo{journal}{\emph{Proceedings of the ACM on Programming Languages}} \bibinfo{volume}{1}, \bibinfo{number}{Oopsla} (\bibinfo{year}{2017}), \bibinfo{pages}{1--29}.
\newblock
\urldef\tempurl%
\url{https://doi.org/10.1145/3133901}
\showDOI{\tempurl}


\bibitem[Kjolstad et~al\mbox{.}(2016)]%
        {kjolstad2016simit}
\bibfield{author}{\bibinfo{person}{Fredrik Kjolstad}, \bibinfo{person}{Shoaib Kamil}, \bibinfo{person}{Jonathan Ragan-Kelley}, \bibinfo{person}{David~IW Levin}, \bibinfo{person}{Shinjiro Sueda}, \bibinfo{person}{Desai Chen}, \bibinfo{person}{Etienne Vouga}, \bibinfo{person}{Danny~M Kaufman}, \bibinfo{person}{Gurtej Kanwar}, \bibinfo{person}{Wojciech Matusik}, {et~al\mbox{.}}} \bibinfo{year}{2016}\natexlab{}.
\newblock \showarticletitle{Simit: A language for physical simulation}.
\newblock \bibinfo{journal}{\emph{ACM Transactions on Graphics (TOG)}} \bibinfo{volume}{35}, \bibinfo{number}{2} (\bibinfo{year}{2016}), \bibinfo{pages}{1--21}.
\newblock
\urldef\tempurl%
\url{https://doi.org/10.1145/2866569}
\showDOI{\tempurl}


\bibitem[Kubelka(1948)]%
        {kubelka1948new}
\bibfield{author}{\bibinfo{person}{Paul Kubelka}.} \bibinfo{year}{1948}\natexlab{}.
\newblock \showarticletitle{New contributions to the optics of intensely light-scattering materials. Part I}.
\newblock \bibinfo{journal}{\emph{Josa}} \bibinfo{volume}{38}, \bibinfo{number}{5} (\bibinfo{year}{1948}), \bibinfo{pages}{448--457}.
\newblock


\bibitem[Kubelka and Munk(1931)]%
        {kubelka1931article}
\bibfield{author}{\bibinfo{person}{Paul Kubelka} {and} \bibinfo{person}{Franz Munk}.} \bibinfo{year}{1931}\natexlab{}.
\newblock \showarticletitle{An article on optics of paint layers}.
\newblock \bibinfo{journal}{\emph{Z. Tech. Phys}} \bibinfo{volume}{12}, \bibinfo{number}{593-601} (\bibinfo{year}{1931}), \bibinfo{pages}{259--274}.
\newblock


\bibitem[Lakens(2017)]%
        {lakens2017equivalence}
\bibfield{author}{\bibinfo{person}{Dani{\"e}l Lakens}.} \bibinfo{year}{2017}\natexlab{}.
\newblock \showarticletitle{Equivalence tests: A practical primer for t tests, correlations, and meta-analyses}.
\newblock \bibinfo{journal}{\emph{Social psychological and personality science}} \bibinfo{volume}{8}, \bibinfo{number}{4} (\bibinfo{year}{2017}), \bibinfo{pages}{355--362}.
\newblock
\urldef\tempurl%
\url{https://doi.org/10.1177/1948550617697177}
\showDOI{\tempurl}


\bibitem[Lam et~al\mbox{.}(2015)]%
        {numba}
\bibfield{author}{\bibinfo{person}{Siu~Kwan Lam}, \bibinfo{person}{Antoine Pitrou}, {and} \bibinfo{person}{Stanley Seibert}.} \bibinfo{year}{2015}\natexlab{}.
\newblock \showarticletitle{Numba: a LLVM-based Python JIT compiler}. In \bibinfo{booktitle}{\emph{Proceedings of the Second Workshop on the LLVM Compiler Infrastructure in HPC}} (Austin, Texas) \emph{(\bibinfo{series}{Llvm '15})}. \bibinfo{publisher}{Association for Computing Machinery}, \bibinfo{address}{New York, NY, USA}, Article \bibinfo{articleno}{7}, \bibinfo{numpages}{6}~pages.
\newblock
\showISBNx{9781450340052}
\urldef\tempurl%
\url{https://doi.org/10.1145/2833157.2833162}
\showDOI{\tempurl}


\bibitem[Lavrov(2020)]%
        {color_science_runtime_complaint_3}
\bibfield{author}{\bibinfo{person}{Dmitry Lavrov}.} \bibinfo{year}{2020}\natexlab{}.
\newblock \bibinfo{title}{Reading 65{\textasciicircum}3 {Iridas} {3D} {LUT} is incredibly slow. · {Issue} \#573 · colour-science/colour}.
\newblock
\newblock
\urldef\tempurl%
\url{https://github.com/colour-science/colour/issues/573}
\showURL{%
\tempurl}


\bibitem[Li et~al\mbox{.}(2017)]%
        {cam16}
\bibfield{author}{\bibinfo{person}{Changjun Li}, \bibinfo{person}{Zhiqiang Li}, \bibinfo{person}{Zhifeng Wang}, \bibinfo{person}{Yang Xu}, \bibinfo{person}{Ming~Ronnier Luo}, \bibinfo{person}{Guihua Cui}, \bibinfo{person}{Manuel Melgosa}, \bibinfo{person}{Michael~H Brill}, {and} \bibinfo{person}{Michael Pointer}.} \bibinfo{year}{2017}\natexlab{}.
\newblock \showarticletitle{Comprehensive color solutions: CAM16, CAT16, and CAM16-UCS}.
\newblock \bibinfo{journal}{\emph{Color Research \& Application}} \bibinfo{volume}{42}, \bibinfo{number}{6} (\bibinfo{year}{2017}), \bibinfo{pages}{703--718}.
\newblock
\urldef\tempurl%
\url{https://doi.org/10.1002/col.22131}
\showDOI{\tempurl}


\bibitem[Lindbloom(2017)]%
        {lindbloom_xyz_2017}
\bibfield{author}{\bibinfo{person}{Bruce Lindbloom}.} \bibinfo{year}{2017}\natexlab{}.
\newblock \bibinfo{title}{{XYZ} to {LAB}}.
\newblock
\newblock
\urldef\tempurl%
\url{http://www.brucelindbloom.com/index.html?Eqn\%5FXYZ\%5Fto\%5FLab.html}
\showURL{%
\tempurl}


\bibitem[Liu et~al\mbox{.}(2022)]%
        {liu2022verified}
\bibfield{author}{\bibinfo{person}{Amanda Liu}, \bibinfo{person}{Gilbert~Louis Bernstein}, \bibinfo{person}{Adam Chlipala}, {and} \bibinfo{person}{Jonathan Ragan-Kelley}.} \bibinfo{year}{2022}\natexlab{}.
\newblock \showarticletitle{Verified tensor-program optimization via high-level scheduling rewrites}.
\newblock \bibinfo{journal}{\emph{Proceedings of the ACM on Programming Languages}} \bibinfo{volume}{6}, \bibinfo{number}{Popl} (\bibinfo{year}{2022}), \bibinfo{pages}{1--28}.
\newblock
\urldef\tempurl%
\url{https://doi.org/10.1145/3498717}
\showDOI{\tempurl}


\bibitem[Marschner and Shirley(2021)]%
        {marschner2021physics}
\bibfield{author}{\bibinfo{person}{Steve Marschner} {and} \bibinfo{person}{Peter Shirley}.} \bibinfo{year}{2021}\natexlab{}.
\newblock \showarticletitle{Chapter 14.6.1 Spectral Energy}.
\newblock In \bibinfo{booktitle}{\emph{Fundamentals of Computer Graphics}}. \bibinfo{publisher}{AK Peters/CRC Press}, \bibinfo{pages}{357--382}.
\newblock
\urldef\tempurl%
\url{https://doi.org/10.1201/9781439865521}
\showDOI{\tempurl}


\bibitem[Massa(2021)]%
        {pytorch_no_type}
\bibfield{author}{\bibinfo{person}{Francisco Massa}.} \bibinfo{year}{2021}\natexlab{}.
\newblock \bibinfo{title}{[feature request] rgb2lab / rgb2hsv / rgb2gray and other color space conversions (maybe upstream from kornia? or colorsys python core module?) · {Issue} \#4029 · pytorch/vision}.
\newblock
\newblock
\urldef\tempurl%
\url{https://github.com/pytorch/vision/issues/4029}
\showURL{%
\tempurl}


\bibitem[McCurdy(2022)]%
        {threejs_color_management}
\bibfield{author}{\bibinfo{person}{Don McCurdy}.} \bibinfo{year}{2022}\natexlab{}.
\newblock \bibinfo{title}{Color management}.
\newblock
\newblock
\urldef\tempurl%
\url{https://threejs.org/docs/#manual/en/introduction/Color-management}
\showURL{%
\tempurl}


\bibitem[Miller and Spicer(2019)]%
        {miller2019color}
\bibfield{author}{\bibinfo{person}{Michael~E Miller} {and} \bibinfo{person}{Spicer}.} \bibinfo{year}{2019}\natexlab{}.
\newblock \bibinfo{booktitle}{\emph{Color in Electronic Display Systems}}.
\newblock \bibinfo{publisher}{Springer}.
\newblock
\urldef\tempurl%
\url{https://doi.org/10.1007/978-3-030-02834-3}
\showDOI{\tempurl}


\bibitem[Myers(1999)]%
        {jflow}
\bibfield{author}{\bibinfo{person}{Andrew~C. Myers}.} \bibinfo{year}{1999}\natexlab{}.
\newblock \showarticletitle{JFlow: practical mostly-static information flow control}. In \bibinfo{booktitle}{\emph{Proceedings of the 26th ACM SIGPLAN-SIGACT Symposium on Principles of Programming Languages}} (San Antonio, Texas, USA) \emph{(\bibinfo{series}{Popl '99})}. \bibinfo{publisher}{Association for Computing Machinery}, \bibinfo{address}{New York, NY, USA}, \bibinfo{pages}{228–241}.
\newblock
\showISBNx{1581130953}
\urldef\tempurl%
\url{https://doi.org/10.1145/292540.292561}
\showDOI{\tempurl}


\bibitem[Okuta et~al\mbox{.}(2017)]%
        {cupy}
\bibfield{author}{\bibinfo{person}{Ryosuke Okuta}, \bibinfo{person}{Yuya Unno}, \bibinfo{person}{Daisuke Nishino}, \bibinfo{person}{Shohei Hido}, {and} \bibinfo{person}{Crissman Loomis}.} \bibinfo{year}{2017}\natexlab{}.
\newblock \showarticletitle{CuPy: A NumPy-Compatible Library for NVIDIA GPU Calculations}. In \bibinfo{booktitle}{\emph{Proceedings of Workshop on Machine Learning Systems (LearningSys) in The Thirty-first Annual Conference on Neural Information Processing Systems (NIPS)}}.
\newblock


\bibitem[Onnx(2018)]%
        {onnx}
\bibfield{author}{\bibinfo{person}{Onnx}.} \bibinfo{year}{2018}\natexlab{}.
\newblock \bibinfo{title}{Open Neural Network Exchange}.
\newblock \bibinfo{howpublished}{\url{https://github.com/onnx/onnx}}.
\newblock


\bibitem[Pharr et~al\mbox{.}(2023)]%
        {pharr2023physically}
\bibfield{author}{\bibinfo{person}{Matt Pharr}, \bibinfo{person}{Wenzel Jakob}, {and} \bibinfo{person}{Greg Humphreys}.} \bibinfo{year}{2023}\natexlab{}.
\newblock \bibinfo{booktitle}{\emph{Physically based rendering: From theory to implementation}}.
\newblock \bibinfo{publisher}{MIT Press}.
\newblock


\bibitem[Poynton(2012)]%
        {poynton2012digital}
\bibfield{author}{\bibinfo{person}{Charles Poynton}.} \bibinfo{year}{2012}\natexlab{}.
\newblock \bibinfo{booktitle}{\emph{Digital video and HD: Algorithms and Interfaces}}.
\newblock \bibinfo{publisher}{Elsevier}.
\newblock
\urldef\tempurl%
\url{https://doi.org/10.1016/B978-0-12-391926-7.50059-X}
\showDOI{\tempurl}


\bibitem[Rowlands(2017)]%
        {rowlands2017physics}
\bibfield{author}{\bibinfo{person}{Andy Rowlands}.} \bibinfo{year}{2017}\natexlab{}.
\newblock \bibinfo{booktitle}{\emph{Physics of digital photography}}.
\newblock \bibinfo{publisher}{IOP Publishing}.
\newblock
\urldef\tempurl%
\url{https://doi.org/10.1088/978-0-7503-2558-5}
\showDOI{\tempurl}


\bibitem[Rowlands(2020)]%
        {rowlands2020color}
\bibfield{author}{\bibinfo{person}{D~Andrew Rowlands}.} \bibinfo{year}{2020}\natexlab{}.
\newblock \showarticletitle{Color conversion matrices in digital cameras: a tutorial}.
\newblock \bibinfo{journal}{\emph{Optical Engineering}} \bibinfo{volume}{59}, \bibinfo{number}{11} (\bibinfo{year}{2020}), \bibinfo{pages}{110801}.
\newblock
\urldef\tempurl%
\url{https://doi.org/10.1117/1.oe.59.11.110801}
\showDOI{\tempurl}


\bibitem[Sabelfeld and Myers(2003)]%
        {info-flow-survey}
\bibfield{author}{\bibinfo{person}{A. Sabelfeld} {and} \bibinfo{person}{A.C. Myers}.} \bibinfo{year}{2003}\natexlab{}.
\newblock \showarticletitle{Language-based information-flow security}.
\newblock \bibinfo{journal}{\emph{IEEE Journal on Selected Areas in Communications}} \bibinfo{volume}{21}, \bibinfo{number}{1} (\bibinfo{year}{2003}), \bibinfo{pages}{5--19}.
\newblock
\urldef\tempurl%
\url{https://doi.org/10.1109/jsac.2002.806121}
\showDOI{\tempurl}


\bibitem[Sampson et~al\mbox{.}(2011)]%
        {enerj}
\bibfield{author}{\bibinfo{person}{Adrian Sampson}, \bibinfo{person}{Werner Dietl}, \bibinfo{person}{Emily Fortuna}, \bibinfo{person}{Danushen Gnanapragasam}, \bibinfo{person}{Luis Ceze}, {and} \bibinfo{person}{Dan Grossman}.} \bibinfo{year}{2011}\natexlab{}.
\newblock \showarticletitle{EnerJ: approximate data types for safe and general low-power computation}.
\newblock \bibinfo{journal}{\emph{SIGPLAN Not.}} \bibinfo{volume}{46}, \bibinfo{number}{6} (\bibinfo{date}{jun} \bibinfo{year}{2011}), \bibinfo{pages}{164–174}.
\newblock
\showISSN{0362-1340}
\urldef\tempurl%
\url{https://doi.org/10.1145/1993316.1993518}
\showDOI{\tempurl}


\bibitem[Sharma(2017)]%
        {sharma2017color}
\bibfield{author}{\bibinfo{person}{Gaurav Sharma}.} \bibinfo{year}{2017}\natexlab{}.
\newblock \showarticletitle{Color fundamentals for digital imaging}.
\newblock In \bibinfo{booktitle}{\emph{Digital color imaging handbook}}. \bibinfo{publisher}{CRC press}, \bibinfo{pages}{1--114}.
\newblock


\bibitem[Sharma and Bala(2017)]%
        {sharma2017digital}
\bibfield{author}{\bibinfo{person}{Gaurav Sharma} {and} \bibinfo{person}{Raja Bala}.} \bibinfo{year}{2017}\natexlab{}.
\newblock \bibinfo{booktitle}{\emph{Digital color imaging handbook}}.
\newblock \bibinfo{publisher}{CRC press}.
\newblock
\showISBNx{0-8493-0900-x}


\bibitem[Slepak et~al\mbox{.}(2014)]%
        {slepak2014array}
\bibfield{author}{\bibinfo{person}{Justin Slepak}, \bibinfo{person}{Olin Shivers}, {and} \bibinfo{person}{Panagiotis Manolios}.} \bibinfo{year}{2014}\natexlab{}.
\newblock \showarticletitle{An array-oriented language with static rank polymorphism}. In \bibinfo{booktitle}{\emph{Programming Languages and Systems: 23rd European Symposium on Programming, ESOP 2014, Held as Part of the European Joint Conferences on Theory and Practice of Software, ETAPS 2014, Grenoble, France, April 5-13, 2014, Proceedings 23}}. Springer, \bibinfo{pages}{27--46}.
\newblock
\urldef\tempurl%
\url{https://doi.org/10.1007/978-3-642-54833-8_3}
\showDOI{\tempurl}


\bibitem[Sochorov{\'a} and Jamri{\v{s}}ka(2021)]%
        {sochorova2021practical}
\bibfield{author}{\bibinfo{person}{{\v{S}}{\'a}rka Sochorov{\'a}} {and} \bibinfo{person}{Ond{\v{r}}ej Jamri{\v{s}}ka}.} \bibinfo{year}{2021}\natexlab{}.
\newblock \showarticletitle{Practical pigment mixing for digital painting}.
\newblock \bibinfo{journal}{\emph{ACM Transactions on Graphics (TOG)}} \bibinfo{volume}{40}, \bibinfo{number}{6} (\bibinfo{year}{2021}), \bibinfo{pages}{1--11}.
\newblock
\urldef\tempurl%
\url{https://doi.org/10.1145/3478513.3480549}
\showDOI{\tempurl}


\bibitem[Stefan(2017)]%
        {scikit_no_types}
\bibfield{author}{\bibinfo{person}{van der~Walt Stefan}.} \bibinfo{year}{2017}\natexlab{}.
\newblock \bibinfo{title}{Finding color space information about image · {Issue} \#2175 · scikit-image/scikit-image}.
\newblock
\newblock
\urldef\tempurl%
\url{https://github.com/scikit-image/scikit-image/issues/2175}
\showURL{%
\tempurl}


\bibitem[Stockman and Brainard(2010)]%
        {stockman2010color}
\bibfield{author}{\bibinfo{person}{Andrew Stockman} {and} \bibinfo{person}{David~H Brainard}.} \bibinfo{year}{2010}\natexlab{}.
\newblock \showarticletitle{Color vision mechanisms}.
\newblock \bibinfo{journal}{\emph{The Optical Society of America Handbook of Optics}}  \bibinfo{volume}{3} (\bibinfo{year}{2010}), \bibinfo{pages}{11--1}.
\newblock


\bibitem[Susungi et~al\mbox{.}(2018)]%
        {susungi2018meta}
\bibfield{author}{\bibinfo{person}{Adilla Susungi}, \bibinfo{person}{Norman~A Rink}, \bibinfo{person}{Albert Cohen}, \bibinfo{person}{Jeronimo Castrillon}, {and} \bibinfo{person}{Claude Tadonki}.} \bibinfo{year}{2018}\natexlab{}.
\newblock \showarticletitle{Meta-programming for cross-domain tensor optimizations}.
\newblock \bibinfo{journal}{\emph{ACM SIGPLAN Notices}} \bibinfo{volume}{53}, \bibinfo{number}{9} (\bibinfo{year}{2018}), \bibinfo{pages}{79--92}.
\newblock
\urldef\tempurl%
\url{https://doi.org/10.1145/3278122.3278131}
\showDOI{\tempurl}


\bibitem[Tate et~al\mbox{.}(2009)]%
        {tate2009equality}
\bibfield{author}{\bibinfo{person}{Ross Tate}, \bibinfo{person}{Michael Stepp}, \bibinfo{person}{Zachary Tatlock}, {and} \bibinfo{person}{Sorin Lerner}.} \bibinfo{year}{2009}\natexlab{}.
\newblock \showarticletitle{Equality saturation: a new approach to optimization}. In \bibinfo{booktitle}{\emph{Proceedings of the 36th annual ACM SIGPLAN-SIGACT symposium on Principles of programming languages}}. \bibinfo{pages}{264--276}.
\newblock
\urldef\tempurl%
\url{https://doi.org/10.1145/1480881.1480915}
\showDOI{\tempurl}


\bibitem[Varkor(2018)]%
        {varkor_types_2018}
\bibfield{author}{\bibinfo{person}{Varkor}.} \bibinfo{year}{2018}\natexlab{}.
\newblock \bibinfo{title}{Types for units of measure}.
\newblock
\newblock
\urldef\tempurl%
\url{https://varkor.github.io/blog/2018/07/30/types-for-units-of-measure.html}
\showURL{%
\tempurl}


\bibitem[Vasilache et~al\mbox{.}(2018)]%
        {vasilache2018tensor}
\bibfield{author}{\bibinfo{person}{Nicolas Vasilache}, \bibinfo{person}{Oleksandr Zinenko}, \bibinfo{person}{Theodoros Theodoridis}, \bibinfo{person}{Priya Goyal}, \bibinfo{person}{Zachary DeVito}, \bibinfo{person}{William~S Moses}, \bibinfo{person}{Sven Verdoolaege}, \bibinfo{person}{Andrew Adams}, {and} \bibinfo{person}{Albert Cohen}.} \bibinfo{year}{2018}\natexlab{}.
\newblock \showarticletitle{Tensor comprehensions: Framework-agnostic high-performance machine learning abstractions}.
\newblock \bibinfo{journal}{\emph{arXiv preprint arXiv:1802.04730}} (\bibinfo{year}{2018}).
\newblock
\urldef\tempurl%
\url{https://doi.org/10.48550/arXiv.1802.04730}
\showDOI{\tempurl}


\bibitem[Vi{\'e}not et~al\mbox{.}(1999)]%
        {vienot1999digital}
\bibfield{author}{\bibinfo{person}{Fran{\c{c}}oise Vi{\'e}not}, \bibinfo{person}{Hans Brettel}, {and} \bibinfo{person}{John~D Mollon}.} \bibinfo{year}{1999}\natexlab{}.
\newblock \showarticletitle{Digital video colourmaps for checking the legibility of displays by dichromats}.
\newblock \bibinfo{journal}{\emph{Color Research \& Application}} \bibinfo{volume}{24}, \bibinfo{number}{4} (\bibinfo{year}{1999}), \bibinfo{pages}{243--252}.
\newblock
\urldef\tempurl%
\url{https://doi.org/10.1002/(SICI)1520-6378(199908)24:4<243::AID-COL5>3.0.CO;2-3}
\showDOI{\tempurl}


\bibitem[Wandell(1995)]%
        {wandell1995foundations}
\bibfield{author}{\bibinfo{person}{Brian~A Wandell}.} \bibinfo{year}{1995}\natexlab{}.
\newblock \bibinfo{booktitle}{\emph{Foundations of vision.}}
\newblock \bibinfo{publisher}{sinauer Associates}.
\newblock


\bibitem[Willsey et~al\mbox{.}(2021)]%
        {willsey2021egg}
\bibfield{author}{\bibinfo{person}{Max Willsey}, \bibinfo{person}{Chandrakana Nandi}, \bibinfo{person}{Yisu~Remy Wang}, \bibinfo{person}{Oliver Flatt}, \bibinfo{person}{Zachary Tatlock}, {and} \bibinfo{person}{Pavel Panchekha}.} \bibinfo{year}{2021}\natexlab{}.
\newblock \showarticletitle{Egg: Fast and extensible equality saturation}.
\newblock \bibinfo{journal}{\emph{Proceedings of the ACM on Programming Languages}} \bibinfo{volume}{5}, \bibinfo{number}{Popl} (\bibinfo{year}{2021}), \bibinfo{pages}{1--29}.
\newblock
\urldef\tempurl%
\url{https://doi.org/10.1145/3434304}
\showDOI{\tempurl}


\bibitem[Wyszecki and Stiles(2000)]%
        {wyszecki2000color}
\bibfield{author}{\bibinfo{person}{G{\"u}nther Wyszecki} {and} \bibinfo{person}{Walter~Stanley Stiles}.} \bibinfo{year}{2000}\natexlab{}.
\newblock \bibinfo{booktitle}{\emph{Color science: concepts and methods, quantitative data and formulae}}. Vol.~\bibinfo{volume}{40}.
\newblock \bibinfo{publisher}{John wiley \& sons}.
\newblock


\bibitem[Yang et~al\mbox{.}(2021)]%
        {tensat}
\bibfield{author}{\bibinfo{person}{Yichen Yang}, \bibinfo{person}{Phitchaya Phothilimthana}, \bibinfo{person}{Yisu Wang}, \bibinfo{person}{Max Willsey}, \bibinfo{person}{Sudip Roy}, {and} \bibinfo{person}{Jacques Pienaar}.} \bibinfo{year}{2021}\natexlab{}.
\newblock \showarticletitle{Equality saturation for tensor graph superoptimization}.
\newblock \bibinfo{journal}{\emph{Proceedings of Machine Learning and Systems}}  \bibinfo{volume}{3} (\bibinfo{year}{2021}), \bibinfo{pages}{255--268}.
\newblock


\end{thebibliography}

\end{document}